\documentclass[11pt]{article}
\usepackage{fancyhdr}
\usepackage{isomath}
\usepackage{amsmath}
\usepackage{amsopn}
\usepackage{amsbsy}
\usepackage{amssymb}
\usepackage{amscd}
\usepackage{amsfonts}
\usepackage{graphicx,color}
\usepackage{verbatim}
\usepackage{euscript}
\usepackage{alltt}
\usepackage{float}
\usepackage{xy}
\usepackage{stmaryrd}
\usepackage{setspace}
\usepackage{pdfpages}
\usepackage[usenames,dvipsnames]{xcolor}
\usepackage{enumitem}
\usepackage{placeins}
\usepackage{standalone}
\usepackage{import}
\usepackage[square,numbers,sectionbib]{natbib}
\usepackage{chapterbib}
\usepackage{titletoc}% http://ctan.org/pkg/titletoc
\usepackage{multirow}
\makeatletter
\newcommand*{\rom}[1]{\expandafter\@slowromancap\romannumeral #1@}
\makeatother
\usepackage{graphicx,color}
\usepackage[parfill]{parskip}
\usepackage[utf8]{inputenc}
\usepackage[english]{babel}
\usepackage{xy}
\usepackage[margin=1in]{geometry}
\usepackage{caption}
\usepackage{subcaption}
\usepackage{url}
\usepackage[titletoc,title]{appendix}
\numberwithin{equation}{section}
\usepackage{float}
\restylefloat{table}
\usepackage{setspace}
\usepackage{xcolor}
\usepackage{enumitem}

\usepackage{grffile}
\usepackage{breqn}
\usepackage{titlesec}

\usepackage{tikz}
\usetikzlibrary{shapes.geometric, arrows}
\usepackage[splitrule]{footmisc}
\usepackage{amsmath}
\usepackage{amsbsy}
\usepackage{amssymb}
\usepackage{amscd}
\usepackage{amsfonts}

\usepackage{isomath}
\usepackage{amsmath}
\usepackage{amsbsy}
\usepackage{amssymb}
\usepackage{amscd}
\usepackage{amsfonts}
\usepackage{graphicx}
\usepackage{verbatim}
\usepackage{euscript}
\usepackage{alltt}

\newcommand{\bfb}{{\mathbold b}}
\newcommand{\bfc}{{\mathbold c}}
\newcommand{\bfd}{{\mathbold d}}
\newcommand{\bfe}{{\mathbold e}}

\newcommand{\bfl}{{\mathbold l}}

\newcommand{\bfn}{{\mathbold n}}

\newcommand{\bft}{{\mathbold t}}
\newcommand{\bfu}{{\mathbold u}}
\newcommand{\bfv}{{\mathbold v}}
\newcommand{\bfw}{{\mathbold w}}
\newcommand{\bfx}{{\mathbold x}}

\newcommand{\bfz}{{\mathbold z}}

\newcommand{\bfA}{{\mathbold A}}
\newcommand{\bfB}{{\mathbold B}}
\newcommand{\bfC}{{\mathbold C}}

\newcommand{\bfL}{{\mathbold L}}

\newcommand{\bfP}{{\mathbold P}}

\newcommand{\bfR}{{\mathbold R}}
\newcommand{\bfS}{{\mathbold S}}
\newcommand{\bfT}{{\mathbold T}}
\newcommand{\bfU}{{\mathbold U}}
\newcommand{\bfV}{{\mathbold V}}

\newcommand{\bfX}{{\mathbold X}}

\newcommand{\beq}{\begin{equation}}
\newcommand{\eeq}{\end{equation}}
\newcommand{\beqs}{\begin{eqnarray}}
\newcommand{\eeqs}{\end{eqnarray}}
\newcommand{\beql}{\begin{equation} \label}
\newcommand{\half}{\frac{1}{2}}

\newcommand{\bfchi}{\mathbold{\chi}}

\newcommand{\bfsigma}{\mathbold{\sigma}}
\newcommand{\bfepsilon}{\mathbold{\epsilon}}

\newcommand{\bfalpha}{\mathbold{\alpha}}

\graphicspath{{./imagedir/}}

\tikzstyle{process} = [rectangle, minimum width=3cm, minimum height=1cm, text width=13 cm, text centered, draw=black] %, fill=orange!30]
\tikzstyle{arrow} = [thick,->,>=stealth]

\DeclareMathOperator*{\argmin}{argmin}

\colorlet{change}{blue}

\title{Plasticity without phenomenology: a first step}

\author{Sabyasachi Chatterjee\thanks{Dept.\ of Civil \& Environmental Engineering, Carnegie Mellon University, Pittsburgh, PA 15213. sabyasac@andrew.cmu.edu.} 
\and Giacomo Po\thanks{Dept. of Mechanical and Aerospace Engineering, University of Miami, Coral Gables, FL 33146. gpo@miami.edu.}
\and Xiaohan Zhang\thanks{Senior Data Scientist, Salesforce.com, Mountain View, CA 94105. xiaohanzhang.cmu@gmail.com.}
\and Amit Acharya\thanks{Dept.\ of Civil \& Environmental Engineering, and Center for Nonlinear Analysis, Carnegie Mellon University, Pittsburgh, PA 15213. acharyaamit@cmu.edu.} 
\and Nasr Ghoniem\thanks{Dept.\ of Mechanical and Aerospace Engineering, University of California, Los Angeles, CA 90095. ghoniem@ucla.edu.}}

\begin{document}
\date{}
\maketitle

\begin{abstract}
\noindent A novel, concurrent multiscale approach to meso/macroscale plasticity is demonstrated. It utilizes a carefully designed coupling of a partial differential equation (pde) based theory of dislocation mediated crystal plasticity with time-averaged inputs from microscopic Dislocation Dynamics (DD), adapting a state-of-the-art mathematical coarse-graining scheme. The stress-strain response of mesoscopic samples at \emph{realistic}, slow, loading rates up to appreciable values of strain is obtained, with significant speed-up in compute time compared to conventional DD. Effects of crystal orientation, loading rate, and the ratio of the initial mobile to sessile dislocation density on the macroscopic response, for both load and displacement controlled simulations are demonstrated. These results are obtained \emph{without using any phenomenological constitutive assumption}, except for thermal activation which is not a part of microscopic DD. The results also demonstrate the effect of the internal stresses on the collective behavior of dislocations, manifesting, in a set of examples, as a Stage I to Stage II hardening transition.
\end{abstract}

\newcommand{\dbtilde}[1]{\accentset{\approx}{#1}}

\section{Introduction}\label{coupling:intro}
We develop and demonstrate a predictive computational tool for microstructure-sensitive mechanical analysis of metallic components subjected to stress and deformation. This is achieved by coupling a realization of Discrete Dislocation Dynamics \cite{po_ghoniem_2014} with a pde based model of meso-macroscopic dislocation mediated crystal plasticity \cite{acharya_roy_2006, Acharya2011} through a coarse-graining scheme for nonlinear ordinary differential equation (ode) called Practical Time Averaging (PTA), the latter described in detail in \cite{cspde_2018}. The challenge is the computation of the plastic strength and associated microstructure at the meso and macroscale \emph{at realistic time scales}, directly from the underlying motion of crystal defects, without using constitutive assumptions. The pde based theory - Mesoscale Field Dislocation Mechanics (MFDM) - contains well-defined place-holders for microscopic dislocation dynamics based input. These inputs are prescribed by a carefully designed coupling, on the `slow' time-scale of meso-macro response, with time-averaged response of `fast', local (on the macroscopic scale) discrete DD simulations. 

The rationale behind using a coupled approach instead of a completely DD based approach is the vast separation in time-scales between plasticity applications that operate at quasi-static loading rates and DD. Thus, it would be impractical to reach appreciable strains using DD alone. Therefore, we apply a modern theory for singularly perturbed ode systems to generate inputs for MFDM from DD. Within this approach, we have been able to obtain the stress-strain response of macroscopic samples at realistic loading rates up to appreciable values of strain, without using any phenomenlogical assumptions beyond those implicit in DD methodology itself (except for thermal activation which is not part of the adopted microscopic model, i.e., DD), and with significant speedup in compute time. This would not be possible using conventional DD alone. Furthermore, our work is fully three-dimensional.

The primary question in coupling dislocation-dynamics with a continuum theory of plasticity is  the determination of the minimum set of space-time averaged variables to be used in the continuum theory that allows capturing the evolution of these average variables \emph{purely in terms of themselves}. Towards achieving this ideal goal, given a large volume $V$, we decompose it into sub-volumes $V_{i}$ (which are called `blocks' as will be explained in Section \ref{sec:ddmfdm_coupling}) and in each sub-volume, a DD box is considered. Space-time averages need to be computed of the fast DD response in the DD box, to couple it with continuum theory. A simpler case to consider is when we assume that the large domain is composed of only one sub-volume/block and we ignore the spatial-averaging. These steps of computing the relevant space-time-averages are further explained in the body of this paper. A primary issue to understand is that we will generally be interested in time-averages of nonlinear state functions and this is not the same thing as evaluating the state functions on time-averages of the state itself. 

The paper is organized as follows. In Section \ref{sec:review}, a literature review of previous work on continuum theory of discrete dislocations is presented. In Section \ref{sec:PTAFORDD}, we briefly describe DD and the constitutive assumption of thermal activation of obstacles that we utilize in this work. This is followed with the definition of coarse variables and their evolution equations. In Section \ref{sec:coarse_grain_dd}, we discuss the setup and outline the algorithm for coarse-graining DD simulations in time using PTA and present results for two loading cases. The pde based model of MFDM is described in Section \ref{sec:ddfdm_coupling}. In Section \ref{sec:ddmfdm_coupling}, we discuss the algorithm for coupling MFDM with DD. This is followed with results obtained using the coupled DD-MFDM strategy. The paper ends with some concluding remarks in Section \ref{sec:coupling_disc}. 

\subsection{Literature review}\label{sec:review}
Plastic deformation of metals depends primarily on the motion and interaction of dislocations. A main goal of crystal plasticity is to develop continuum constitutive relations from the underlying dynamics of a system of discrete dislocations. A statistical approach for the kinetic evolution of idealized dislocation systems on a single slip system in 2-d has been developed. Groma and collaborators \cite{groma_1997, groma1999investigation} derived a continuum description for a system of straight parallel dislocations from the equations of motion of individual dislocations, work that has also received mathematically rigorous attention, see, e.g., \cite{briani2009time,garroni2020convergence}.  A primary result of \cite{garroni2020convergence} is that the core radius has to go to 0 at a slower rate than the rate at which the  number of dislocations go to $\infty$ for the Groma-Balogh equations to result as the limiting set of continuum evolution equations from microscopic 2-D dislocation dynamics. `Short range' dislocation interactions, interpreted as the effect of dislocation dipoles with small separation, are neglected in all of the above results. El-Azab \cite{elazab_2000, elazab_2006} developed a continuum description of the dynamics of a system of curved dislocation in 3D using a different statistical mechanics framework. This work suffers from an inadequate accounting, at the mesoscale, of the connectedness of dislocation lines, a shortcoming that has been remedied in later work \cite{xia_elazab} that does not have a  statistical mechanical underpinning. Groma, Zaiser and Csikor \cite{groma_2003} demonstrated the influence of short range dislocation-dislocation correlations by a local flow stress which scales like the square root of dislocation density and a plastic strain gradient term, introduced on an ad-hoc basis, motivated by spatial correlations of 2-d straight discrete dislocation distributions at equilibrium.

Hochrainer et al. \cite{hoch_sand_2014} developed Continuum Dislocation Dynamics (CDD) which consists of solving a complicated set of evolution equations of internal variables for each slip system. This system is derived, by averaging over the line direction variable, from a kinetic theory like description for line direction and curvature probability density functions (\cite{hoch_sand_2014}). The evolution equations for these density functions, i.e. the microscopic dynamics, are \textit{postulated}, much like in the kinetic theory of gases, \emph{without being derived from discrete dislocation dynamics}; thus such a model accounts for dislocation interactions in an approximate manner, much like the restrictions posed by collision operator approximations in the kinetic theory of gases, and such approximations taking into account dislocation interactions, even in the most rudimentary ways, has not appeared in the so-called `kinematically-closed' versions of CDD. CDD also does not include physics of dislocation interactions on different slip systems and out of plane motion of dislocations. 

Berdichevsky \cite{berdichevsky_2019} developed a phenomenological thermodynamic framework for plastic deformation in FCC metals at slow strain rates and temperature. The theory involves thermodynamic parameters like dislocation polarization (Kroupa's dislocation loop density \cite{kroupa_1962}), and new ideas of entropy and temperature of microstructure. Constitutive assumptions are required, including a history-\emph{dependent} elastic energy density dependent on the difference of (history-dependent) total strain and the (history-independent) polarization\footnote{We note that there are many examples of dislocation distributions, without involving boundary segments, which can arise from two different histories of plastic deformation due to dislocation motion and, consequently, total strain histories.}. The framework is motivated by the study of a set of edge dislocations in 2D \cite{soutyrine2018statistical}, where states encountered in the `evolution' are explicitly restricted to local equilibrium states.  The stress strain curves show intervals of slow deformation followed by slip avalanches. Under the assumed protocol for evolution of the discrete dislocation assembly, it is found that practically all dissipation is generated at avalanches.

Kooiman et al. \cite{kooiman2015effective,kooiman2016viscoplastic} assume the GENERIC framework of Grmela and Ottinger \cite{ottinger2005beyond} to describe the microscopic behavior of dislocation assemblies, which requires \emph{defining/assuming} energy and entropy functionals for the microscopic system. On that basis, and with further simplifying assumptions about dislocation dynamics, they arrive at a power-law stress exponent for effective dislocation velocity of $3.7 > 1.0$, the latter embodied in the microscopic Peach-Koehler force of individual dislocations.

Yasin, Zbib and Khaleed \cite{yasin_zbib_2001} developed a numerical model coupling 3D discrete dislocation dynamics with a continuum finite element model in which the plastic strain rate is obtained from DD. However they do not develop the theoretical and computational infrastructure for averaging in time, so their coupled theory in effect operates at the time scale of DD. Using the superposition principle, dislocation-surface interactions are computed numerically which are shown to have effects on the results. Zbib, Rubia and Bulatov \cite{zbib_rubia_2002} used a similar hybrid continuum-discrete framework to investigate a wide range of small scale plasticity phenomena such as formation of deformation bands and surface distortions under dynamic loading conditions. Groh and Zbib \cite{groh_zbib_2009} reviewed the use of dislocation dynamics to replace the constitutive equations in continuum plasticity models. They also addressed issues related to image stresses when dislocations exist in in finite volumes.

Lemarchand \cite{lemarchand_2001} proposed the Discrete-Continuum Model (DCM) which is similar to the approach followed by Zbib et. al. (\cite{yasin_zbib_2001}, \cite{zbib_rubia_2002}) in the sense that it uses a coupled DD-finite element approach in which DD is used as a substitute for the constitutive form used in usual finite element frameworks, while the finite element code is used to test the conditions of mechanical equilibrium. However, the difference in this approach from Zbib et. al. (\cite{yasin_zbib_2001}, \cite{zbib_rubia_2002}) is that the stress at the Gauss points of the finite element mesh are interpolated to the midpoint of the dislocation segments to solve for the motion of dislocation segments. This is different from the approach in Zbib et. al. (\cite{yasin_zbib_2001}, \cite{zbib_rubia_2002}) in which the dislocation-dislocation interaction is computed for all dislocations present in the same element to obtain a homogenized internal stress, while the stress induced by dislocation segments not present in the same element is obtained using a multipole expansion.

Acharya and Roy \cite{acharya_roy_2006} proposed Phenomenological Mesoscale Field Dislocation Mechanics to study initial-boundary value problem of small-scale plasticity. It is obtained by space-time averaging of the equations of Field Dislocation Mechanics (FDM) to obtain MFDM, and phenomenologically specifying some of its `non-closed' inputs. These inputs are a model of (local) space-time averaged plastic strain rate due to dislocations which are averaged out (statistically stored dislocations or SSDs) and similar averages of the microscopic, vectorial dislocation velocity. The resulting coarse model has only one extra material parameter over and above macroscopic continuum plasticity. Finite-element based computational predictions of this theory are presented in \cite{acharya_roy_2006, puri_multicrystalline, arora_2019}, where size effects, strong inhomogeneity in simple shear of plastically constrained grains and non-locality in elastic straining leading to Bauschinger effect are demonstrated.
\section{PTA for DD simulation}
\label{sec:PTAFORDD}
The framework and implementation of a scheme called Practical Time Averaging (PTA) which is used to coarse-grain nonlinear ordinary differential equations in time is discussed in detail in \cite{cspde_2018}. Here, we discuss why this scheme is relevant for coarse graining DD simulations in time and then describe some specifics of its application to DD. To proceed, we briefly describe DD following  \cite{po_ghoniem_2014}. 

\subsection{Discrete Dislocation Dynamics}\label{sec:DD}

Discrete Dislocation Dynamics (DD) refers to the collective dynamics of dislocation ensembles which is used to predict plastic properties of materials. The goal of DD is to evolve a dislocation configuration based on the local stress. This includes self-stress of the loop, the stress due to other loops and other sources of stress, including externally applied stress. The Cauchy stress tensor due to a dislocation loop \cite{eshelby_1957} is given by
\begin{align*}
\sigma_{ij}= C_{ijkl} \big(  u_{k,l} - {\beta}^P_{kl} \big) =  S_{ijkl} \bfR * \alpha_{kl},
\end{align*}
where $C$ is the fourth order, possibly anisotropic tensor of spatially constant linear elastic moduli, $\bfu$ is the displacement field due to the dislocation loop, $\beta^P$ is the plastic distortion tensor, $\bfS$ is a linear differential operator acting on the Euclidean distance $\bfR$ (given by $\bfR=\bfx-\bfx'$, where $\bfx$ is the point where the stress field is being calculated and $\bfx'$ is a point on the dislocation loop), $\bfalpha$ is the dislocation density tensor and the symbol $*$ indicates convolution in three dimensional space. The force acting on a dislocation segment of infinitesimal length $d \ell$ due to the stress field is given by $df_k=\epsilon_{kjm} \sigma_{ji} b_i d\ell_m$ and is called the \emph{Peach-K\"{o}ehler} force (\cite{peach_koehler}). 

The velocity field $\bfw$ is defined on the dislocation curves, and discretely on the nodes that discretize the curves. It is obtained by the solution of the following: 
\begin{align}\label{eq:DD_evol}
\oint_{\mathcal{L}} \Big[ \tilde{w_i}  B_{ij} w_j  + \tilde{\lambda_2} \epsilon_{ijk} w_i b_j \hat{\xi_k} \Big] d\ell=\oint_{\mathcal{L}} \Big[ \tilde{w_i} \Big( \frac{1}{\theta} \epsilon_{ijk} \sigma_{jm} b_m \hat{\xi_k}  - \lambda_2 \epsilon_{ijk} b_j \hat{\xi_k} \Big)  \Big] d\ell,
\end{align} 
which must be satisfied for arbitrary variations $\tilde{w_i}$ and $\tilde{\lambda_2}$. Here, $\bfsigma$ is the Cauchy stress tensor, $\bfb$ is the Burgers vector of the dislocation loop, $\boldsymbol \xi$ is the unit tangent to the dislocation line, $\bfB$ is a positive definite tensor subject to Onsager's symmetry relations, $\bfepsilon$ is the third order alternating tensor, $\lambda_2$ is the chemical force preventing climb, $\mathcal{L}$ is the closed line bounding any surface spanned by the dislocation loop during its motion, $\theta$ is the absolute temperature and $d \ell$ is the length of infinitesimal segment of $\mathcal{L}$. 

The position of each node $p$ is updated as 
\[
\bfx_p (t+\Delta t)= \bfx_p(t) + \bfw_p(t) \Delta t, 
\]
where $t$ is the current time and $\Delta t$ is the DD time step.

\subsection{Thermal activation}\label{sec:thermal_activation} Discrete Dislocation Dynamics is described in Section \ref{sec:DD}.  However, when we use DD, we face a problem which is described next. The local plastic distortion rate $\bfL^p_{seg}$ produced by the motion of a dislocation segment is given by $\bfL^p_{seg}=\frac{\bfb}{A} \otimes ( \hat{\bfl} \times {\bfV})$, where $\bfb$ is the Burgers vector, $A$ is the core cross-section area, $\hat{\bfl}$ is the line direction and $\bfV$ is the velocity of the segment (denoted as $\bfw$ in \eqref{eq:DD_evol}). If a single straight dislocation running from one boundary to another of the DD simulation box is considered and its motion is unimpeded by any obstacles and driven only by the applied stress, then this stress determines the magnitude of $\bfV$ in the expression for $\bfL^p_{seg}$ (accounting for phonon drag). The value of $|\bfL^p_{seg}|$ due to such a segment, at an applied stress of $10\, MPa$, is around $10^{11} s^{-1}$, which is extremely high. 

In order to approach realistic magnitudes of  strain rates under slow loading, let the DD box be populated with many straight mobile and sessile dislocation segments running from boundary to boundary of the box. The setup and its justification are provided later in Section \ref{subsec:dd_setup}. The mobile segments move and intersect with the sessile segments and such an intersection is called a junction (to be precise, it should be called a sessile junction because this type of junction does not move). The volume averaged plastic distortion rate is given by ${\bfL}^p_{avg}={1 \over |B_x|} \sum {\bfL^{p,i}_{seg}} \, l_i ~ A_i$, where $l_i$ and $A_i$ are the length of the segment and area of core cross section (see Fig. \ref{fig:disloc_segment}) of dislocation segment $i$ respectively, $\bfL^{p,i}_{seg}$ is the local plastic strain rate (defined as $\bfL^p_{seg}$ above) produced by dislocation segment $i$, and $|B_x|$ is the volume of the DD simulation box. When the dislocations are moving freely, the volume averaging reduces the magnitude of the volume averaged plastic distortion rate to around $10^3 s^{-1}$. However, even at very realistic, practical values of applied stress, the configuration gets stuck, i.e. there is no dislocation motion and $|\bfL^p_{avg}|$ is found to vanish. This is due to the high sessile density and low mean spacing between obstacles, so that the applied stress necessary for the mobile segments to break past barriers (the junctions formed at the intersection of mobile and sessile segments) is much higher than the applied stress. So, $|\bfL^p_{avg}|$ is $0$ or $10^{3} s^{-1}$, and nothing in between. 

Therefore, we implement thermal activation of obstacles by breaking junctions (intersection of mobile and sessile dislocation segments) randomly in time, a physically realistic feature of plasticity at relatively low stresses which has the effect of reducing the time-averaged value of $|\bfL^p_{avg}|$. This is not a part of conventional DD explained in Section \ref{sec:DD} but is an important constitutive assumption in our approach, based in the modeling of reality. A dislocation is an arrangement of an atomic configuration that is constantly jiggling and when there is enough temperature - i.e. kinetic energy of atomic motion - coordinated motions can happen for a dislocation to break past barriers. This can be addressed somewhat fundamentally using  Transition State theory and Molecular Dynamics, ideas that have been used in developing the phenomenology of kinetics of plasticity \cite{kocks19thermodynamics, kocks_mecking_2003}. We defer the fundamental modeling of thermal activation for later work, adopting in its place the simplest possible constitutive assumption to qualitatively represent it, as described below. 

The breaking time, $t_b$, of a junction is the elapsed time between its formation and its breaking. In the absence of a fundamental characterization of thermal activation from MD, we adopt a very simple functional form for $t_b$:
\begin{align}\label{eq:tb}
t_b=f~a, 
\end{align}
where $a$ is the maximum breaking time (in the results presented in Section \ref{sec:sgp_results} and Section \ref{sec:coupling_results}, $a$ was set as $10^{-3} \,s.$) and $f$ is a fraction generated using a uniformly distributed floating point random number generator. The corresponding attempt frequency of junction breaking may be defined as $\frac{1}{a}$.

With thermal activation enabled (with an attempt frequency of $10^3 s^{-1}.$), the time-averaged value of $|\bfL^p_{avg}|$ comes out as $10^{-2} s^{-1}$. It is important to note that the timescale set by the time-averaged value of $|\bfL^p_{avg}|$ (i.e. $10^2 \, s.$) is not directly related to (and orders of magnitude larger) than the timescale set by $t_b$, and the achieved overall strain rates in the simulations are a truly \emph{emergent} feature of our work that allows us to simulate realistic slow loading-rate regimes of behavior.

\subsection{Application of PTA for coarse-graining DD simulation} \label{sec:pta_dd}
The PTA framework is described in \cite{cspde_2018}, applicable to understanding the behavior of equations of the form given by \eqref{eq:coarseC-def} which have a separation of fast and slow dynamics governed by the small parameter $\epsilon$ (which is defined as the ratio of the time period of the fast and the slow dynamics). The problem of studying the slow behavior of DD also has a separation into fast and slow dynamics. The fast dynamics is the evolution of the dislocation segments, whose characteristic time period $T_f$ is set by the drag, which is on the order of \emph{nanoseconds}. The time period of slow dynamics is governed by the applied loading, which often ranges between $1$ to $1000 s$, corresponding to applied strain rates of $1\,s^{-1}$ to $10^{-3}\,s^{-1}$ and slower. Hence there is a vast separation in time scale of the fast and slow evolution (the parameter $\epsilon=\frac{T_f}{T_s} \approx \frac{10^{-9}}{10^3}=10^{-12}$), which justifies the application of PTA to this problem in order to study the slow time scale behavior of the fast dynamics (i.e.DD). The slow time-scale $t$, which corresponds to the time-scale of applied loading, is defined as $t=\frac{t^*}{T_s}$, where $t^*$ is the dimensional time. The dimensional DD equations (in time-scale $t^*$) can be posed on the slow time-scale $t$ as
\begin{equation}
\label{eq:coarseC-def}
\begin{aligned}
&\epsilon \frac{d\bfX}{dt}(t) =H({\bfX};l) \\
&\frac{dl}{dt} = L(l),\\
%&C(t)=\frac{1}{\overline{\Delta}}\int_{t}^{t+{\Delta}}\Lambda({\bf X}(p);l(p))~dp
\end{aligned}
\end{equation}
where $\bfX$ is a $n$-dimensional vector of position of the nodes. \textit{Here, $n$ is assumed to be fixed for now although, as we will discuss in Section \ref{sec:modifications_in_DD}, the number of degrees of freedom (dofs) in DD is not fixed}. $H$ is a function of the state, $L$ is the loading program employed and $l(t)$ represents the load (corresponding to the magnitude of the applied stress) on the DD box. The evolution of a single dislocation loop is given by \eqref{eq:DD_evol}. The evolution of a system of dislocation loops can be posed in the form of \eqref{eq:coarseC-def}, where the function $H$ is composed of the forces experienced by the segments and is composed of the contributions from the rhs of \eqref{eq:DD_evol} corresponding to the segments comprising each dislocation loop in the system. 
The slow time-scale, $t$, is related to the fast time-scale $\sigma$ through
\[
t = \epsilon\, \sigma ,\, 0<\epsilon= \frac{T_f}{T_s}\ll 1.
\]

The fast time equation, obtained by changing the time scale in \eqref{eq:coarseC-def} to $\sigma=\frac{t}{\epsilon}$, is
\begin{align}\label{eq:fast_dd}
\frac{d\bfX}{d\sigma}(\sigma) =H({\bfX};l). 
\end{align}

We define the running time average $R^\Lambda_t$, of the state function $\Lambda$,  as
\begin{align}\label{eq:R_lambda_t}
R^\Lambda_t := {1 \over {\sum_{i=1}^{N_t} {\Delta \sigma}_i} } \sum_{i=1}^{N_t} \Lambda( \bfX(\sigma_i),l_t) {\Delta \sigma}_i,
\end{align}
where ${\Delta \sigma}_i$ are the DD time steps on the fast time scale and $N_t$ is the number of increments required for the value of $R^\Lambda_t$ to converge up to a specified value of tolerance. The successive values of $\Lambda( \bfX(\sigma_i),l_t)$ are obtained by solving the fast equation \eqref{eq:fast_dd} with initial condition $\bfX^0_t$ and fixed load $l_t$ at time $t$ on the slow time-scale. 

The coarse variable/observable $\overline{\Lambda}$ is defined as the average of $R^\Lambda_t$ over the interval [$t-\Delta,t$]: 
\begin{align}\label{eq:def-C}
\overline{\Lambda} (t)=\int_{t-\Delta}^{t} R^\Lambda_{t'} \, dt', 
\end{align}
where $\Lambda$ is a general state function of $\bfX$ and the nondimensional time interval $\Delta$ is an interval in the slow time-scale $t$, and is defined as $\Delta:=\frac{\Delta^*}{T_s}$, where $\Delta^*$ is a fraction of the slow characteristic time, $T_s$. The coarse variable $\overline{\Lambda}$ depends on the ``history'' of $R^\Lambda_t$, namely its value over an interval prior to $t$. Hence, it is called an \textit{H-observable}, where $H$ stands for ``history" \footnote{\label{ft:history}The general form of $H$-observables is defined in \cite{cspde_2018}. Following that definition, the \textit{H-observable}, is defined as
\begin{align}\label{eq:def-C-genl}
\overline{\Lambda}(t)=\int_{t-\Delta}^{t} \int_{\mathbb{R}^N} \Lambda (\gamma) \mu_{t, l(t), \bfX^{0}} (d\gamma). 
\end{align}
The Young measure $\mu_{(.)}$ corresponding to a sequence of solutions of \eqref{eq:coarseC-def}, parametrized by $\epsilon \to 0$, is a probability measure-valued map of the time, $t$, whose values are invariant measures of the fast time equation \eqref{eq:fast_dd}. In \eqref{eq:def-C-genl}, $\mu_{t,l(t),{\bfX}^0}$ denotes Young measure at time $t$, with applied load $l(t)$, starting from initial state $\bfX^0$. 

The term $\int_{\mathbb{R}^N} \Lambda (\gamma) \mu_{t, l(t), \bfX^{0}} (d\gamma)$ is the average of the state function $\Lambda$ with respect to the Young measure at time $t$.}.  

The evolution of $\overline{\Lambda}$, obtained by differentiating \eqref{eq:def-C} in time, is given by
\begin{equation}
\label{eq:def-barc}
\frac{d \overline{\Lambda}}{dt} = \frac{1}{\Delta}\left( R^\Lambda_t - R^\Lambda_{t-\Delta} \right).
\end{equation}
%2. Followed the convention in CSPDE for naming the $R^\Lambda_t$'s. The load and initial conditions don't show up in $R^\Lambda_t$ expression in CSPDE. Rather, in the definition of $R^\Lambda_t$ itself, the IC and load to run the fast dynamics starting at time $t$, which goes into calculation of $R^\Lambda_t$'s is provided. I think this reduces unnecessary subscripts.

%{\color{change}\subsubsection{The general form of \textit{H-observable}}\label{sec:h_obs_genl}
 %}

%where the Young measures are approximated as averages of $M$ Dirac masses at $M$ values of $\bfX$, i.e., 
%
%\begin{equation}
%\mu_{t,l(t),\bfX^{0}} \approx \frac{1}{M} \sum^{M}_{i=1} \delta(\bfX^{l(t), \bfX^{0}} (\sigma_i)).
%\end{equation}
%Here, $\sigma_i$ are the discrete time instants in the fast run with the load $l(t)$ fixed. $M$ is chosen to be large for the averages to converge. The reasoning for arriving at Eq. (\ref{eq:def-barc}) is given in Section 5 of \cite{cspde_2018}. %The function $\overline{C}(t)$ is a Lipschitz function of $t$ (the Lipschitz constant may be large when $\Delta$ is small) and, in particular, is almost everywhere differentiable. The almost everywhere derivative of the slow variable is expressed at the point $t$, where $\mu_{(.)}$ is continuous at $t$ and $t-\Delta$, by \eqref{eq:def-barc}.

\subsection{Examples of $\Lambda$ functions}\label{sec:examples_lambda}
Here we discuss a few choices of state functions $\Lambda$ specific to DD and their evolution. 
\begin{enumerate}
\item Let $\Lambda({\bfX};l)=\bfX_n$, where $n$ counts segments or nodes in the representation of dislocations and $\bfX$ without an index means the whole array of positions of nodes. Then:
$$\dot{\overline{{\bfX_n}}}(t)=\frac{1}{\Delta}\left(  R^{\bfX_n}_{t} - R^{\bfX_n}_{t-\Delta} \right),$$
\item Let $\Lambda({\bfX};l)=l^{\bfx}({\bfX}(\sigma),l(\sigma))$ be the total line length per unit volume of the dislocations present in the DD box at point $\bfx$ and at time $\sigma$, then:
$$\dot{\overline{l^{\bfx}}}(t)=\frac{1}{\Delta}\left(  R^{l^{\bfx}}_{t} - R^{l^{\bfx}}_{t-\Delta} \right).$$

\begin{figure}[!h]
\centering
  \includegraphics[width=0.3\linewidth]{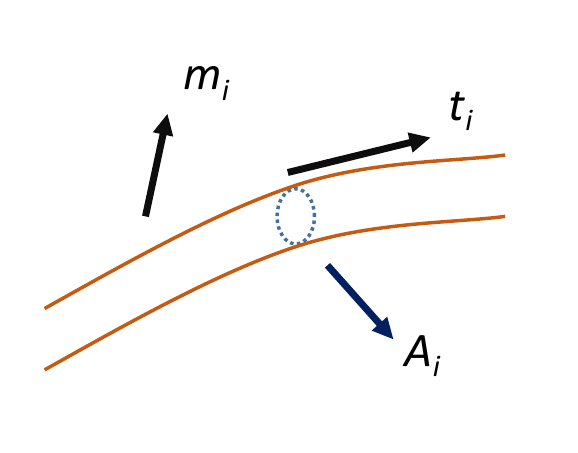}
  \caption{\textit{$Dislocation~ segment~ modeled~ as~ cylindrical~ tube$.} }
  \label{fig:disloc_segment}
\end{figure}

Here:
\[
l^{\bfx}=\frac{1}{|B_{\bfx}|}\int_{B_{\bfx}}\boldsymbol{\alpha}:\boldsymbol{\alpha}~dB_{\bfx} = \frac{1}{|B_{\bfx}|} \sum_i \boldsymbol{\alpha}_{i}:\boldsymbol{\alpha}_{i} \  l_{i} A_{i},
\]
where dislocation segments are modeled as cylindrical tubes as shown in Fig. \ref{fig:disloc_segment}. Here, $\bfalpha_i$ is the dislocation density tensor, $B_{x}$ is the cube centered around spatial point $\boldsymbol{x}$, and $A_{i}$ is the core area of the segment $i$ (which is assumed to be $|\bfb_i|^2$ up to a constant), $\boldsymbol{m_{i}}$ is unit Burgers vector direction,  $\boldsymbol{t_{i}}$ is the unit line direction,  and ${\boldsymbol{\alpha}}_i$ is the contribution to the dislocation density tensor due to segment $i$. Using the fact that $\boldsymbol{\alpha}_{i}=\frac{|\boldsymbol{b}_{i}|}{A_{i}}\boldsymbol{m}^{i}\otimes \boldsymbol{t}^{i}$, the expression \\ $\frac{1}{|B_{x}|} \sum_i \boldsymbol{\alpha}_{i}:\boldsymbol{\alpha}_{i} \  l_{i} A_{i}={1 \over |B_x|} \sum_{i} \frac{|\bfb_i|^2}{A_i^2}l_i A_i={1 \over |B_x|} \sum_{i} \frac{|\bfb_i|^2}{|\bfb_i|^4}l_i |\bfb_i|^2={1 \over {|B_x|}} \sum_i l_i$, which shows that $l^x$ is the total dislocation line length per unit volume, i.e. the total dislocation density.
\item The plastic strain rate of a microscopic dislocation segment is given by $\bfalpha \times \bfV$ (a detailed explanation is provided in \cite{Acharya2011}). The average plastic strain rate, denoted by $\bfL^p$, gives the rate of the plastic slip distortion tensor $\bfU^p$:
\begin{equation*}
\dot{\bfU}^{p,\bfx}(t)={\bfL}^{p,\bfx}(,t), 
\end{equation*}
for a spatial point $\bfx$. If $\tau_{i}$ is the resolved shear stress on segment $i$, 
%\begin{subequations}
\begin{align}
(\boldsymbol {\alpha}\times{{\bfV}})^{x}({\bf X}(\sigma),l(\sigma)) =& \frac{1}{|B_{x}|} \sum_{i}\frac{\tau^{i}\ |\boldsymbol{b}_{i}|}{B}\boldsymbol{m}^{i}\otimes \boldsymbol{n}^{i}\frac{|\boldsymbol{b}_{i}|}{A_{i}}\ l_{i} A_{i} = \frac{1}{|B_{x}|} \sum_{i}\frac{\tau^{i}\ |\boldsymbol{b}_{i}|^2 \ l_{i}}{B}\boldsymbol{m}^{i}\otimes \boldsymbol{n}^{i} \label{eq:Lp_dd} 
\end{align}
\begin{align}\label{eq:Lp_evol}
\left({\bfL}^{p,\bfx}\right)^{\cdot}(t)=\frac{d{\bfL}^{p,\bfx}}{dt}(t) &:=\dot{\overline{(\boldsymbol{\alpha}\times{\bfV})^{\bfx}}}(t) =\frac{1}{\Delta}\left(  R^{\bfL^{p,\bfx}}_{t} - R^{\bfL^{p,\bfx}}_{t-\Delta} \right). 
\end{align}
%\end{subequations}
%And for all $\boldsymbol{x}$, evolve the quantities $\overline{L}^{x}$, $\overline{\bfL}^{p,x}$, and ${\bf \overline{X}}_{n}$ on the slow time scale with Eqs. \ref{evolution} and \ref{newton}.
\item Let ${\bfV}^{\bfx}({\bf X}(\sigma),l(\sigma))$ be the volume-averaged dislocation velocity around $\boldsymbol{x}$, defined as 
\begin{align}\label{eq:V_dd}
{\bfV}^{\bfx}&=\frac{1}{|B_{\bfx}|}\sum_{i}\frac{\tau_{i}}{B}\left\{(\boldsymbol{Tm_{i}})\times\boldsymbol{t_{i}}\right\}_{//} l_i A_i ,
\end{align}
where $\bfT$, henceforth, represents the Cauchy stress (denoted by $\bfsigma$ in \eqref{eq:DD_evol}) and $//$ represents projection to slip plane and one needs to adjust for cross-slipping segments. 
Then:
\begin{equation}\label{eq:V_evol}
\dot{\overline{{\bfV}^{\bfx}}}(t)=\frac{1}{\Delta}\left(  R^{\bfV^{\bfx}}_{t} - R^{\bfV^{\bfx}}_{t-\Delta} \right).
\end{equation}

\end{enumerate}

\subsection{Adaptation of PTA in application to DD}\label{sec:modifications_in_DD}
Everything explained in the previous sections (Section \ref{sec:pta_dd} and \ref{sec:examples_lambda}) are for fixed number of dofs. However, the number of dofs in DD is not fixed. PTA is applicable to ode systems while DD is not an ode system (because of the non-fixed number of dofs) and does not have a fixed phase space in time. Thus, the notion of Young measure, as discussed in footnote \ref{ft:history}, does not apply directly in the case of DD. However, the notion of the running time average $R^\Lambda_t$, defined in \eqref{eq:R_lambda_t} survives and can be determined.  

%This limitation of a non-fixed number of dofs can be addressed using a microscopic model based on pde, like FDM, in which the state space can be divided into parts. A 2D model of FDM has been defined in \cite{xiaohan_2014}, however an extension of the model to 3D has not yet been developed. 

The application of PTA to DD does not include every step in the PTA algorithm described in \cite[Sec. 9]{cspde_2018}. In particular, there are two main exclusions: 
\begin{itemize}
\item The closest point projection of a fixed point in state space on the measure at different instants of slow time \cite[Step 3, Sec. 9]{cspde_2018}, in order to obtain appropriate guesses for fine initial conditions, is not determined. This is because the microstructure involved in the DD simulations has non-fixed number of dofs. Instead, the final microstructure of the previous run (at time $t-h'+\Delta$) is used as the initial condition for the current run (at time $t$). 
\item The criteria of accepting the measure at discrete times \cite[Step 4, Sec. 9]{cspde_2018} is relaxed as this constraint is too hard, especially for the coarse variable $\bfL^p$. Instead, the value of the coarse variable obtained from the extrapolation rule is accepted, unless there is a jump (as per Step 4 of Sec. \ref{subsec:sgp_algo} below). 
\end{itemize}

These exclusions significantly weaken the power of the coarse graining scheme we employ in comparison to PTA, but, unfortunately, this is the price that has to be paid for the application to DD.

\section{Coarse graining DD simulations in time}\label{sec:coarse_grain_dd}
In this section, we consider a single DD box and apply the algorithm to coarse-grain DD simulations in time, in order to obtain the stress-strain response of the box at slow loading rates. We describe the algorithm of applying PTA in this case, describe the setup and then present results. 
\subsection{Algorithm}\label{subsec:sgp_algo}
%\[
%v(t):= \frac{1}{\Delta}\int_0^{\Delta} m(\tau) d\tau
%\]
\par
{\bf Given} at time $t-h'$: predicted density $\rho_{pred}(t-h')$, rate of change of density $\dot{\rho}(t-h')$, predicted plastic distortion rate ${\bfL}^p_{pred}(t-h')$, rate of change of plastic distortion rate $\dot{{\bfL}}^p(t-h')$.

\textbf{Remark.} The evolution equation in the model is for $\dot{\bfL^p}$ ( as given by \eqref{eq:Lp_evol}). Usually, the evolution equation is for plastic strain $\bfU^p$. However, in our case, only $\bfL^p$ can be defined as a state variable but not  $\bfU^p$.

We know the step size: $h'$ and the loading rate: $L(t)=c_1 $, where $c_1$ is a constant. The initial loading is $l(-\Delta)=0$, where $\Delta$ is a fraction of the time period of the slow time-scale, $T_s$ (which can be obtained as ${1 \over c_1}$).\\
Also given is the predicted density at time $t$: 
\[
\rho_{pred}(t)=\rho_{pred}(t-h') + \dot{\rho}(t-h') h', 
\]
and the predicted plastic distortion rate at time $t$:  
\[
{\bfL}^p_{pred}(t)={\bfL}^p_{pred}(t-h') + \dot{{\bfL}^p}(t-h') h' .
\]

The tolerance for convergence of $R^{\rho}_t$ and $R^{\rho}_{t-\Delta}$ is denoted as $tol_{\rho}$ while the tolerance for convergence of $R^{\bfL^p}_t$ and $R^{\bfL^p}_{t-\Delta }$ is denoted as $tol_{\bfL^p}$. The maximum allowed value of $|\dot{\bfL^p}|$ is given by the threshold $|\dot{\bfL}^p_{max}|$, and if $|\dot{\bfL^p}| > |\dot{\bfL}^p_{max}|$, a `jump', on the slow-time scale, in the state of the system is said to have occurred at time $t$. The value of $|\dot{\bfL}^p_{max}|$ is chosen such that it is not so large such that no jump is ever detected, and it is neither so small that almost all $\bfL^p$ obtained by the algorithm using DD and the library MoDELib result in a jump in state. The value of $|\dot{\bfL}^p_{max}|$ listed in Table \ref{tab:simulation_details} satisfies these requirements for the simulations presented in this paper.

We need to obtain: $\dot{\rho}(t)$, $\dot{{\bfL}}^p(t)$.   
\par
The steps are:
\begin{enumerate}
\item 
We use the microstructure obtained at the end of $t-h'+\Delta$ and apply stress $l(t-\Delta)$ to obtain $R^{\rho}_{t-\Delta}$ and $R^{{\bfL}^p}_{t-\Delta}$ (up to tolerance of $tol_{\rho}$ and $tol_{\bfL^p}$ respectively).
\item
With the same microstructure as at the end of Step 1 and with stress $l(t)$ and obtain $R^{\rho}_{t}$  and $R^{{\bfL}^p}_{t}$. 
\item
We obtain $\dot{{\bfL}^p}(t)$ from $R^{{\bfL}^p}_{t-\Delta}$ and $R^{{\bfL}^p}_{t}$ as $\dot{{\bfL}^p}(t)={1 \over \Delta} (R^{{\bfL}^p}_{t} - R^{{\bfL}^p}_{t-\Delta})$. 
\item If $|\dot{{\bfL}^p}(t)|> |\dot{\bfL}^p_{max}|$, as mentioned above, a jump in state is said to have occurred at time $t$. We take final state (of the dislocation system) at time $t$ as the initial state and go back to Step 1 and repeat all the steps. 
\item If $|R^{{\bfL}^p}_{t-\Delta}| > |R^{{\bfL}^p}_{t}|$, we do not accept $R^{\bfL^p}_{t-\Delta}$ as the converged value of the running time average of $\bfL^p$ at time $t-\Delta$. In this case, we keep running the time-average $R^{\bfL^p}_{t-\Delta}$, till $|R^{\bfL^p}_{t-\Delta}| \leq |R^{{\bfL}^p}_{t}|$, in which case we accept the value of $R^{\bfL^p}_{t-\Delta}$. If $|R^{\bfL^p}_{t-\Delta}| > |R^{{\bfL}^p}_{t}|$ after running the time-average $R^{\bfL^p}_{t-\Delta}$ for a very long period of time ($N^{t-\Delta}$ in \eqref{eq:R_lambda_t} is large enough so that there is essentially negligible change in the value of $R^{\bfL^p}_{t-\Delta}$ with increasing $N^{t-\Delta}$, so that $|R^{\bfL^p}_{t-\Delta}| \leq |R^{{\bfL}^p}_{t}|$ is unlikely to be true in this case), we accept the value of $R^{\bfL^p}_{t-\Delta}$ as the converged value of the running time average of $\bfL^p$ at time $t-\Delta$. 
%\item If $|\dot{{\bfL}^p}(t)|$ is small, we obtain the step size $h$ using the following time step control: 
%\begin{enumerate}[label=(\roman*)]
%\item $ |{\bfL}^p(t)| h \leq 0.002 $ 
%\item $\left( |{\bfL}^p(t)| + 0.5 |\dot{{\bfL}}^p(t)| h \right) h <=0.002 $
%\item $ |\dot{\rho}(t)| ~ h <= 0.01 \rho $. 
%\end{enumerate} 
%\end{enumerate}
\item The current time step $h$ is subjected to the following time step control: 
\[
 |{\bfL}^p(t)|  \leq \frac {0.002}{h} 
\]
\item 
We store $\dot{\rho}(t)$ and $\dot{{\bfL}}^p(t)$. 
We repeat steps 1 to 4 but now at time $t+h$.
\end{enumerate}

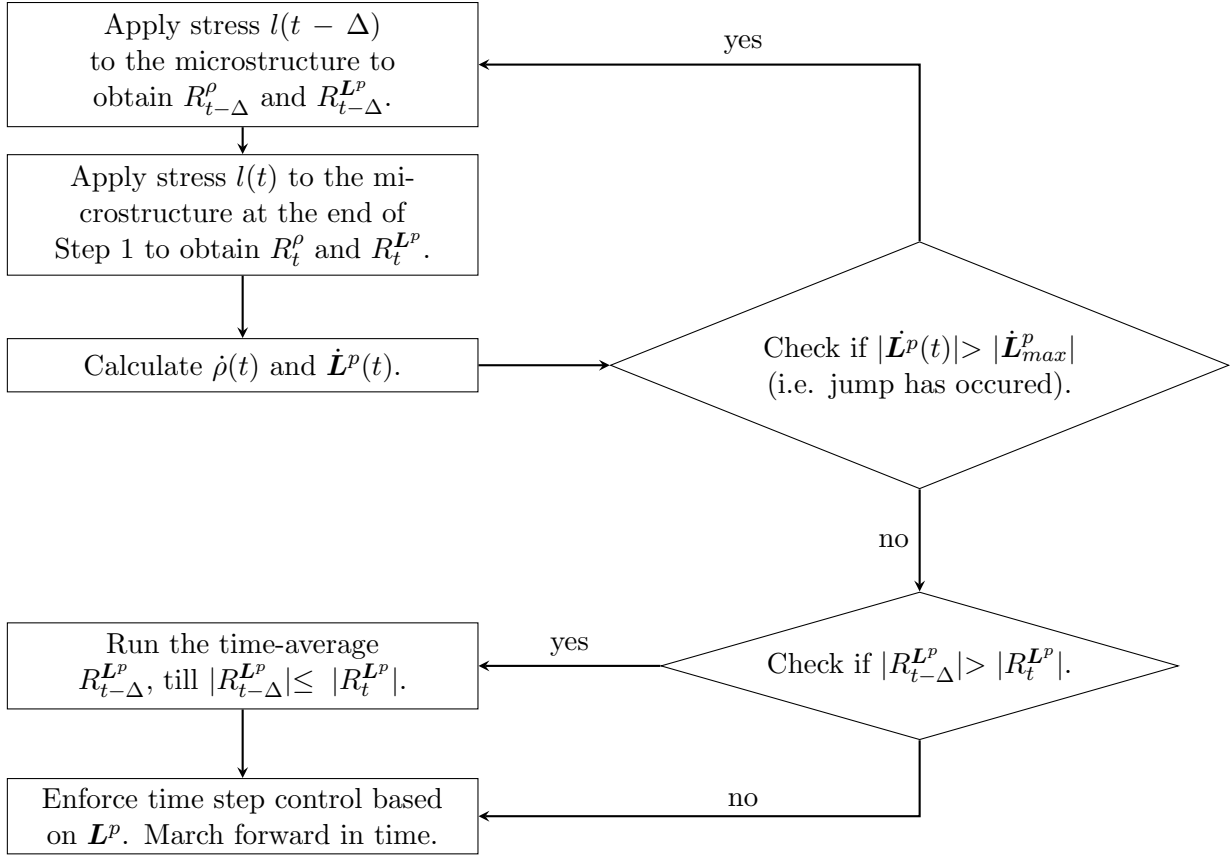
\begin{figure}[h!]
\begin{center}
\begin{tikzpicture}[node distance=2cm]
\tikzstyle{process} = [rectangle, text centered, text width= 6cm, draw=black]

\node (pro1) [process] {Apply stress $l(t-\Delta)$ to the microstructure to obtain $R^{\rho}_{t-\Delta}$ and $R^{\bfL^p}_{t-\Delta}$.};
\node (pro2) [process, below of=pro1] {Apply stress $l(t)$ to the microstructure at the end of Step 1 to obtain $R^{\rho}_t$ and $R^{\bfL^p}_t$.};  
\node (pro3) [process, below of=pro2] {Calculate $\dot{\rho}(t)$  and $\dot{\bfL}^p(t)$.};
\node (pro4) [draw, diamond, aspect=2.5, text width= 5cm, right of=pro3, xshift=7cm,text width=5cm,text centered] {Check if  $\vert \dot{{\bfL}^p}(t) \vert >$ $\vert \dot{\bfL}^p_{max} \vert $ (i.e. jump has occured).};
\node (pro5) [draw, diamond, aspect=3.5, below of=pro4, yshift=-2cm] {Check if $\vert R^{{\bfL}^p}_{t-\Delta}\vert >$ $ \vert R^{{\bfL}^p}_{t}\vert$.};
\node (pro6) [process, below of=pro3, yshift=-2cm] {Run the time-average $R^{\bfL^p}_{t-\Delta}$, till $\vert R^{\bfL^p}_{t-\Delta}\vert \leq \vert R^{{\bfL}^p}_{t}\vert$.};
\node (pro7) [process, below of=pro6] {Enforce time step control based on $\bfL^p$. March forward in time.};

\draw [arrow] (pro1) -- (pro2);
\draw [arrow] (pro2) -- (pro3);
\draw [arrow] (pro3) -- (pro4);
\draw [arrow] (pro4) -- node[anchor=east] {no} (pro5);
\draw [arrow] (pro5) -- node[anchor=south] {yes} (pro6);
\draw [arrow] (pro6) -- (pro7);
\draw [arrow] (pro5) |- node[anchor=south, pos=0.7] {no} (pro7);
\draw [arrow] (pro4) |- node[anchor=south, pos=0.7] {yes} (pro1);

\end{tikzpicture}
\end{center}
\caption{Overview of coarse graining DD simulations in time.}
\label{fig:ddsgp_flowchart}
\end{figure}

A flowchart comprising the above steps is shown in Fig. \ref{fig:ddsgp_flowchart}. 

\subsection{DD simulation setup}\label{subsec:dd_setup}

%\begin{figure}[H]
%\centering
%  \includegraphics[width=0.25\linewidth]{figures_coupling/DD_setup.pdf}
%  \caption{\textit{DD simulation box} }
%  \label{fig:dd_setup}
%\end{figure}

We use the library MoDELib (Mechanics of Defect Evolution Library) \cite{po_ghoniem_2014} to run the DD simulations. We generate a microstructure with a specified value of mobile and sessile density of dislocation segments. The mobile and sessile segments run from boundary to boundary of the DD simulation box. The mobile segments form junctions with the sessile segments, which act as pinning points, around which they expand. The sessile segment density is much larger than the mobile segment density and the sessile segments essentially act as obstacles to the motion of the mobile segments. 

The sessile segments are constructed as Lomer Cottrell (LC) locks, therefore their Burgers vector do not lie in their glide plane. However, majority of sessile segments in FCC crystals do not have this property (i.e. their Burgers vector lie in their slip plane). Therefore, a more physically appropriate case is when the Burgers vector of the sessile segments lie in the slip plane. We have presented results for that case as well, to show that such simulations can be performed. %However, this approach produces sessile segments, which act as obstacles to the motion of the mobile segment. Hence, we construct the sessile segments as LC locks in this first exercise. %However, this is an easier way of modeling sessile segments using $MODEL$ (rather than forcibly assigning particular segments, whose Burgers vector lie in the slip plane, as sessiles), and since our main objective is to create obstacles to the motion of mobile segments, we use LC locks to model sessile segments in this first exercise.  

The preference for using LC locks in this paper is not fundamental but is related to the limitation of the version of MoDELib that was used when this work was started. 

\subsubsection{Construction of initial microstructure} \label{sec:construction_mc}We populate the domain with mobile and sessile segments as follows: we assume a certain target density of mobile and sessile segments denoted by $\rho_m$ and $\rho_s$ respectively (with $\rho_s \gg \rho_m$). We insert the mobile density $\rho_m$ in the ratio of the Schmid factor of the slip system $i$ (denoted as $f_{s,i}$), i.e. the target mobile density of slip system $i$ is $\rho_{m,i}=\rho_m ~ \frac{|f_{s,i}|}{\Sigma_{k=1}^N |f_{s,k}|}$, where $N$ is the total number of slip systems in the crystal. The Schmid factor of slip system $i$ is calculated as 
\begin{align}\label{eq:schmid}
f_{s,i}= \frac{{\bfb}_i \cdot (\bfsigma_{e}{\bfn}_i)}{|\bfsigma_e|}, 
\end{align}
where $\bfsigma_{e}$ is the externally applied stress and $|\bfsigma_e|$ is its norm and $\bfb_i$ and $\bfn_i$ are the Burgers vector and slip plane normal of slip system $i$. The reason behind this kind of insertion is that segments in slip systems with small Schmid factor are expected to move less compared to those belonging to slip systems with higher Schmid factor, and hence their contribution to the coarse variables $\bfL^p$ and $\bfV$ are less. 

To insert segment $n$ (which lies in slip system $i$), we construct a candidate segment as follows. We choose a random point $\bfP_{0,n}$ in the domain and then construct a ray from this point along a direction $\bfd_n$, which lies in the slip plane and is rotated at an angle $\theta_n$ from the Burgers vector $\bfb_i$ of its slip system, till it intersects the boundary at point $\bfP_{1,n}$. We also construct a ray from $\bfP_{0,n}$ in the opposite direction $-\bfd_n$ till it intersects the boundary at point $\bfP_{2,n}$. In this way, a candidate segment with end points on the boundary, given by $\bfP_{1,n}$ and $\bfP_{2,n}$ is constructed. If the density of the candidate segment is very close to $\rho_{m,i}$ (up to a specified tolerance), it is inserted as segment $n$, otherwise the process of finding a candidate segment is repeated until a suitable candidate is obtained. %If we consider a {\color{change}slip plane parallel to the slip plane of slipsystem $i$ (with normal $\bfn_i$)}, and which passes through a specified point in the box, say {\color{change}$(1 1 1)$}, then {\color{change}the distance of the inserted segment $n$ from this  parallel slip plane is denoted as $\bfs_n$.} 

We construct another segment $n+1$ from another random point $\bfP_{0,n+1}$ using the approach mentioned above, which belongs to the same slip system and is on the same slip plane but has opposite line direction. Thus, we have two segments which have the same density and belong to the same slip system and are on the same slip plane but have opposite line directions. This is to ensure that the net mobile dislocation density is very close to 0. Similarly, we construct a pair of segments on the other slip systems. 

After this, we construct a number of sessile segments of total density $\rho_s$ distributed isotropically across all slip systems and with zero net dislocation density i.e. every segment constructed has a corresponding segment in the same slip system at a different position and with same density but opposite line direction. %Thus, the constructed microstructure is given by the mobile and sessile target densities $\rho_m$ and $\rho_s$.
%Ideally we should insert $\rho_m$ in all slip systems equally, but doing so makes the computations very expensive. 

\subsubsection{Reinsertion of segments}\label{subsec:reinsertion} As the system of dislocation segments evolves, some mobile segments exit the box, leading to a reduction in the density of mobile segments. To compensate for this, there is a possibility of reinsertion of segments. Possible strategies for reinsertion is discussed in the Remark of Section \ref{sec:key_changes}. 
However, in the results that we present in the next sections, reinsertion of segments is not done.

\subsection{Results}\label{sec:sgp_results}
We present the results of coarse graining DD simulations in time. Traction boundary conditions are applied and the boundaries of the DD box are considered \emph{open} (i.e. dislocation segments that exit are not reinserted and the infinite medium stress fields of individual dislocation segments are employed without correction for finite boundaries - this is simply an approximation, and not an essential restriction in MoDELib).
\subsubsection{Uniaxial tension}\label{sec:uniaxial_tension}
We consider a cubic box and apply tensile loading (traction boundary condition) in the $y$-direction ($t_{22}$ loading), with the crystal in the symmetric double slip orientation  (see, e.g.,  \cite{pierce_bifurc}). The details of the crystallographic setup are in the Appendix. 
%%\ref{app:c2g}

\begin{figure}[!h]
\centering
  \includegraphics[width=0.4\linewidth]{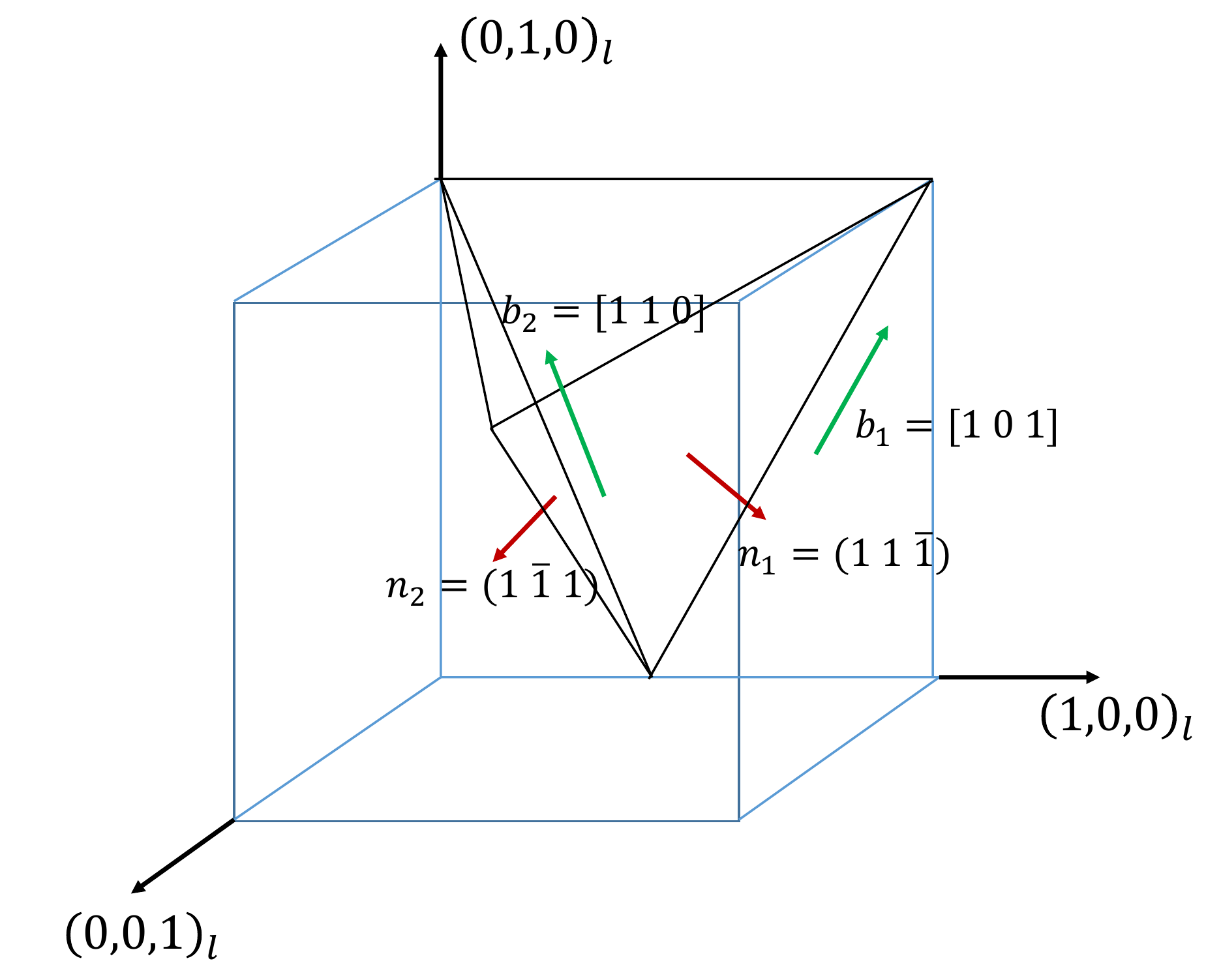}
  \caption{\textit{\small Rotated Thompson tetrahedron of the crystal in tension, the primary and conjugate slip systems are given by $\{\bfb_1,\bfn_1\}$ and $\{\bfb_2,\bfn_2\}$ respectively. The fixed laboratory axes are marked with subscript $l$.}}
  \label{fig:t22_orientation}
\end{figure}

We choose $\rho_m=5\times 10^{12} m^{-2}$ and $\rho_s=2\times 10^{14} m^{-2}$. We insert mobile segments in two slip systems, called the primary and the conjugate slip systems are $[1 0 1] (1 1 \bar{1})$ and $[1 1 0](1 \bar{1} 1)$ respectively (see Fig. \ref{fig:t22_orientation}).   

All simulation parameters are provided in Table \ref{tab:simulation_details}. 

\begin{table}[H]
\centering
\begin{tabular}[H]{|c|l|l|}
\hline
Name &   Physical definition & Values  \\
\hline
$E$     &  Young's modulus     &   $110\,\mathit{GPa}$  \\
$\mu$   &  shear modulus &  $48\, \mathit{GPa}$  \\
$b$   &  Burgers vector   &  $2.55 \times 10^{-10}m$   \\
$B$   &   Drag    &   $6.30 \times 10^{-5} Pa.s$      \\
$A$ &   Box size  &  $4000\,b$ \\
$\Delta^*$  &  time interval in $t^*$  &   $0.1 s$  \\
$|e_1|$  & tolerance for convergence for $\rho_x$  & $10^{-2}$\\
$|e_2|$  & tolerance for convergence for $L^p_x$  & $3 \times 10^{-2}$\\
$|\dot{\bfL}^p_{max}|$  &Threshold for $|\dot{\bfL}^p|$ to detect a jump   & $0.05 s^{-2}$\\
$L$ & loading rate & $1~ MPa/s$\\
$\rho_m$ & Mobile density & $5 \times 10^{12} ~m^{-2}$ \\ 
$\rho_s$ & Sessile density & $2 \times 10^{14} ~m^{-2}$ \\ 
\hline
\end{tabular}
\caption{Simulation parameters for the problem of coarse graining DD simulations in time.}
\label{tab:simulation_details}
\end{table}    

The following are the results obtained in this setting:  
\begin{figure}[!h]
\centering
\begin{minipage}{.4\textwidth}
  \centering
  \includegraphics[width=\linewidth]{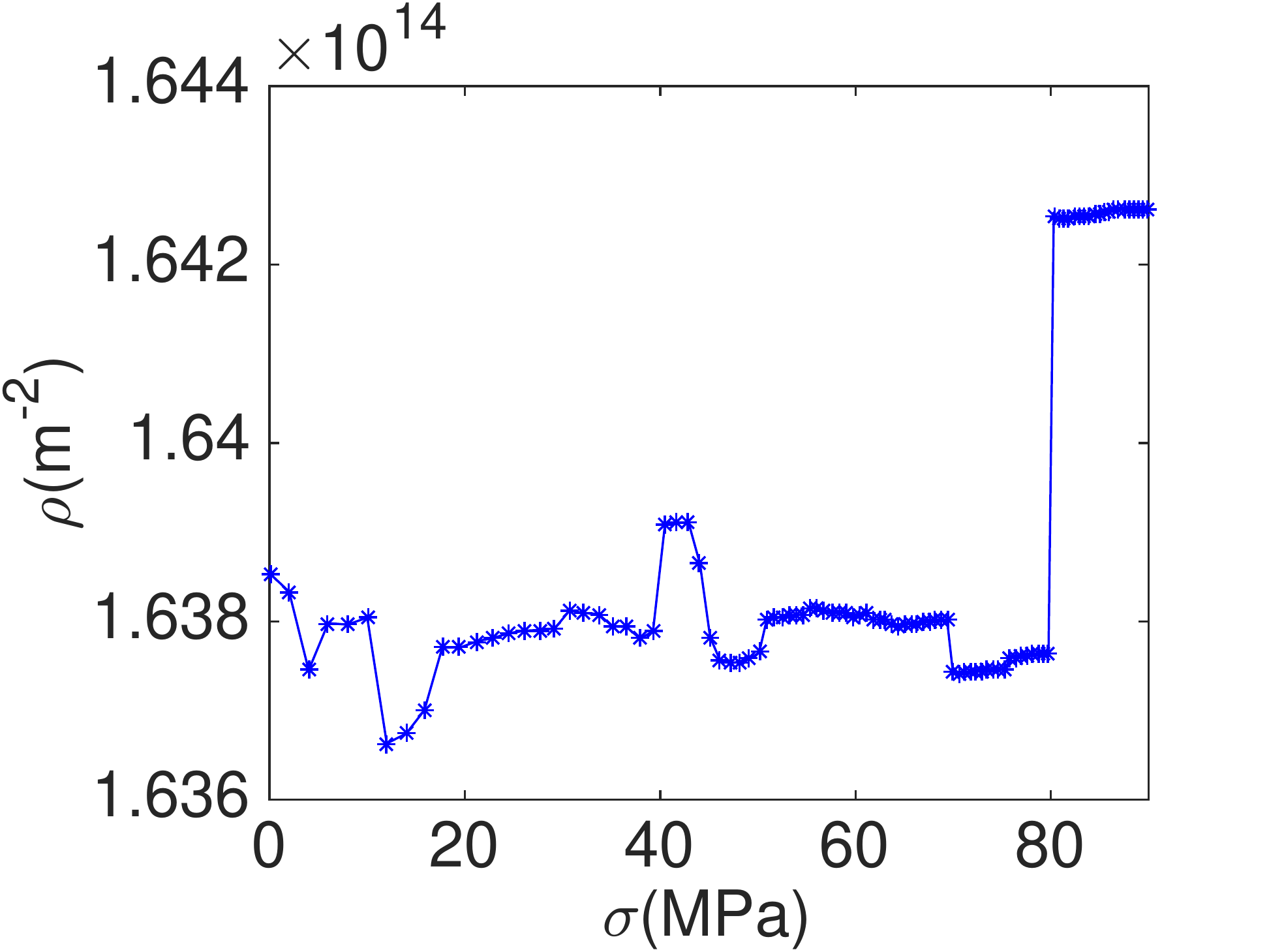}
  \caption{\textit{Evolution of $\rho$.} }
  \label{fig:t22_rhox}
\end{minipage}%
\hfill
\begin{minipage}{.4\textwidth}
  \centering
  \includegraphics[width=\linewidth]{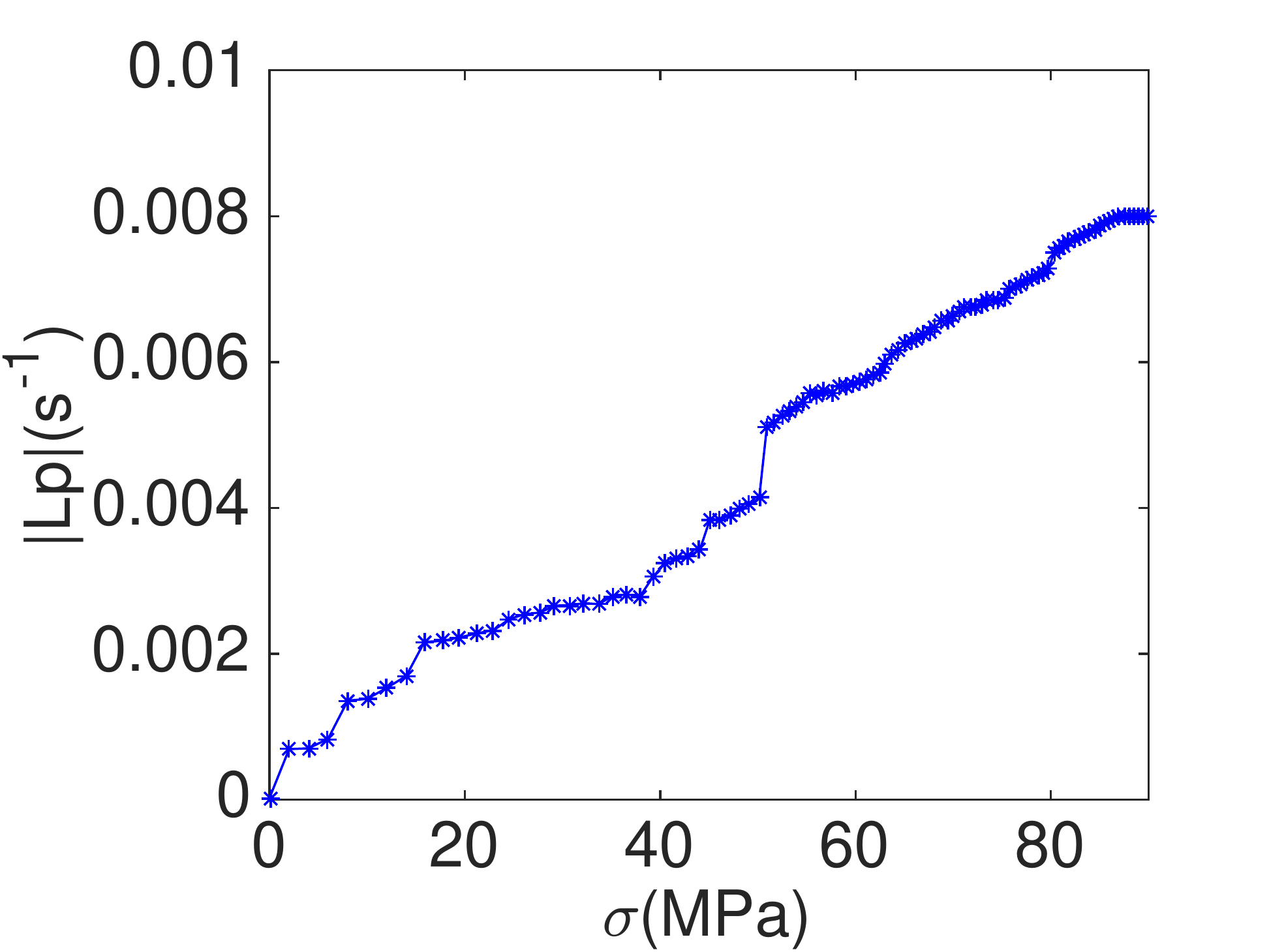}
  \caption{\textit{Evolution of ${\bfL}^p$.} }
  \label{fig:t22_lpx}
\end{minipage}
\end{figure}

\begin{figure}[!h]
\centering
\begin{minipage}{.4\textwidth}
  \centering
  \includegraphics[width=\linewidth]{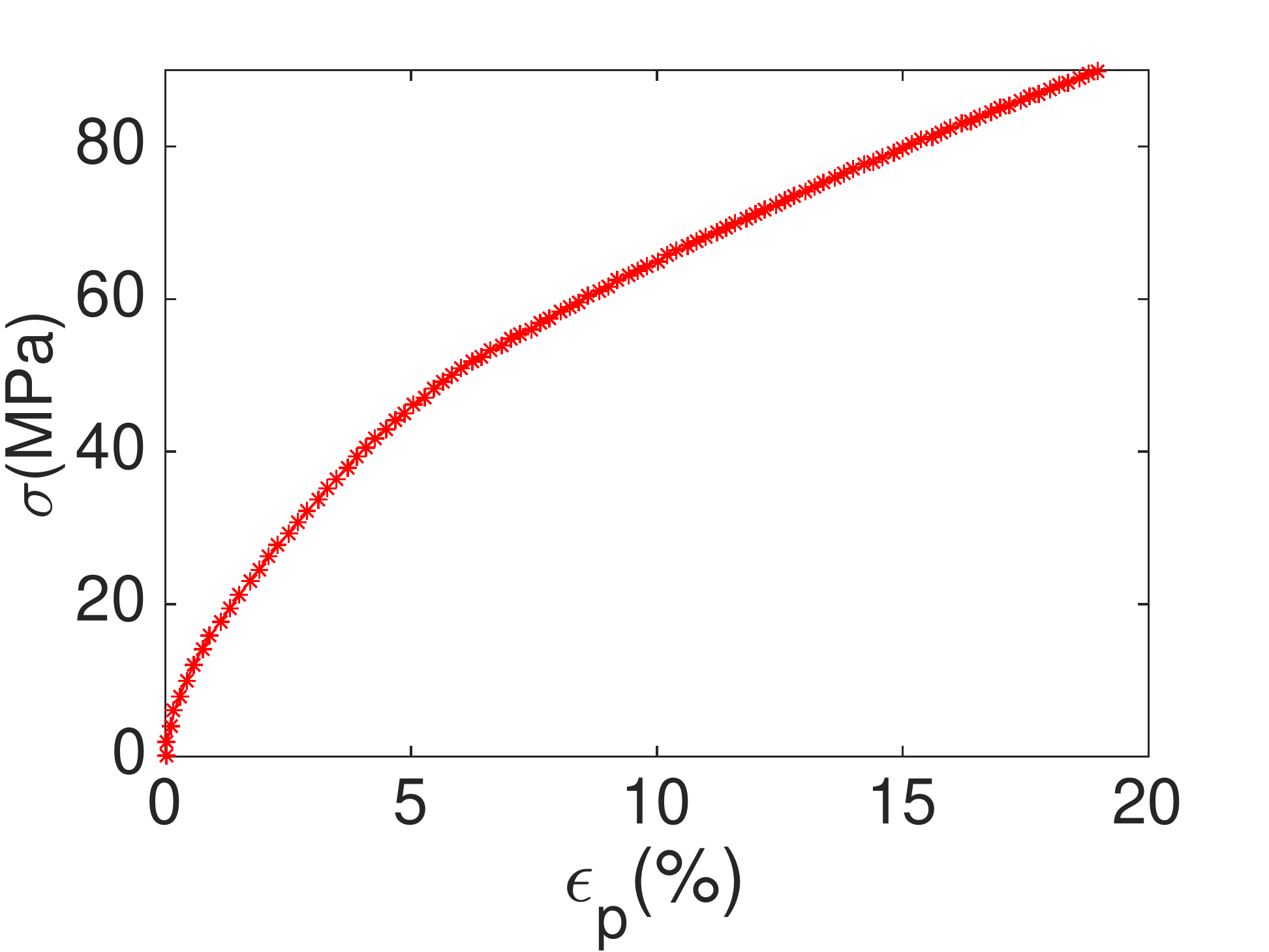}
  \caption{\textit{Stress-strain profile.} }
  \label{fig:t22_stress_strain}
\end{minipage}%
\hfill
\begin{minipage}{.4\textwidth}
  \centering
  \includegraphics[width=\linewidth]{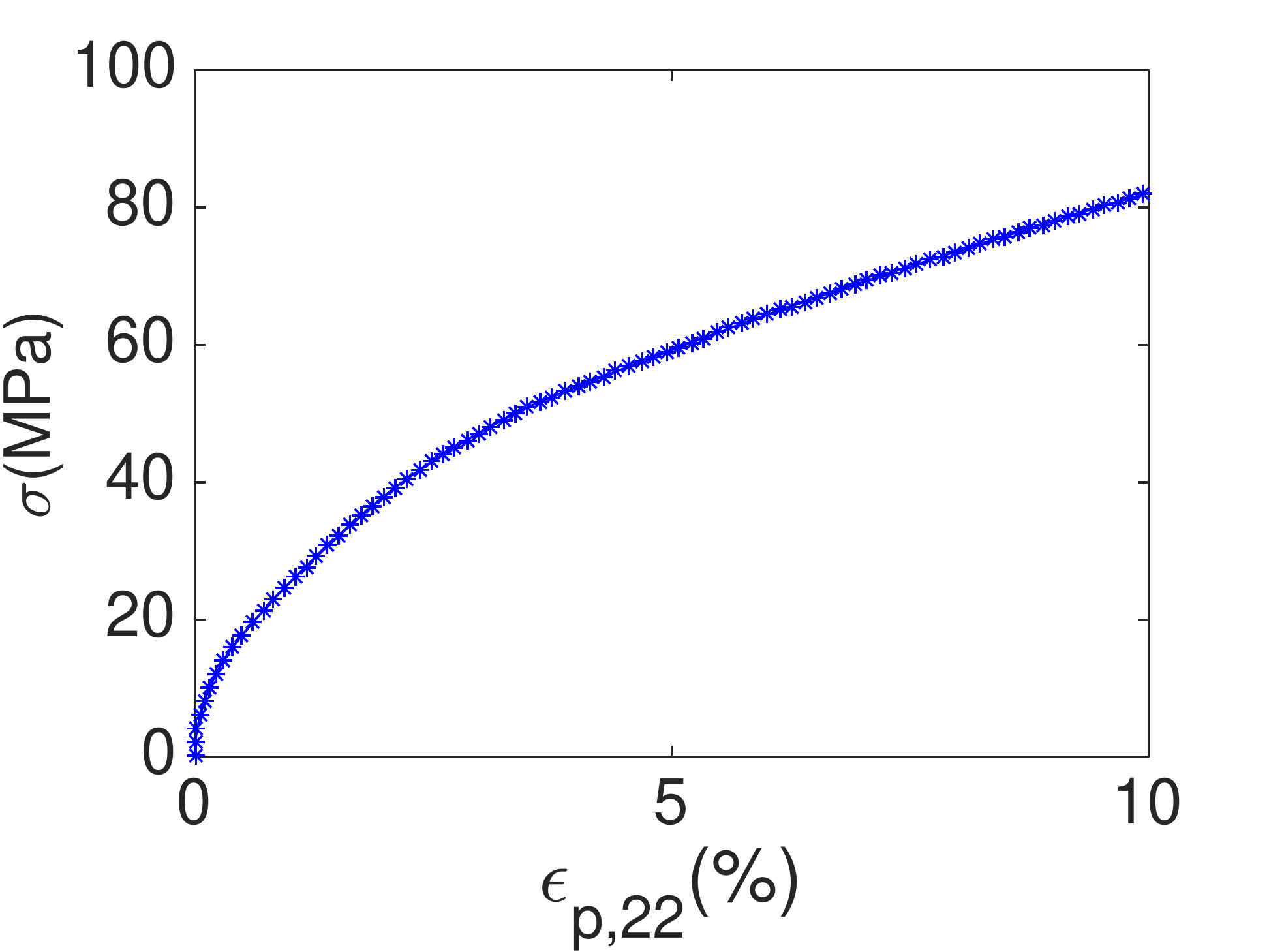}
  \caption{\textit{Stress vs $\epsilon_{p,22}$.} }
  \label{fig:t22_directional_stress_strain}
\end{minipage}
\end{figure}

Figure \ref{fig:t22_rhox} shows how $\rho$ evolves with increasing stress. It increases as the mobile segments form junctions with the sessile segments around which they expand and grow in length. Figure \ref{fig:t22_lpx} shows that ${\bfL}^p$ is increased with increasing stress.  The plastic strain norm $\epsilon_p$ is obtained by integrating $|\bfL^p|$ in time, i.e. $\bfepsilon_p(t)=\int_0^t |\bfL^p(t')|dt'$. The plastic strain components, which are also called the directional plastic strain, are obtained as $\epsilon_{p,ij}=\int_0^t {(L^p)}^{sym}_{ij}(t')dt'$, where ${(\bfL^p)}^{sym}={1 \over 2} (\bfL^p+{(\bfL^p)}^T)$ is the symmetric part of $\bfL^p$. The stress-strain profile is shown in Figure \ref{fig:t22_stress_strain}. The hardening in the stress-strain profile depends on the mobile and sessile segment density of the initial microstructure. In general, hardening increases with increase in sessile density and decreases with increase in mobile density. It also depends on the applied loading rate and increases with increase in the loading rate. These factors are discussed in more detail in Section \ref{sec:coupling_results}. In Figure \ref{fig:t22_directional_stress_strain}, the directional plastic strain ${\epsilon}_{p,22}$ stays positive with increasing stress as it should. This is not guaranteed to happen since we do not have a primary slip plane in this case. However, our algorithm can robustly predict the correct direction of ${\epsilon}_{p,22}$. 

\begin{figure}[H]
\centering
  \includegraphics[width=0.5\linewidth]{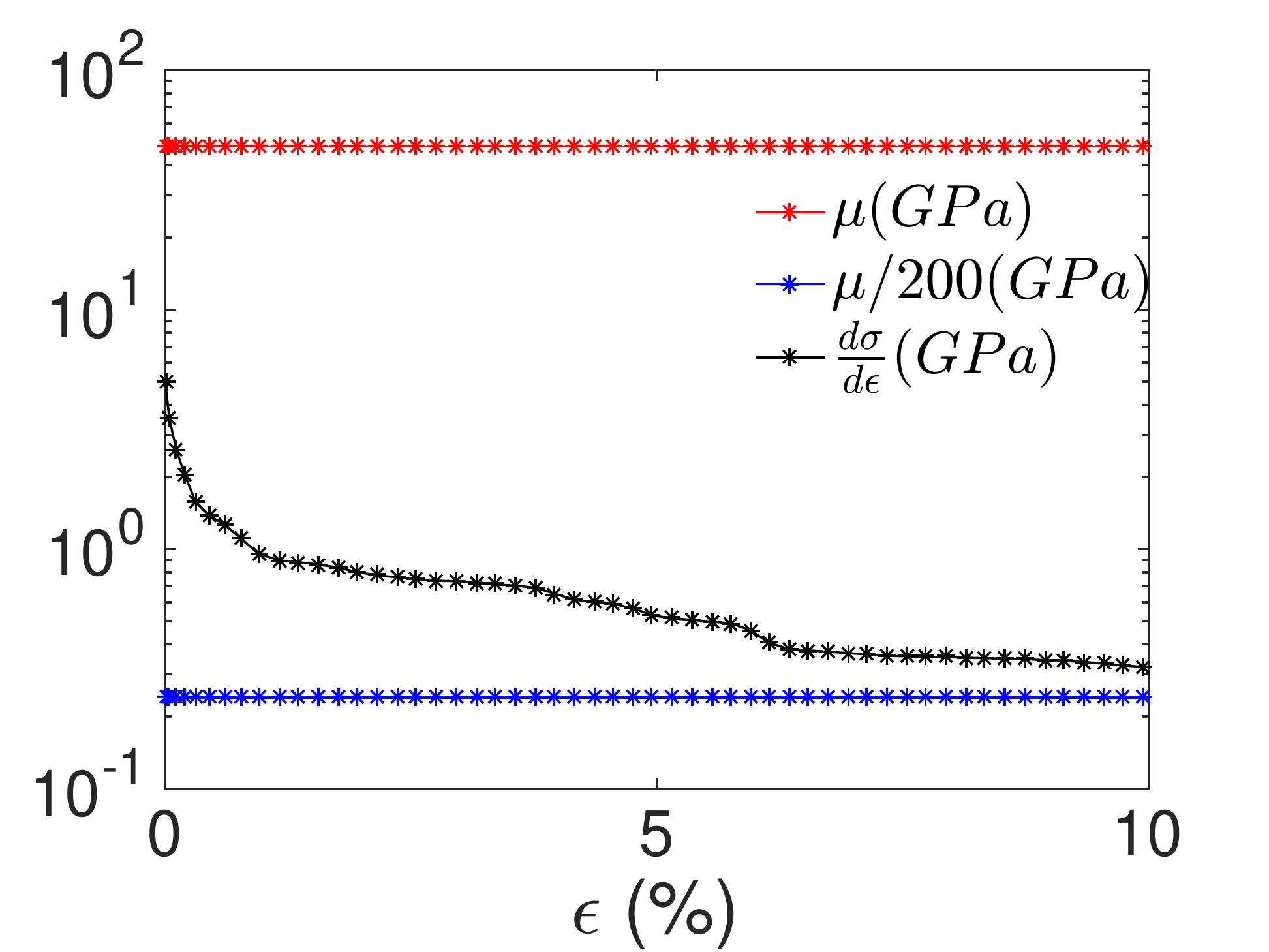}
  \caption{\textit{Tangent modulus.} }
  \label{fig:t22_tangent_mod}
\end{figure}

The total strain $\epsilon$ is determined as $\epsilon=\frac{\sigma}{\mu}+\epsilon_p$, where $\frac{\sigma}{\mu}$ is the elastic strain. The slope of the stress versus total strain curve (tangent modulus) is shown in Figure \ref{fig:t22_tangent_mod} and it is approximately $\frac{\mu}{200}$, which is the slope that we expect to see in Stage II hardening \cite{kocks_mecking_2003} (we expect Stage II hardening behavior as we start with a high density of sessile segments). In general, the tangent modulus decreases with the increase in the ratio of the mobile to sessile segment density of the initial microstructure. We see that we are able to reach appreciable values of strain at realistic loading rates, at which experiments can be performed on macroscopic samples to study their plastic response. Performing simulations at these loading rates using DD simulations alone, for the given domain size and dislocation density, is very expensive and practically impossible. 

The speedup in compute time, $S$, of conventional DD to PTA is obtained as follows. The compute time $t^{cpu}_{DD}$ to run DD up to a time  $t_{DD}$ on the slow time-scale is determined. The compute time $t^{cpu}_{PTA}$ to run PTA up to a time $t_{PTA}$, which is chosen to be the slow time at $\epsilon_p=10\%$, is also determined. Then, the speedup in compute time, $S$, is obtained as $S={\left(\frac{t^{cpu}_{DD}}{t_{DD}} \right)} \div {\left(\frac{t^{cpu}_{PTA}}{t_{PTA}}\right)}$. The value of $S$ is around $5000$ for this loading case.

\subsubsection{Simple shear}\label{sec:simple_shear}
We consider the same setting as in uniaxial tension but apply shear loading (traction boundary condition) in $t_{12}$ direction. We rotate the crystal such that the crystallographic direction $[1 \bar{1} 1]$ lies along the global $Y$ axis and the slip direction $[0 1 1]$ lies along the global $X$ direction. The corresponding details of the crystallographic setup are in the Appendix. %%\ref{app:c2g}. 

\begin{figure}[!h]
\centering
  \includegraphics[width=0.4\linewidth]{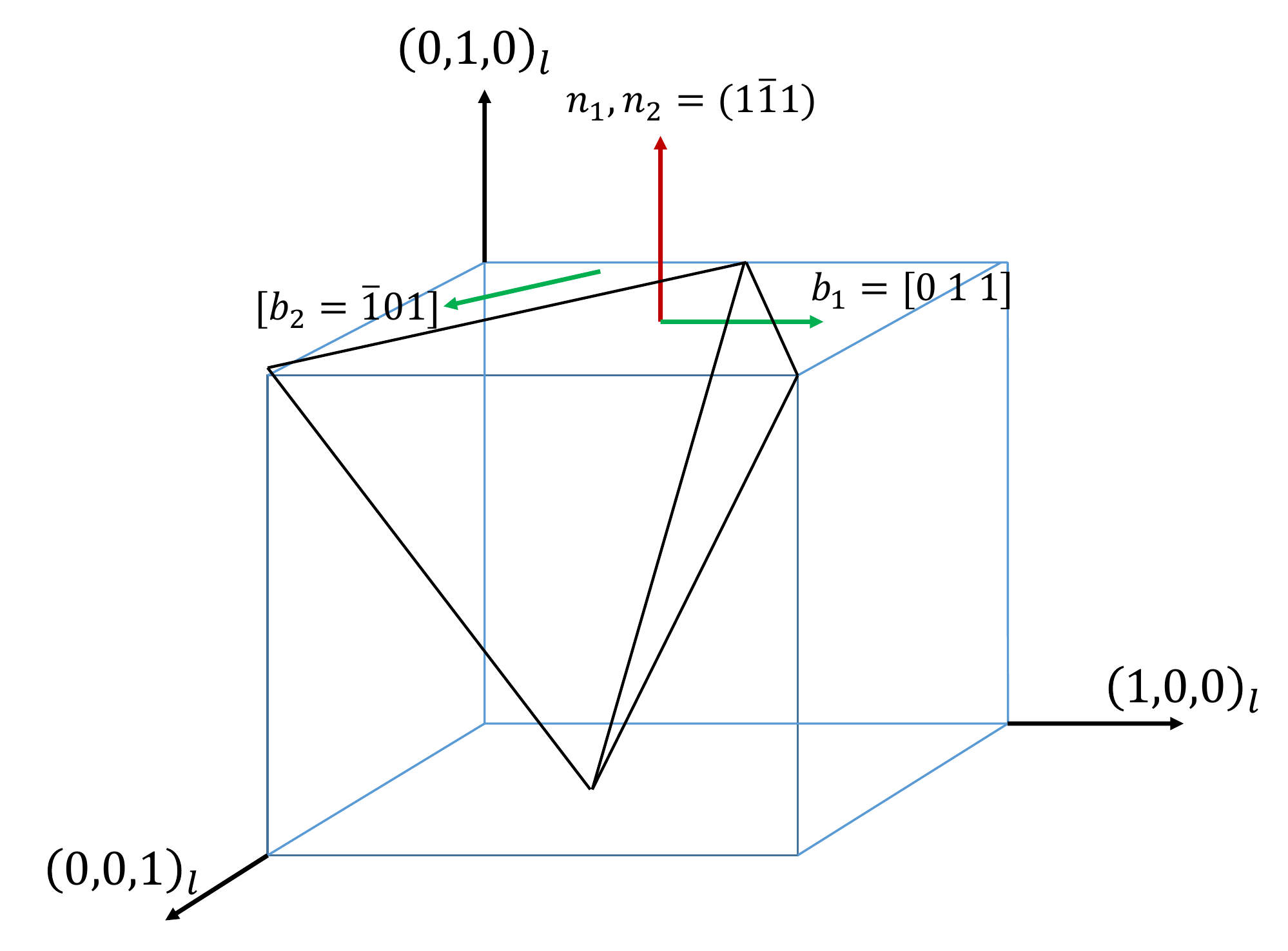}
  \caption{\textit{\small Rotated Thompson tetrahedron of the crystal in shear.
  %%the primary and conjugate slip systems are given by $\{\bfb_1,\bfn_1\}$ and $\{\bfb_2,\bfn_2\}$ respectively.
 The fixed laboratory axes are marked with subscript $l$.} }
  \label{fig:t12_orientation}
\end{figure}

In this case also, we insert segments on two slip systems: $[0 1 1] (1 \bar{1} 1)$ and $[\bar{1} 0 1] (1 \bar{1} 1)$. The former is the primary slip system as after rotation, its normal is along the global $Y$ axis and we shear along its slip direction (global $X$ axis).  The rotated crystal is shown in Fig. \ref{fig:t12_orientation}. 

The results are presented below: 
\begin{figure}[H]
\centering
\begin{minipage}{.4\textwidth}
  \centering
  \includegraphics[width=\linewidth]{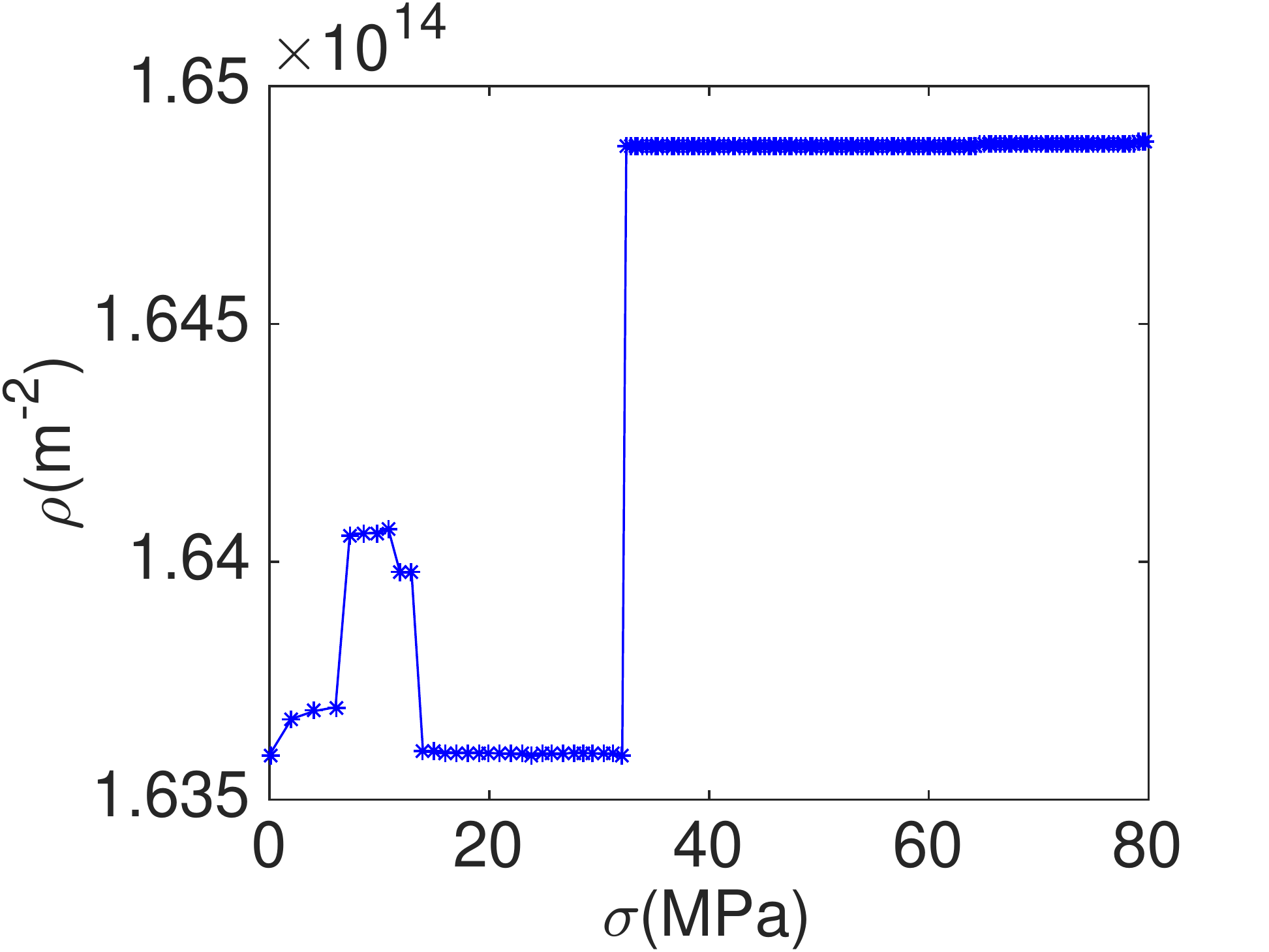}
  \caption{\textit{Evolution of $\rho$.} }
  \label{fig:t12_rhox}
\end{minipage}%
\hfill
\begin{minipage}{.4\textwidth}
  \centering
  \includegraphics[width=\linewidth]{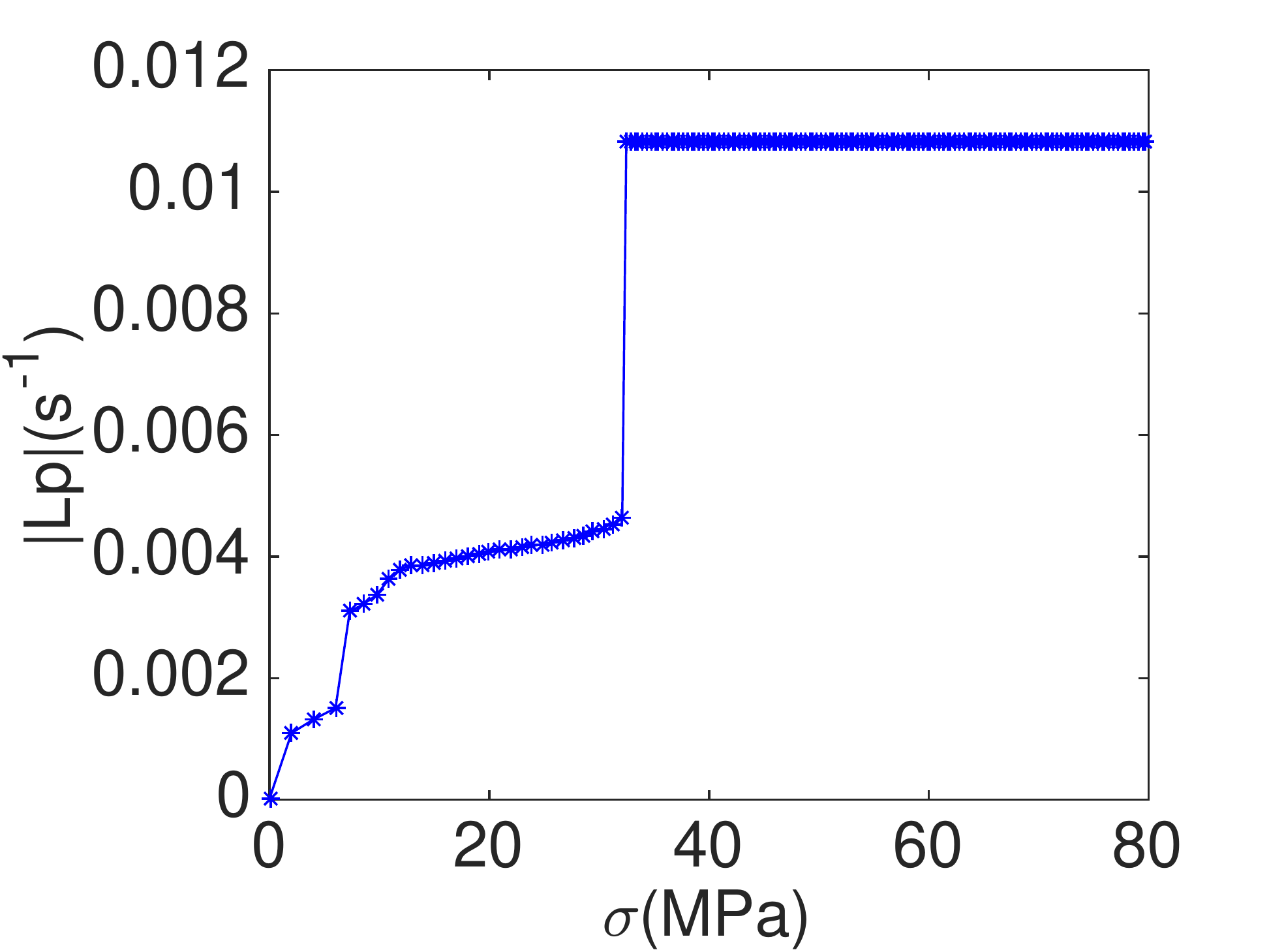}
  \caption{\textit{Evolution of ${\bfL}^p$.} }
  \label{fig:t12_lpx}
\end{minipage}
\end{figure}

\begin{figure}[H]
\centering
\begin{minipage}{.4\textwidth}
  \centering
  \includegraphics[width=\linewidth]{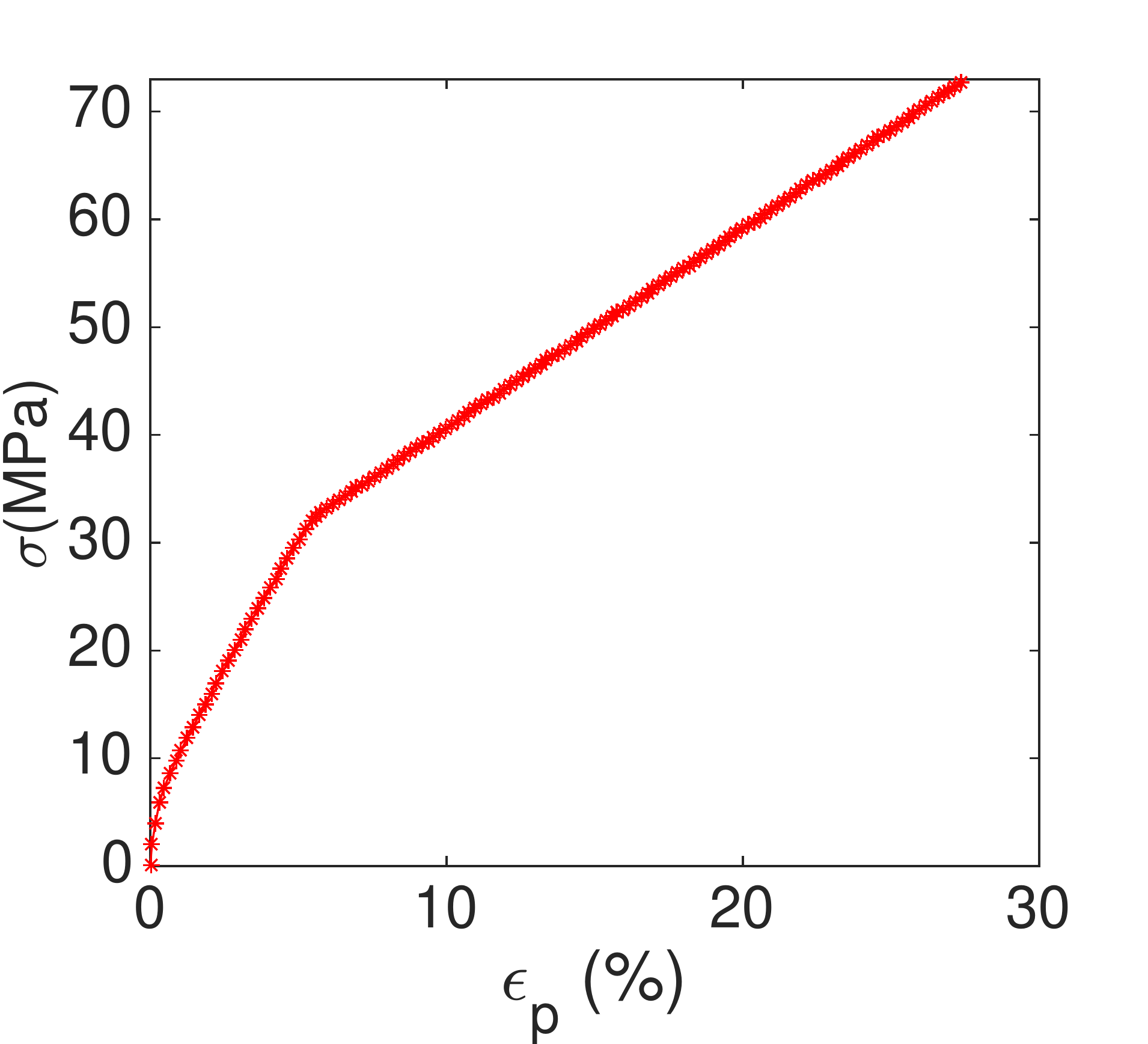}
  \caption{\textit{Stress-strain profile.} }
  \label{fig:t12_stress_strain}
\end{minipage}%
\hfill
\begin{minipage}{.5\textwidth}
  \centering
  \includegraphics[width=0.9\linewidth]{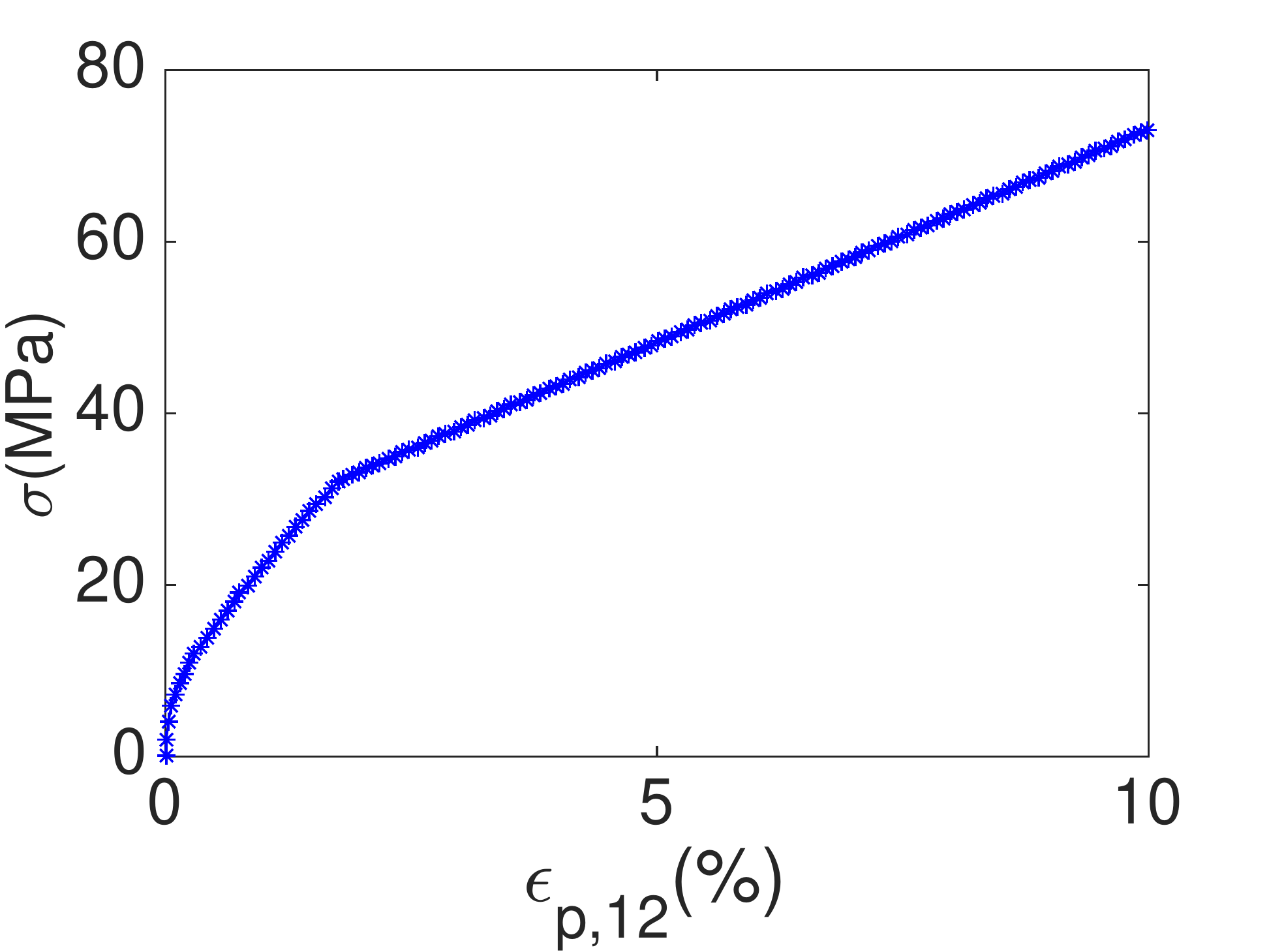}
  \caption{\textit{Stress vs $\epsilon_{p,12}$.} }
  \label{fig:t12_directional_stress_strain}
\end{minipage}
\end{figure}

These results follow a similar trend as in the uniaxial loading case presented in Section \ref{sec:uniaxial_tension}. We see in Figure \ref{fig:t12_rhox} that dislocation density $\rho$ increases with stress. The definition of plastic strain norm $\epsilon_p$, directional plastic strain $\epsilon_{p,ij}$ and the total strain $\epsilon$ are provided in the previous section (Section \ref{sec:uniaxial_tension}). In Figure \ref{fig:t12_directional_stress_strain}, we see that the directional plastic strain strain ${\epsilon}_{p,12}$ remains positive with increasing stress, as it is supposed to. In Figure \ref{fig:t12_stress_strain}, the slope of the stress-total strain curve (Figure \ref{fig:t12_stress_strain}) comes close to $\frac{\mu}{200}$, which is the slope we observe in Stage II hardening. The speedup in compute time, $S$, defined in Section \ref{sec:uniaxial_tension}, is around $2000$. 

\begin{figure}[H]
\centering
  \includegraphics[width=0.4\linewidth]{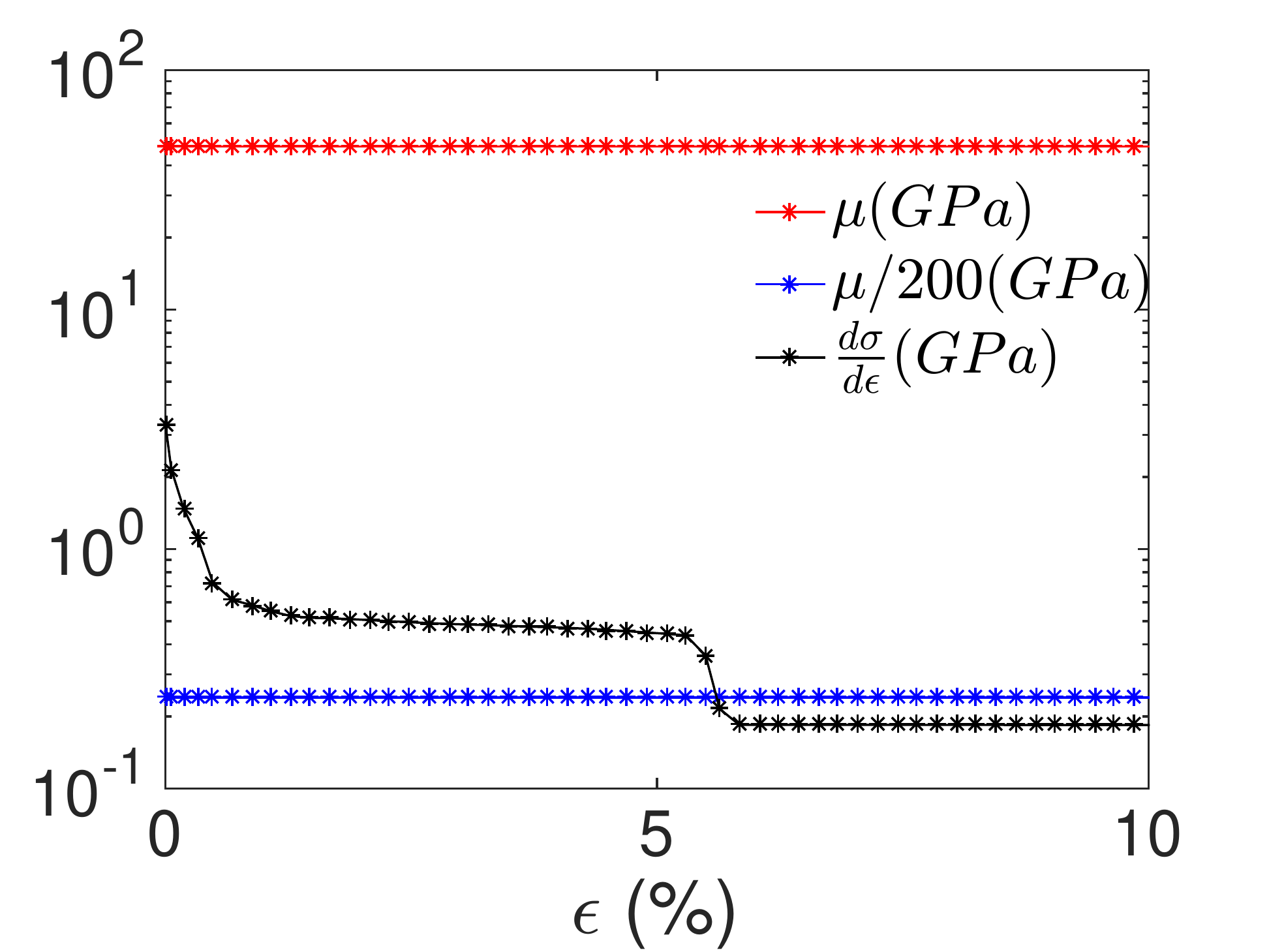}
  \caption{\textit{Tangent modulus.} }
  \label{fig:t12_stress_strain}
\end{figure}

\section{DD-continuum plasticity coupling}\label{sec:ddfdm_coupling}

All the discussions in the previous sections were for DD simulations in one box. Now we think of many boxes being part of a larger domain in which we want to do regular plasticity calculations and couple this with a larger plasticity calculation in the body, in which equilibrium equations are solved. The pde-based theory which represents time averaged Dislocation Dynamics is MFDM, with its typically `non-closed' time averaged inputs now specified from DD, which are obtained using the framework of PTA as outlined in Section \ref{sec:pta_dd} and Section \ref{sec:examples_lambda}.  %We can define $\overline{\bfL}^{p,x}$, ${\overline L}^{x}$, and ${\bfu}^{p}$ in Section \ref{sec:examples_lambda}. Note $\rho$ in continuum plasticity is nothing but ${\overline L}^{x}$ defined as in Section \ref{sec:examples_lambda}. 

MFDM \cite{acharya_roy_2006} involves the evolution of the following system of pdes:
\begin{subequations}\label{eq:MFDM}
\begin{align}
%\begin{split}
&\dot{\overline{\bfalpha}}=-curl~(\overline{ \bfalpha}\times\overline{\bfV}+{\bfL}^{p})\label{eq:alpha_evol}\\
%&{\bf \dot U}^{p}={\bfL}^{p}+ \boldsymbol {\overline \alpha}\times\overline{\bfV}\\
&curl \overline{{\bfchi}}=\overline{{\bfalpha}}\nonumber\\
&div \overline{{\bfchi}}=0 \label{eq:chi_evol}\\
&div(grad \, \dot{\overline{{\bfz}}})=div(\overline{\bfalpha}\times \overline{{\bfV}} + {\bfL}^p) \label{eq:z_evol}\\
&\overline{{\bfT}}={\bfC}:(grad (\overline{{\bfu}}-\overline{{\bfz}}) + \overline{\bfchi} \nonumber\\
&div \overline{{\bfT}}=0 \label{eq:u_evol}. 
%\end{split}
\end{align}
\end{subequations}
%and we need to satisfy $\left[|{\bfL}^{p}+ \boldsymbol {\overline \alpha}\times\overline{\bfV}|\right]=0$ (jump condition) everywhere (i.e. at element boundaries in FEM). 

The tensor $\overline{\bfalpha}$ is the mesoscopic space-time averaged dislocation density tensor, $\overline{\bfV}$ is the averaged dislocation velocity vector, ${\bfC}$ is the fourth-order, possibly anisotropic, tensor of linear elastic moduli, $\overline{\bfu}$ is the averaged total displacement vector, $\overline{\bfchi}$ is the incompatible part of the averaged elastic distortion tensor, $\overline{\bfu}- \overline{\bfz}$ is a vector field whose gradient is the compatible part of the averaged elastic distortion tensor and $\overline{\bfT}$ is the averaged stress tensor. The averaged slipping distortion tensor $\bfS$ is $\overline{\bfalpha} \times \overline{\bfV} + \bfL^p$. When the constitutive inputs $\bfL^p$ and $\overline{\bfV}$ are phenomenologically specified, the model is called Phenomenological MFDM (PMFDM) \cite{acharya_roy_2006} . 

From here onwards, \textit{fields without overhead bars refer to averaged fields}.

\subsection{DD-MFDM coupling}\label{sec:ddmfdm_coupling}
Here, we aim to obtain the constituitve inputs of MFDM theory using $PTA$.The values of the plastic distortion rate, ${\bfL}^p$ and the polar dislocation velocity, ${\bfV}$ need to be defined at every Gauss point of the MFDM FE mesh. %%should be obtained using \eqref{eq:Lp_evol} and \eqref{eq:V_evol} respectively using the value of the stress, ${\bfT}$, at the Gauss point.
For this, we divide the domain, $\Omega$, into $n \times n$ blocks. For example, in Fig. \ref{fig:blocksAndElements}, the domain is divided into $5 \times 5$ blocks. In turn, each block is a collection of a number of FEM elements that are used in the solution of the MFDM equations. Please note that the thickness of the block is the same as the thickness of the sample, which implies a state of plane stress. 

Every block contains a DD box of a fixed size in which DD simulations are performed. We assume the initial DD microstructure to be the same for the DD simulations corresponding to all the blocks. The stress that is used as input to perform the DD simulation in each block is the volume average of the stress obtained from the solution of the MFDM equations, over the block. The (linear, in-plane) dimension of the block, denoted by $B$, is therefore called the \emph{stress-averaging} size. Let the entire domain be denoted by $\Omega$ and the set of all points that lie within block $i$ be denoted as $\Omega_i$. The averaged stress for block $r$ is denoted as $\langle \bfT \rangle^B_r$, and is given by
\begin{align}
\langle \bfT\rangle^B_r= \frac{\int_{\Omega_r} T dv}{  |\Omega_r|},
\end{align}
where $|\Omega_r|:=B\times B \times a$, where $a$ is the thickness of the block/sample. The \emph{stress-averaging} size $B$ plays a crucial role in convergence of the solution for imposed inhomogeneous deformation. It is shown later in section \ref{sec:coupling_results} that the in-plane \emph{stress-averaging} size is limited approximately by the DD box size, in order to obtain a converged solution in such cases. However, for imposed homogeneous deformation, convergence in results occur for relatively large \emph{stress-averaging} sizes.

\textit{Since the size of the block remains fixed for the results in Section \ref{sec:coupling_results} (except for the convergence studies in Section \ref{sec:res_conv}), the superscript $B$ in $\langle\bfT\rangle^B_r$ is dropped from here onwards for notational convenience.}

DD simulations for block $r$ using $\langle\bfT\rangle_r$ at time $t$ and $t-\Delta$ are run to obtain measures of the plastic strain rate and dislocation velocity for that block by integrating \eqref{eq:Lp_evol}-\eqref{eq:V_evol}, which we denote as ${\bfL}^{p}_{r}$ and $\overline{\bfV}_{r}$, respectively. 

\subsubsection{Obtaining $\bfL^p$, $\overline{\bfV}$ at Gauss point of element} \label{sec:lpAtElem} 
Let the characteristic function of block $i$ is given by 
\[
\chi^i(\bfx)=
\begin{cases}
1, & \text{if} \quad \bfx \in \Omega_i \\
0, & \text{if} \quad \bfx \notin \Omega_i
\end{cases}
\]

Define $\widetilde{\bfL}^p(\bfx)$  by 
\[
\widetilde{\bfL}^p(\bfx)=\sum_{i=1}^{N_B} {\bfL}^{p}_i \chi^i (\bfx), \quad x \in \Omega
\]
where $N_B$ is the total number of blocks.

The field $\widetilde{\bfL}^p$ is discontinuous across blocks. To obtain a ($C^0$) continuous field in $\Omega$, we perform the following operations. We obtain an $L^2$ projection of $\widetilde{\bfL}^p$ on the finite dimensional space, $C^{0,B}$, formed by the linear span of globally continuous, piecewise smooth finite element shape functions corresponding to a FE mesh for $\Omega$, \emph{comprising the blocks} of size $B$ (the MFDM calculations involve another finer FE mesh that further discretizes the blocks). This projection, after discretization, gives the values of the plastic strain rate at the nodes of the blocks. Each block, in turn, contains many elements for the MFDM calculations, and we interpolate using the isoparametric shape functions for the blocks and for the elements within them to obtain the value of $\bfL^p$ at the MFDM elemental Gauss points. 

The above operations can be stated as follows. Define 
\[
\widehat{\bfL}^p :=  \argmin_{\bfL \in C^{0,B}(\Omega) }\int_{\Omega} \half |\bfL - \widetilde{\bfL}^p|^2 dv.
\]
To keep the debauch of indices to a minimum in what follows, we rename $\widehat{\bfL}^p:=\bfA$. The above definition translates to the following discrete statement:
\[
\sum_{R=1}^{{\sf N}} \sum_{Q=1}^{{\sf N}} \delta {A}^{R}_{ij} \Big[ \int_{\Omega_i} N^R \delta_{ik} \delta_{jl} N^Q dv \Big] {A}^{Q}_{kl} = \sum_{R=1}^{{\sf N}} \delta {A}^{R}_{ij} \int_{\Omega_i} N^R \delta_{ik} \delta_{jl} \widetilde{\bfL}^p_{kl} dv. 
\]
(note that $p$ is not an index). Here $\delta {\bfA}^{R}$ is a test function and $R$ and $Q$ are indices representing nodes of the $n \times n$ `block' FE mesh with $N^{R}$ and $N^Q$ denote global shape functions of the mesh. ${\sf N}$ denotes the total number of nodes of the block mesh. This results in a linear solve for the nodal values of $\bfA$ on the block FE mesh.

\begin{figure*}[t!]
    \centering
        \includegraphics[height=3.5in]{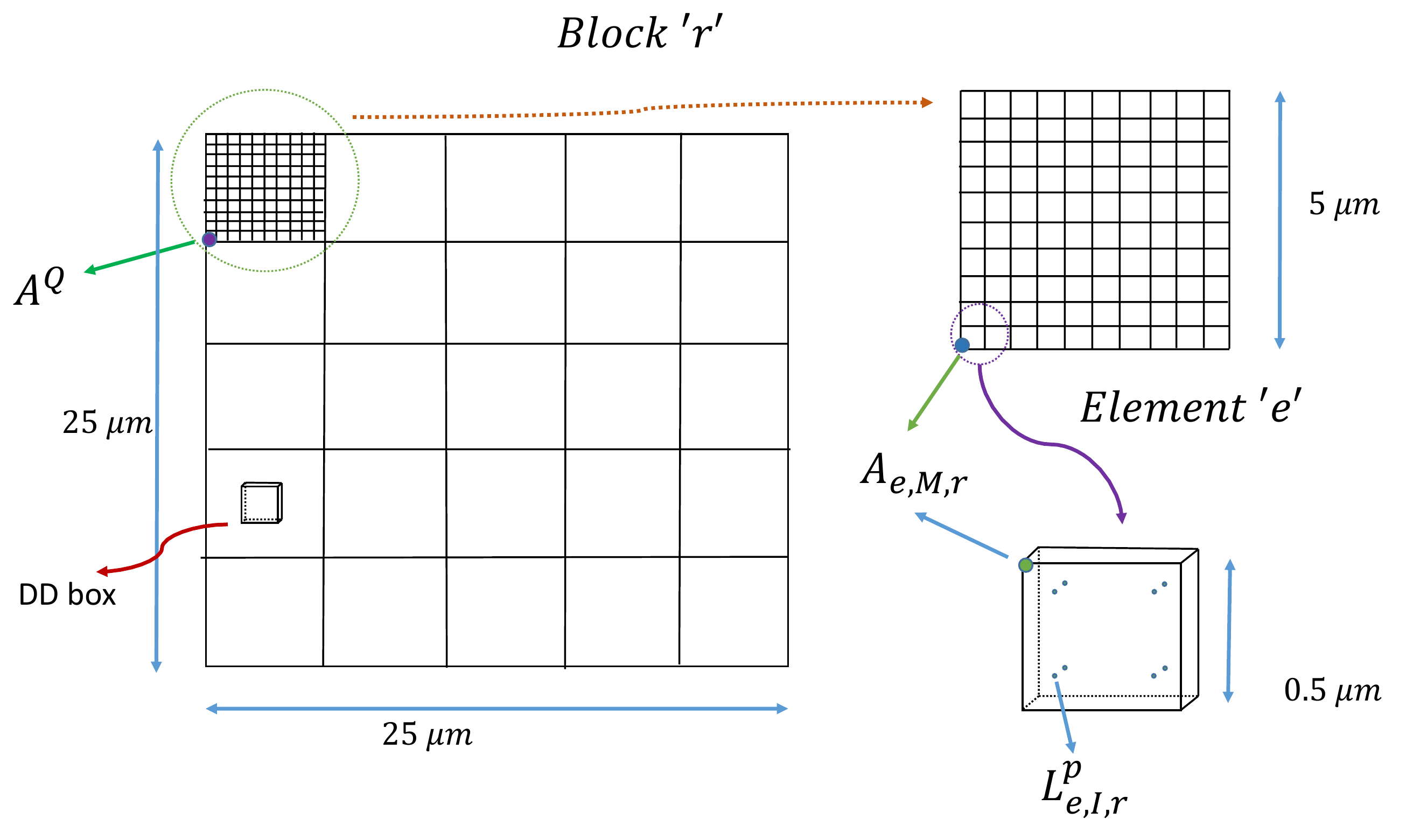}
        \caption{\textit{This figure shows the decomposition of the domain of size $25 (\mu m)^2$ into $5 \times 5$ blocks. Each block contains a DD box. Each block also consists of a number of elements $(10 \times 10$ in this figure$)$.}}    
    \label{fig:blocksAndElements}
\end{figure*}

With the nodal values of $\bfA$ determined so that it is a globally continuous function on the domain, we now determine the values of this continuous function at the Gauss points of the finite elements comprising the FEM mesh for the MFDM calculations (where $\bfA$ is needed as an input). This is done as follows. Let $M$ be a node of element $e$ that is contained in block $r$, whose isoparametric coordinate (with respect to the containing block $r$ that is an element of the block-FE solve) is denoted as $\xi^{r}_{e,M}$. Then $\bfA$ at node $M$ of element $e$ can be obtained as 
${\bfA}_{e,M,r}=\sum_{Q=1}^{N^v} {\bfA}^Q N^Q (\xi^r_{e,M})$, where $N^v$ is the number of nodes on a block (e.g. 8 for a hexahedral brick element).  Finally, $\bfL^p$ at Gauss point $I$ of element $e$ in block $r$ can be obtained as $\bfL^p_{e,I,r}=\sum_{K=1}^{N^v} {\bfA}_{e,K,r} N^K (\xi^e_I)$, where $K$ is a node of element $e$ (see Fig. \ref{fig:blocksAndElements}) and $\xi^e_I$ is the isoparametric coordinate of Gauss point $I$ in element $e$ (and we have made the (non-essential) assumption that the each element of the block-mesh and MFDM-mesh have the same number of nodes). 

We obtain the polar dislocation velocity at the Gauss point $I$ of element $e$, of block $r$, $\overline{\bfV}_{e,I,r}$ in the same way.

%% elements and blocks

\subsubsection{Ensuring non-negative dissipation}
Let the $\bfL^p$ and $\bfV$ (we revert here to dropping overhead bars) obtained at a Gauss point of an element (for MFDM calculations) as described above be denoted as $\bfL^p_{gp}$ and $\bfV_{gp}$, respectively. In order to ensure that the dissipation is non-negative , we redefine the ${\bfL}^p$ and ${\bfV}$ as
\begin{enumerate}
\item 
If ${\bfT} : {\bfL}^{p}_{gp}=d$ and $d < 0$, we take the component of ${\bfL}^{p}_{gp}$ given by
\begin{align}\label{eq:Lp_gauss}
{\bfL}^p={\bfL}^{p}_{gp} - d \frac{{\bfT}}{{|{\bfT}|}^2}.
\end{align} 
\item 
If $\beta={\bfV}_{gp} \cdot \left({\bfX}{\bfT} {\bfalpha} \right) <0$, we take the component of ${\bfV}_{gp}$ given by 
\begin{align}\label{eq:V_gauss}
{\bfV}={\bfV}_{gp} - \beta \frac{{\bfX}{\bfT} {\bfalpha}}{{|{\bfX}{\bfT} {\bfalpha}|}^2}. 
\end{align}
\end{enumerate}
Here, $\bfT$ and $\bfalpha$ are the stress and the dislocation density tensor at the Gauss point respectively, while $\bfX$ is the third order alternating tensor. The dissipation resulting from the components of $\bfL^p$ and $\bfV$ given by  \eqref{eq:Lp_gauss}  and \eqref{eq:V_gauss} is 0, which can be verified by taking an inner product of \eqref{eq:Lp_gauss}  and \eqref{eq:V_gauss} with $\bfT$ and $\bfX \bfT \bfalpha$, respectively. These $\bfL^p$ and $\bfV$ are used to solve the MFDM equations which is described in detail in Section \ref{sec:ddfdm_modifications}. 

\subsection{ Numerical Aspects}\label{sec:key_changes}
In this section, we discuss the various numerical aspects that are considered in the coupled DD-MFDM strategy. 
\begin{itemize}
\item \textbf{Setting the time interval $\Delta^*$ and the FDM time step $\Delta t$} \\For the average of the state functions ($R^\Lambda_t$ defined in \eqref{eq:R_lambda_t}) to converge, the fast dynamics, which is DD in this case, has to be run long enough. During this period, many junctions are formed and broken, as part of thermal activation described in Section \ref{sec:thermal_activation}. The period $t'$ should be much smaller than the interval $\Delta^*$ (a fraction of time period of slow time-scale $T_s$; the running time average $R^\Lambda_t$, of state functions of DD, are averaged over the nondimensional interval $\Delta=\frac{\Delta^*}{T_s} $ to generate coarse variables, see \eqref{eq:def-C}), due to the vast separation in the time-scale of the fast and the slow dynamics. Therefore, it is reasonable to say that $a \ll t' \ll \Delta^*$ (where the maximum junction breaking time $a$ is defined in the discussion following \eqref{eq:tb} in Section \ref{sec:thermal_activation}) and we fix $\Delta^*$ as $\Delta^*=n~a$, where $n$ is a positive integer (for the results presented in Section \ref{sec:coupling_results}, $n$ was set as $100$).  

The MFDM time step is denoted as ${\Delta t}$, and given by ${\Delta t}=m~\Delta^*$, where $m$ is a positive integer (for the results presented in Section \ref{sec:coupling_results}, $m$ was set as $10$). This implies the condition $\Delta^* \ll {\Delta t}$, which is a necessary constraint for the application of PTA (see Section 7 in \cite{cspde_2018}). 

The above discussion can be summarized as the following constraint: 
\begin{align}\label{eq:constraint_delta*}
a \ll \Delta^* \ll {\Delta t}.
\end{align}

\item \textbf{The \emph{limit} load}\\The MFDM system evolves in a stable way when the plastic strain increment is less than a threshold of 0.2\% in a given time increment, which is stated as
\begin{align}\label{eq:ddfdm_stability}
\Delta t \leq \frac{0.002}{|\bfalpha \times \bfV| + |\bfL^p|}.   
\end{align}

Equations \eqref{eq:ddfdm_stability} and \eqref{eq:constraint_delta*} have to be always satisfied and form the constraints of the DD-MFDM coupling problem. However, when $|\bfL^p|$ is high, such that 
\begin{align}\label{eq:limit_cond}
\frac{0.002}{|\bfalpha \times \bfV| + |\bfL^p|} \leq \Delta^*,
\end{align}
for one or more blocks, \eqref{eq:ddfdm_stability} and \eqref{eq:limit_cond}, when combined together, may violate \eqref{eq:constraint_delta*}. For instance, if $\Delta^*=0.01~s$ and $|\bfalpha \times \bfV|+|\bfL^p|=0.03~s^{-1}$, \eqref{eq:limit_cond} is satisfied and $\Delta t \leq 0.0067<\Delta^*$ by \eqref{eq:ddfdm_stability}. Thus, \eqref{eq:constraint_delta*} is violated. In such situations, since $\Delta t$ is free to choose, we explicitly set it as $\Delta t=\Delta^*$. When $\bfL^p$ is high, it is physically expected that the local flow stress either stays fixed or decreases. Based on this, we assume that the local stress at time  $t-\Delta^* (=t-\Delta t)$ and  $t$ are the same, which implies $\dot{\bfL}^p=0$ and $\dot{\bfV}=0$ by \eqref{eq:Lp_evol}. When such a plastic instability happens at any point, we declare that the system has reached a \emph{limit} load and do not allow the external loading to increase, i.e., $L =0$ in \eqref{eq:coarseC-def} (we consider that the simulation is performed in a sophisticated loading apparatus). 

However, if $\dot{\bfL^p}$ following \eqref{eq:Lp_evol} is such that it reduces $|\bfL^p|$ to a value such that \eqref{eq:limit_cond} is not true, $\bfL^p$ is allowed to evolve using $\dot{\bfL^p}$ for that block. If it happens at any time that none of the blocks satisfy \eqref{eq:limit_cond}, then the system is no longer in the state of \emph{limit} load. In that case, the loading rate is set back to the prescribed non-zero value for the problem. Hence, the system is allowed to get out of the \emph{limit} load condition in a consistent manner. 
\end{itemize}

We next outline the algorithm of the coupled strategy, which is based on PMFDM algorithm but is modified to incorporate the above features. 

%As the value of $|\bfL^{p,k}|$ keeps increasing with the progress of the simulation, the value of ${\Delta t}^hk$ will keep dropping (due to the constraint \eqref{eq:mfdm_stability}) and it will approach $\Delta^*$. 
%When ${\Delta t}^k \approx \Delta^*$, we consider that the \emph{limit} load has been reached. In this situation, $|\dot{\bfL}^{p,k}| \to 0$ and we enable that by setting the loading rate, $l \to 0$. 

\subsubsection{Algorithm of DD-MFDM coupling}\label{sec:ddfdm_modifications}

%\vspace{1cm}

\begin{figure}[h!]
\begin{center}
\begin{tikzpicture}[node distance=2cm]

\node (pro1) [process] {Average stress over all gauss points in a block.};
\node (pro2) [process, below of=pro1] {Apply PTA using averaged stress to calculate $\dot{\bfL}^p$ and $\dot{\bfV}$ and $\bfL^p$ and $\bfV$.};
\node (pro3) [process, below of=pro2] {Use $\bfL^p$ and $\bfV$ to solve MFDM to get $\bfalpha$, $\bfchi$, $\bfz$ and $\bfu$.};
\node (pro4) [process, below of=pro3] {Enforce constraint on MFDM time step based on plastic strain increment and $\bfV$. March forward in time and repeat above steps.};

\draw [arrow] (pro1) -- (pro2);
\draw [arrow] (pro2) -- (pro3);
\draw [arrow] (pro3) -- (pro4);

\end{tikzpicture}
\end{center}
\caption{Overview of the DD-MFDM coupling strategy}
\label{fig:ddfdm_flowchart}
\end{figure}
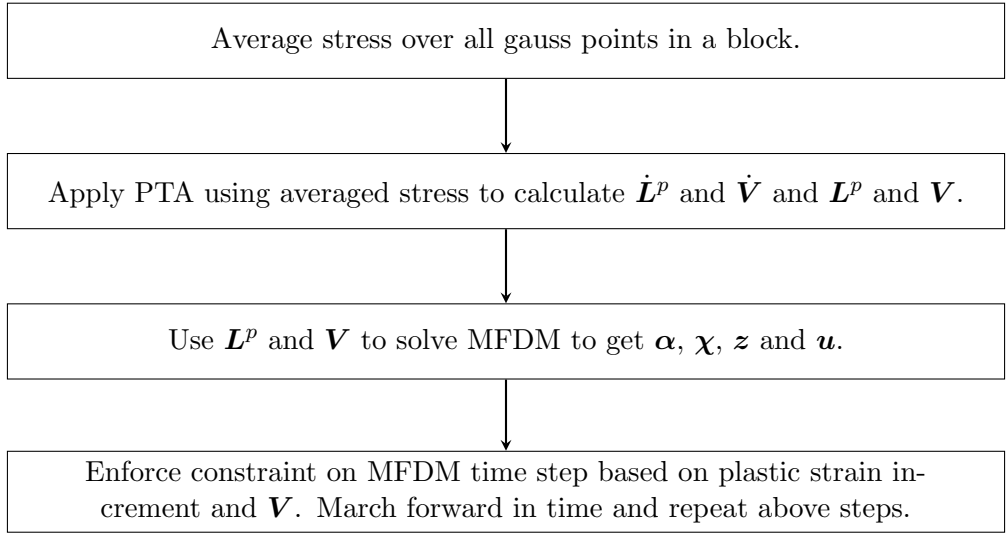

A flowchart comprising the key steps in the coupled DD-MFDM strategy is shown in Fig. \ref{fig:ddfdm_flowchart}. In the following part of this section, we describe the above steps in more detail.  

%{\color{change}Fig.  \ref{fig:blocksAndElements} shows the decomposition of the domain into blocks and elements. The domain is decomposed into $n \times n$ blocks(where $n=5$ in Fig. \ref{fig:blocksAndElements}). Each block consists of $n' \times n'$ elements ($n'=10$ in Fig. \ref{fig:blocksAndElements}). Each block contains a DD box in which PTA calculations are performed to obtain }
Let $BC$ and $IC$ be the abbreviation for \emph{Boundary Condition} and \emph{Initial Condition} respectively.  

{\bf Step 1}: 

BC : $\bfz$ constrained to prevent rigid body motion; $ {\bfchi}  {\bf n} =\bf0$, where $\bfn$ is the outward unit normal at the boundary surface, 

IC : $ {\bfalpha}_{0} $ is prescribed. 

Solve for ${\bfchi}_{0}$.  
Solve for $\bfz$ and the initial state of stress due to ${\bfalpha}_0$. 
%There is no $\bfz$ in this step and we use $\bfu$ as a proxy for $\bfz$. 

%{\bf Step 2} : 

%Calculate the state of stress under no loads and zero displacement. 

%Slip and elastic distortion at the end of this step are adjusted to include the effect of using $\bfu$ as a proxy for $\bfz$ in Step 1, and are given by: \\
%
%$({\bfU}^e)^2=({\nabla {\bfu}})^2 + ({\nabla {\bfz}})^2 + ({\bfchi})^2 + ({\nabla {\bfu}})^1 + ({\nabla {\bfz}})^1 $ \\
%
%$({\bfU}^p)^2=-(({\nabla {\bfu}})^2 + ({\nabla {\bfz}})^2 + ({\bfchi})^2 + ({\nabla {\bfu}})^1 + ({\nabla {\bfz}})^1 )$. \\
%
%Hence, $({\bfU}^p)^2 = -({\bfU}^e)^2$ and $\bfu$ is $\bf 0$ everywhere. \\

{\bf Step 2} : 

In case a problem on the elastic theory of continuously distributed dislocations (ECDD) needs to be solved for the distribution $\bfalpha_0$ with applied displacement and/or traction boundary conditions, impose displacement and traction BCs as per the physical problem we are trying to solve. Superpose the initial state of stress due to ${\bfalpha}_0$ with the stress due to displacement and traction BCs. This is done by solving the MFDM problem with $\bfV$ and ${\bfL}^p$ set to $\bf 0$. \\

{\bf Step 3} : 
Now initialize the MFDM problem. \\
IC: $ {\bfu}, {\bfz}, {\bfalpha}$ and $\bfchi$ to be retrieved from the previous step results. \\
BC: $ {\bfchi}  {\bf n} =0$ at the boundary, which implies that the incompatible part of $\bfU^e$ is $\bf0$ if $\bfalpha=\bf0$. $\bfz$ is to be specified at one point to get a unique solution. \\
The BC on $\bfalpha$ can be specified in two ways, which are called the \emph{constrained} and \emph{unconstrained} cases. In the \emph{constrained} case, the body is plastically constrained on the boundaries and dislocations cannot exit the body, but can only move in a tangential direction at the boundary. The BC for this case is $(\bfalpha \times \bfV + \bfL^p) \times \bfn = 0$ on the boundary. A less restrictive BC which corresponds to the \emph{unconstrained} case is the imposition of the dislocation flux $\bfalpha (\bfV \cdot \bfn)$ on the inflow points on the boundary (where $\bfV \cdot \bfn=0$), along with the specification of $\bfL^p \times \bfn$ on the entire boundary. This condition allows the free exit of GNDs at the outflow points. 

The time step at the first increment is ${\Delta t}^0={\Delta t}_{pres}$, where ${\Delta t}_{pres}$ is the prescribed time step for the problem. The total time of the simulation is $T_s$. \\
The steps are as follows. For every increment $k$ (while $t^k < T_s$),
\begin{enumerate}
\item The time step ${\Delta t}^k$ is subjected to the time step controls in items \ref{item:deltk}, \ref{item:time_step} and \ref{item:stressBased} below. 
\end{enumerate}
~ ~ For each block $r$: 
\begin{enumerate}[resume]
{\setlength{\itemindent}{0.25in}
\item Obtain the averaged stress at the start of increment $k$, $\langle{\bfT}\rangle^{k}_r$ (the values of $\bfu^k$, $\bfz^k$, $\bfalpha^k$ and $\bfchi^k$ are known at all Gauss points at the start of increment $k$). Next, we pass the stress $ \widehat{\bfT}^{k}_{r,t-\Delta}=\langle{\bfT}\rangle^{k}_r$ and $\widehat{\bfT}^{k}_{r,t}=\langle{\bfT}\rangle^{k}_r + \frac{\Delta^*}{ {\Delta t}^{k-1}} \left (\langle{\bfT}\rangle^{k}_r - \langle{\bfT}\rangle^{k-1}_r \right)$ to run PTA at block $r$. The time interval $\Delta^*$ was defined in Section \ref{sec:key_changes}. 
\item \label{step:stress_threshold} If  $|\widehat{\bfT}^k_{r,t-\Delta}|$ and  $|\widehat{\bfT}^k_{r,t}|$ are close to each other ($|  \widehat{\bfT}^k_{r,t} -  \widehat{\bfT}^k_{r,t-\Delta} |$ is less than a threshold, which was found to be around $0.5 ~ MPa$), the numerator on the rhs of \eqref{eq:Lp_evol} (which gives $\bfL^p$) becomes small and DD cannot resolve it, which is a \emph{limitation} of DD and the library MoDELib which we used to implement DD. In that case, since $ \widehat{\bfT}^k_{r,t}$ is the only variable we are free to modify, we change it, while keeping  $\widehat{\bfT}^k_{r,t-\Delta}$ fixed, such that the difference is 0.5 MPa. It is obtained as: 
$mag=\frac{0.5}{| \widehat{\bfT}^k_{r,t} - \widehat{\bfT}^k_{r,t-\Delta}|}$, if $mag > 1$,  $\widehat{\bfT}^k_{r,t}=\widehat{\bfT}^k_{r,t-\Delta} + mag ~( \widehat{\bfT}^k_{r,t}-\widehat{\bfT}^k_{r,t-\Delta})$. 
\item \label{item:lp_V_ddfdm} Obtain $\dot{{ \bfL}}^{p,k}_r$ and $\dot{\overline{\bfV}}^{k}_r$ using PTA (using \eqref{eq:Lp_evol} and \eqref{eq:V_evol} respectively). 
\item If $mag>1$, then scale $\dot{{\bfL}}^{p,k}_r$ and $\dot{\overline{\bfV}}^{k}_r$ down by $mag$ i.e. $\dot{{ \bfL}}^{p,k}_r=\frac{\dot{{ \bfL}}^{p,k}_r}{mag}$ and $\dot{\overline{\bfV}}^{k}_r=\frac{\dot{\overline{\bfV}}^{k}_r}{mag}$. This is because ideally $\dot{{ \bfL}}^{p,k}_r$ should be generated from  $\widehat{\bfT}^{k}_{r,t-\Delta}$ and  $\widehat{\bfT}^{k}_{r,t}$ using PTA as per Step \ref{item:lp_V_ddfdm} above. However, due to the restriction imposed by DD on the minimum threshold of the difference  $|\widehat{\bfT}^k_{r,t}-\widehat{\bfT}^k_{r,t-\Delta}|$,  the value of  $\widehat{\bfT}^k_{r,t+\Delta}$ was modified in order to scale up the difference to 0.5 MPa, as outlined in Step \ref{step:stress_threshold}. Thus, the resulting $\dot{{\bfL}}^{p,k}_r$ must be scaled down such that it corresponds to the original stress difference between  $\widehat{\bfT}^k_{r,t}$ and  $\widehat{\bfT}^k_{r,t-\Delta}$. 
\item Calculate ${\bfL}^{p,k}_r={\bfL}^{p,k-1}_r + \dot{{\bfL}}^{p,k-1}_r \, {\Delta t}^{k-1}$ and $\overline{\bfV}^{k}_r=\overline{\bfV}^{k-1}_r + \overline{\bfV}^{k-1}_r \, {\Delta t}^{k-1}$.}
\item \label{item:deltk} We check if the \emph{limit} load has been reached by checking if $\frac{0.002}{|\bfalpha^k \times \bfV^k| + |{\bfL}^{p,k}_r|} \leq \Delta^*$ (equation \eqref{eq:limit_cond}) at any block $r$. If yes, we set the loading rate $L$ to 0, otherwise we keep it at the prescribed value for the problem. We also set ${\Delta t}^k=\Delta^*$. Moreover, if for any block $r$, equation \eqref{eq:limit_cond} is satisfied, we set ${\bfL}^{p,k+1}_r={\bfL}^{p,k}_r$ and $\overline{\bfV}^{k+1}_r=\overline{\bfV}^{k}_r$  (which is equivalent to setting $\dot{\bfL}^{p,k}_r=0$ and $\dot{\overline{\bfV}}^k_r=0$). The justification for these assignments is provided in Section \ref{sec:key_changes}. %This also implies that $\dot{\overline{\bfL}}^{p,k}_r=0$ and $\dot{\overline{\bfV}}^{k}_r=0$, following the argument surrounding \eqref{eq:limit_cond}. 
\end{enumerate}
~ ~ For all Gauss points, 
\begin{enumerate}[resume]
%\begin{description}
%%{\setlength{\itemindent}{0.25in}
\item \label{item:calc_lp_v} Calculate ${\bfL}^{p,k}$ and ${\bfV}^k$ as follows: 
\begin{enumerate}
\item Obtain ${\bfL}^{p,k}_{gp}$ and $\bfV^k_{gp}$ at Gauss points of elements in block $r$ from ${\bfL}^{p,k}_r$ and $\overline{\bfV}^{k}_r$ using the procedure described in Section \ref{sec:lpAtElem}. 
\item Modify ${\bfL}^{p,k}$ and ${\bfV}^k$ using \eqref{eq:Lp_gauss} and \eqref{eq:V_gauss} respectively, to ensure non-negative dissipation at every Gauss point. 
\end{enumerate}
%\begin{enumerate}[resume]
\item \label{item:time_step} The numerical stability condition is given by: 
\begin{align}\label{eq:mfdm_stability}
{\Delta t}^k \leq min_{gp} \Big(\frac{0.002}{|{\bfalpha}^k \times {\bfV}^k|+ |\bfL^{p,k}|}, f  \frac{d}{|\bfV|} \Big), ~ f \sim 0.1
\end{align}
where $gp$ is the set of all Gauss points in the sample and $d$ is a minimum element edge length. This reflects a conservative choice between a Courant condition and a maximum bound of 0.2\% on the plastic strain increment. 
\item Impose displacement and/or traction boundary condition. 
%%\end{enumerate}
%%\begin{enumerate}[resume]
%%\setlength{\itemindent}{0.25in}
\item \label{item:alphaSolve}Solve $\bfalpha$ equation \eqref{eq:alpha_evol} for ${\bfalpha}^{k+1}$: ${\bfalpha}^{k+1} = {\bfalpha}^k - {\Delta t}^k curl({\bfalpha}^{k+1} \times {\bfV}^k + {\bfL}^{p,k})$. 
\item Solve $\bfchi$ equation \eqref{eq:chi_evol} for ${\bfchi}^{k+1}$: ${\bfalpha}^{k+1}= curl {\bfchi}^{k+1}$ and $div {\bfchi}^{k+1}=0$. \
\item Solve $z$ equation \eqref{eq:z_evol} for ${\bfz}^{k+1}$: $ div (grad  ~ \dot{\bfz}^k) = div ({\bfalpha}^{k+1} \times {\bfV}^k + {\bfL}^{pk}) $. 
\item \label{item:uSolve}Solve $\bfu$ equations \eqref{eq:u_evol} for ${\bfu}^{k+1}$: $div {\bfT}^{k+1} = 0 $, $ {\bfT}^{k+1}={\bf C}:{\bfU}^{e(k+1)} $ , $ {\bfu}^{e(k+1)}=grad({\bfu}^{k+1} -{\bfz}^{k+1}) + {\bfchi}^{k+1} $. %\end{description}

If ${\Delta t}^k$ does not satisfy
\[
{\Delta t}^k <= min_{gp} \left(\frac{0.002}{|{\bfalpha}^{k+1} \times {\bfV}^{k+1}|+ |\bfL^{p,k}|}, f  \frac{d}{|\bfV^{k+1}|} \right),
\]
then it is likely that the computed state at increment $k+1$ gives rise to a large plastic strain rate, and the increment from $k$ to $k+1$ should be done with a smaller time step to have better control on the evolution. Therefore, as a preemptive measure, set it as 
\[
{\Delta t}^k = min_{gp} \left(\frac{0.002}{|{\bfalpha}^{k+1} \times {\bfV}^{k+1}|+ |\bfL^{p,k+1}|}, f  \frac{d}{|\bfV^{k+1}|} \right). 
\]
 Then reinitialize the increment $k$ and go to Item 1 of Step 4. This process of rerunning the increment is called \emph{cutback}. 
 
 Special algorithms are required to solve the MFDM equations (items \ref{item:time_step} through \ref{item:uSolve} above). These algorithms can be found in \cite{acharya_roy_2005, acharya_roy_2006}. 

 \item \label{item:stressBased}An additional stress-based time step control is placed due to the introduction of DD in the MFDM problem. It is implemented as follows. Compute $|  \widehat{\bfT}^{k+1}_{r,t-\Delta} - \widehat{\bfT}^k_{r,t-\Delta} |$, if it is greater than a threshold (assumed to be $3 ~ \textrm{MPa}$), then reduce ${\Delta t}^k$, calculated using item \ref{item:deltk} above, by half, and rerun the current increment. If in this process, ${\Delta t}^k$ comes out less than $\Delta^*$, then put ${\Delta t}^k=\Delta^*$. Restricting the value of $| \widehat{\bfT}^{k+1}_{r,t-\Delta} - \widehat{\bfT}^k_{r,t-\Delta} |$ to within a threshold by reducing the time step has been found to make the evolution of the DD-MFDM coupled problem more stable, as the DD microstructure is not subjected to high variation in the applied stress that goes into the PTA calculation, between consecutive time steps. 
\item If it happens that ${\Delta t}^k < {\Delta t}_{step}$ and $min_{gp} \left( \frac{0.002}{|{\bfalpha}^k \times {\bfV}^k|+ |\bfL^{p,k}|} , f \frac{d}{\bfV} \right) \geq {\Delta t}^k$ (where $gp$ is the set of all Gauss points in the sample) for two consectutive  steps, then double ${\Delta t}^k$. This increases the time step when plastic strain rate reduces.
\end{enumerate}

\textbf{Remark.} There is only stress-coupling between DD and MFDM in this first exercise. The DD microstructure can also be coupled to other descriptors and the density in the DD boxes can be adjusted through reinsertion of segments (which was discussed briefly in Section \ref{subsec:reinsertion}) in tune with such descriptors. One such descriptor is the GND density $\overline{\bfalpha}$ predicted by the coupled MFDM-DD strategy, whose field equation is provided in \eqref{eq:MFDM}. 

More importantly, another descriptor - the averaged total dislocation density $\overline{\rho}$ (whose evolution equation is derived in \cite{hierarchy}\footnote{The evolution of $\overline{\rho}$ is given by
\begin{align*}\label{eq:rhobardot}
\dot{\overline{\rho}}=& - grad~ \overline{\rho} \cdot \overline{\bfV} - 2 ~\overline{\rho}~ div \overline{\bfV} + 2 ~\overline{\bfalpha} : ( div \overline{\bfalpha} \otimes \overline{\bfV} ) + 2~ \overline{\bfalpha}: \{ \overline{\bfalpha} ~ grad \overline{\bfV} \} - \overline{ {\Sigma}^{grad {\rho}} \cdot {\Sigma}^{\bfV}}  \nonumber \\
& - 2 \overline{{\Sigma}^{\rho} {\Sigma}^{div V}} + 2 ~\overline{\bfalpha} : ( \overline{ \Sigma^{div \bfalpha}  \otimes \Sigma^{\bfV}} ) + 2 \overline{ \Sigma^{\bfalpha} : \Sigma^{ div \bfalpha \otimes \bfV} } +  2~ \overline{\bfalpha} : \overline{ {\Sigma}^{\bfalpha} ~ {\Sigma}^{grad{\bfV}} } \nonumber\\
& + 2~ \overline{ {\Sigma}^{\bfalpha} :  {\Sigma}^{ {\bfalpha} ~ {grad{\bfV}} }  },
\end{align*}
where $\Sigma^{(\cdot)}$ represents the fluctuation of the quantity $(\cdot)$ and is defined as
\[
\Sigma^{(\cdot)}=(\cdot) - \overline{(\cdot)},
\]
where the space-time averaged field $\overline{(\cdot)}$ is obtained using an averaging procedure utilized in the literature for multiphase flows (see \cite{babic1997}). }), where the microscopic total dislocation density is defined as $\rho:=\bfalpha:\bfalpha$, needs to be solved and evolved as an additional equation in MFDM-DD coupling, thus augmenting its current structure. These descriptors will act as feedback for the initialization of the DD microstructure at discrete time steps. 

\subsection{Results and discussion}\label{sec:coupling_results}
In this section, we present results on the 
\begin{itemize}
\item convergence
\item orientation effect
\item rate effect
\item effect of initial DD microstructure
\end{itemize}
for the DD-MFDM coupled problem under load and displacement control. 

Following the discussion in Section \ref{subsec:dd_setup}, there are two cases into which the results can be categorized:
\begin{itemize} 
\item \textbf{Case 1.} The sessile segments are constructed as Lomer Cottrell (LC) locks, with their Burgers vector out of the slip plane.
\item \textbf{Case 2.} The sessile segments are constructed such that their Burgers vector lie in the slip plane. 
\end{itemize}
Most of the results presented in this Section correspond to Case 1, while a few results for Case 2 have also been presented. The justification for the preference of Case 1 in has been provided in Section \ref{subsec:dd_setup}. 

\subsubsection{Case 1 with Load Control}
\begin{figure*}[t!]
        \centering
        \includegraphics[width=0.5\linewidth]{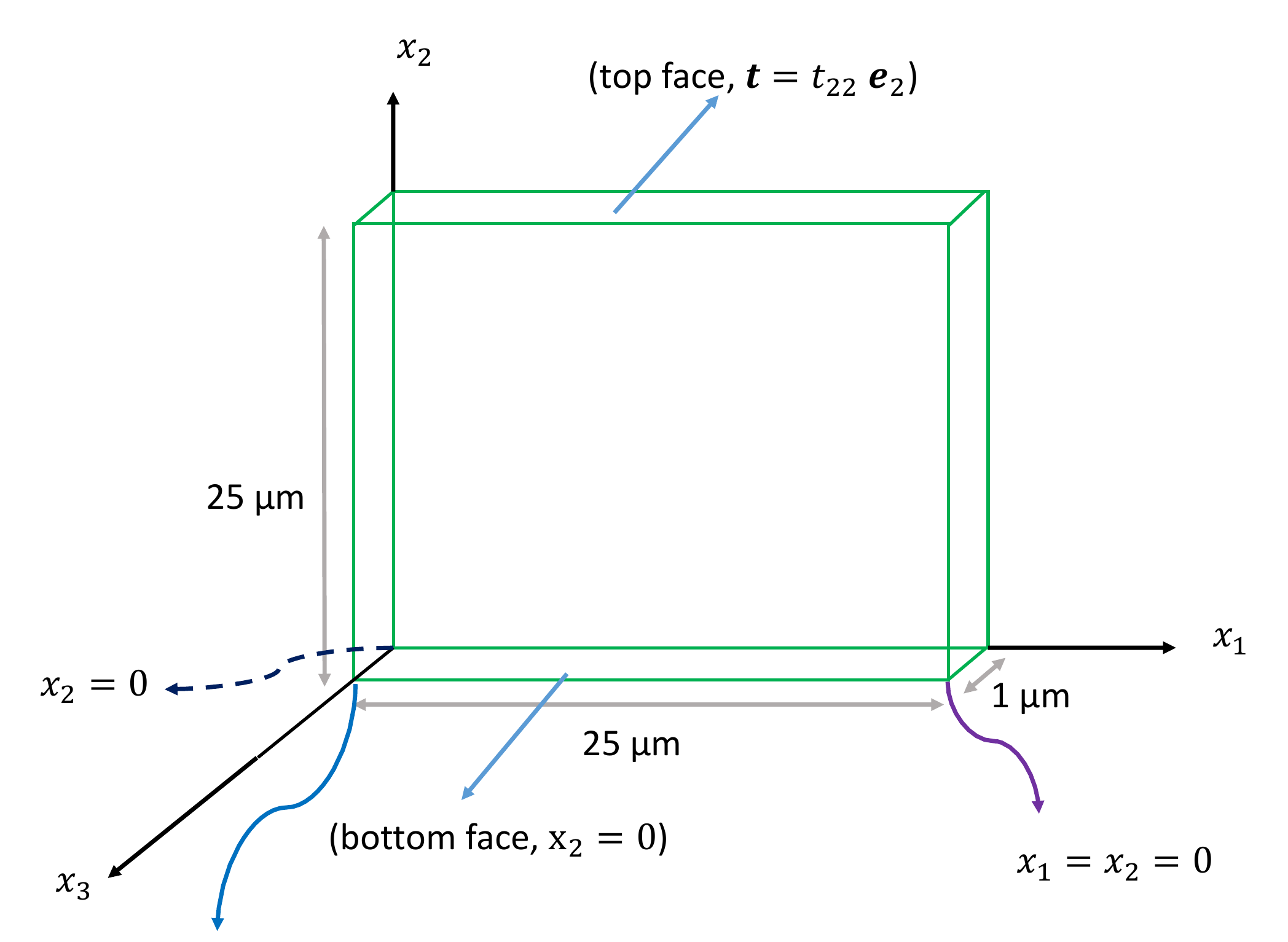}
        \caption{\textit{\small Boundary conditions for uniaxial tension}.}
         \label{fig:uniaxial_tension_bc}
\end{figure*}

We apply two load cases of simple shear and uniaxial tension. The boundary conditions for the two loading cases are as follows. Standard displacement boundary condition to prevent rigid body motion is applied. For uniaxial tension, we apply the traction $\bft=t_{22} \bfe_2$ on the top face and keep the bottom face fixed in the $Y$ direction ($x_2=0$), as shown in Fig. \ref{fig:uniaxial_tension_bc}. For the shear problem, we apply the traction $\bft=t_{12} \bfe_2$ and $\bft=t_{12} \bfe_1$ on the top and right face respectively, and $\bft=-t_{12} \bfe_1$ and $\bft=-t_{12} \bfe_2$ on the left and bottom faces respectively. The load ($t_{12}$ for the shear problem and $t_{22}$ for the tension problem) depends on the loading rate $l$, which is set as $1 MPa/s$ unless the \emph{limit} load is reached, in which case it is set to 0. All simulation details are mentioned in Table \ref{tab:simulation_details} in Section \ref{sec:sgp_results}. 

\textbf{Convergence} \label{sec:res_conv}
We choose a $25 \mu m \times 25 \mu m \times 1 \mu m$ sample and divide it into $2500$  (tri)linear brick elements each of size $0.5 \mu m \times 0.5 \mu m \times 1 \mu m$. As introduced and explained in Section \ref{sec:ddfdm_coupling}, we  divide the domain into $5 \times 5$, $7 \times 7$ and $10 \times 10$ blocks with \emph{stress-averaging} size of $5 \mu m$, $3.5 \mu m$ and $2.5 \mu m$ respectively. We perform DD simulations in each such block (in parallel). 

%%boundary conditions

%%convergence_25mic_t22
\begin{figure}[!h]
\centering
\begin{minipage}{.45\textwidth}
  \centering
  \includegraphics[width=\linewidth]{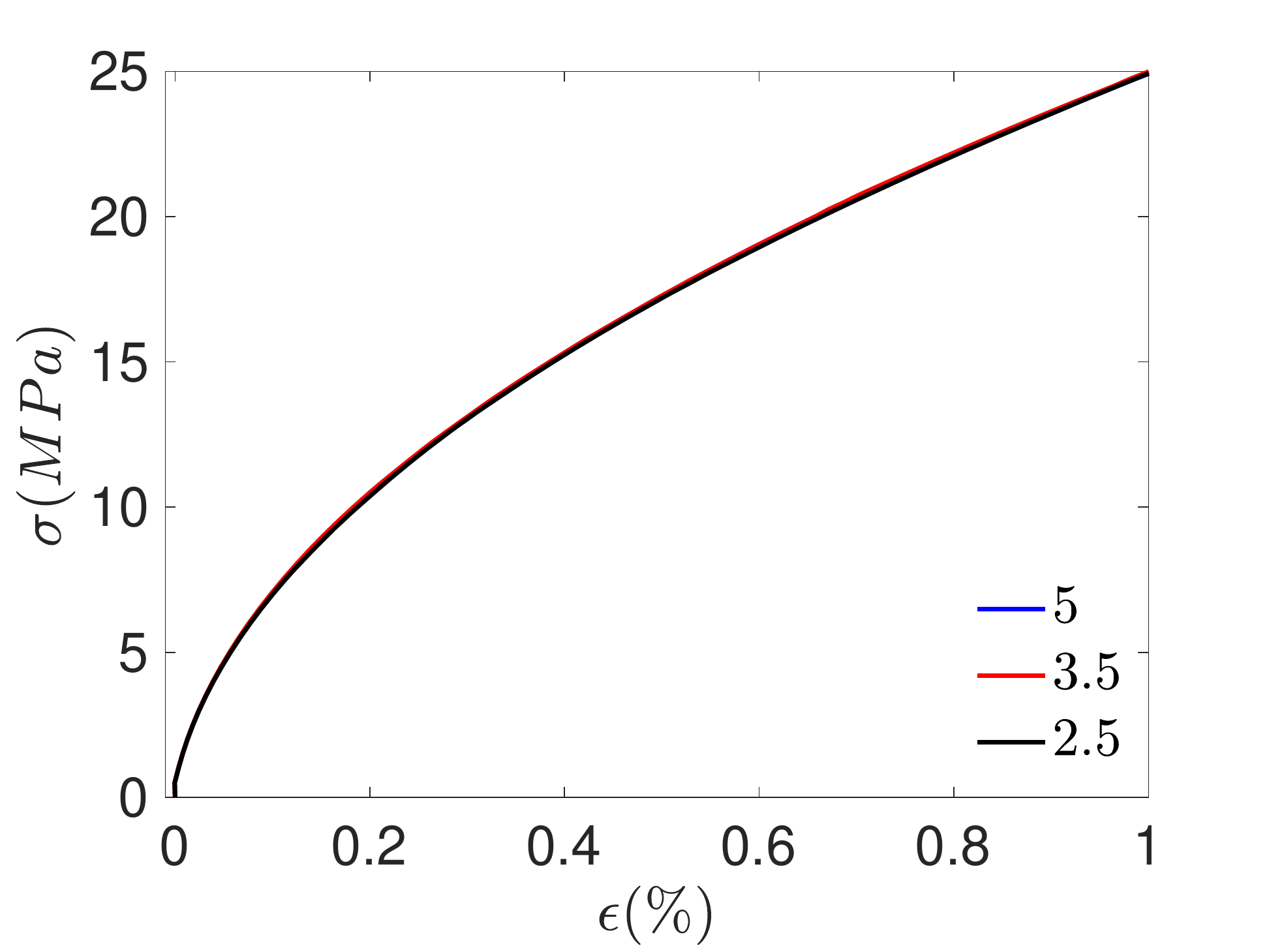}
  \caption{\textit{\small Convergence in stress-strain response for 25 $\mu m$ sample in tension for different stress-averaging sizes. The averaging blocks are squares with edge-lengths in $\mu m$.}}
  \label{fig:conv_25mic_t22}
\end{minipage}%
\hfill
\begin{minipage}{.5\textwidth}
  \centering
  \includegraphics[width=\linewidth]{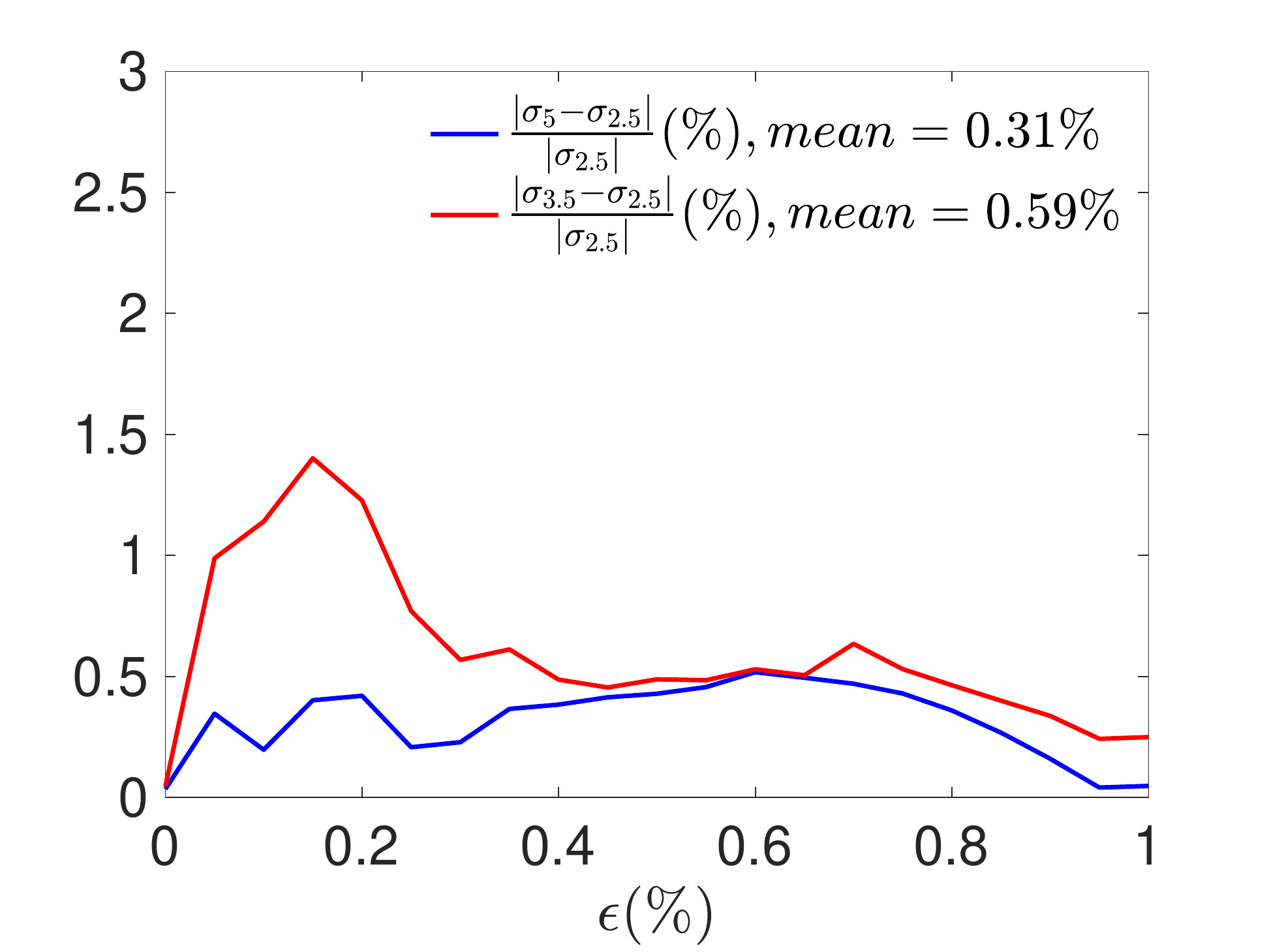}
  \caption{\textit{\small Relative error between the responses in uniaxial tension}.}
  \label{fig:conv_25mic_t22_rel_err}
\end{minipage}
\end{figure}

%%convergence - 400 mic
\begin{figure*}[t!]
        \centering
        \includegraphics[width=0.4\linewidth]{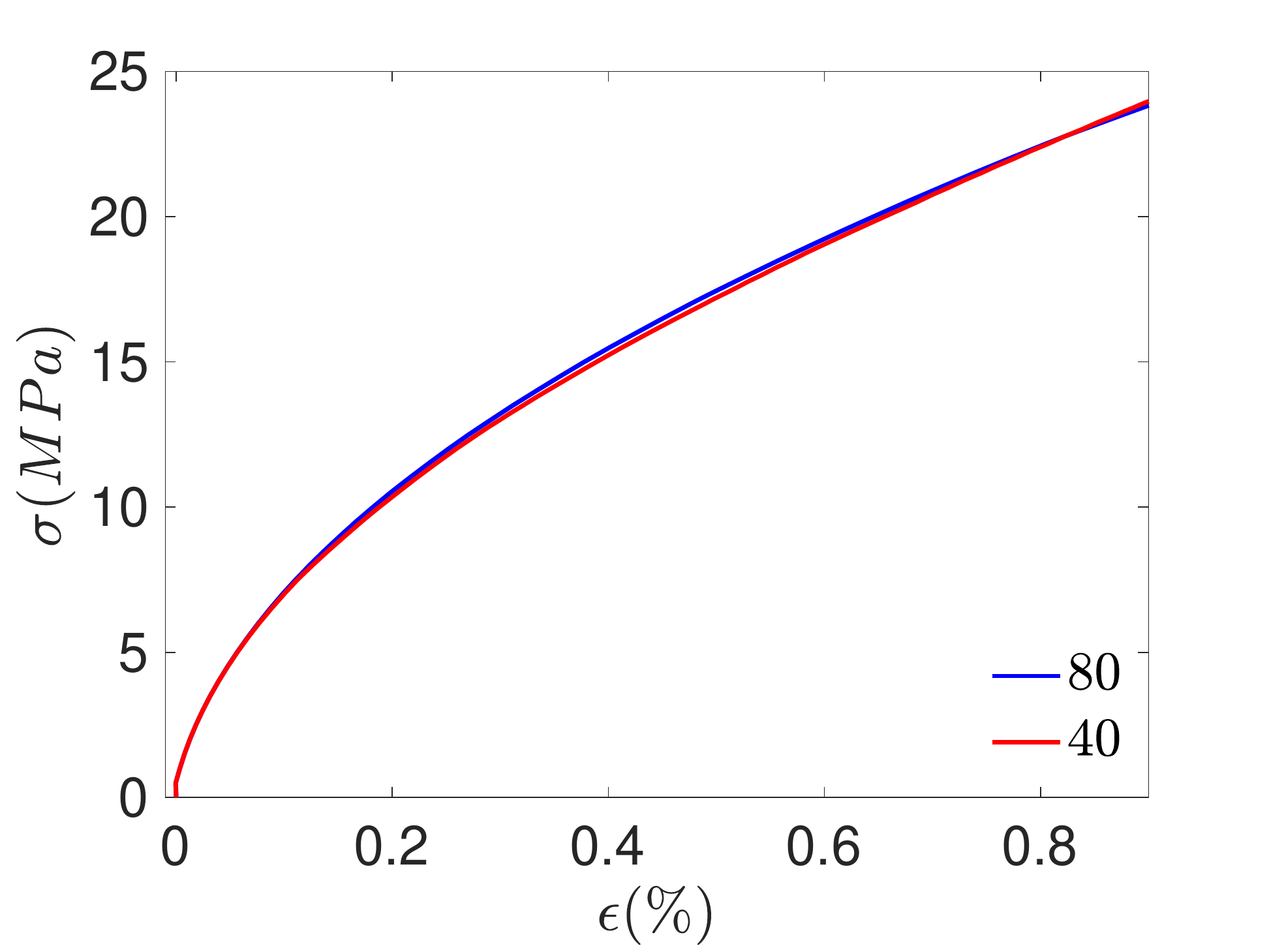}
        \caption{\textit{\small Convergence in stress-strain response for 400 $\mu m$ sample in tension for different stress-averaging sizes \emph{(in $\mu m$)}.}}
         \label{fig:conv_400mic_t22}
\end{figure*}

%%convergence_8by32mic_t12
\begin{figure}[!h]
\centering
\begin{minipage}{.45\textwidth}
  \centering
  \includegraphics[width=\linewidth]{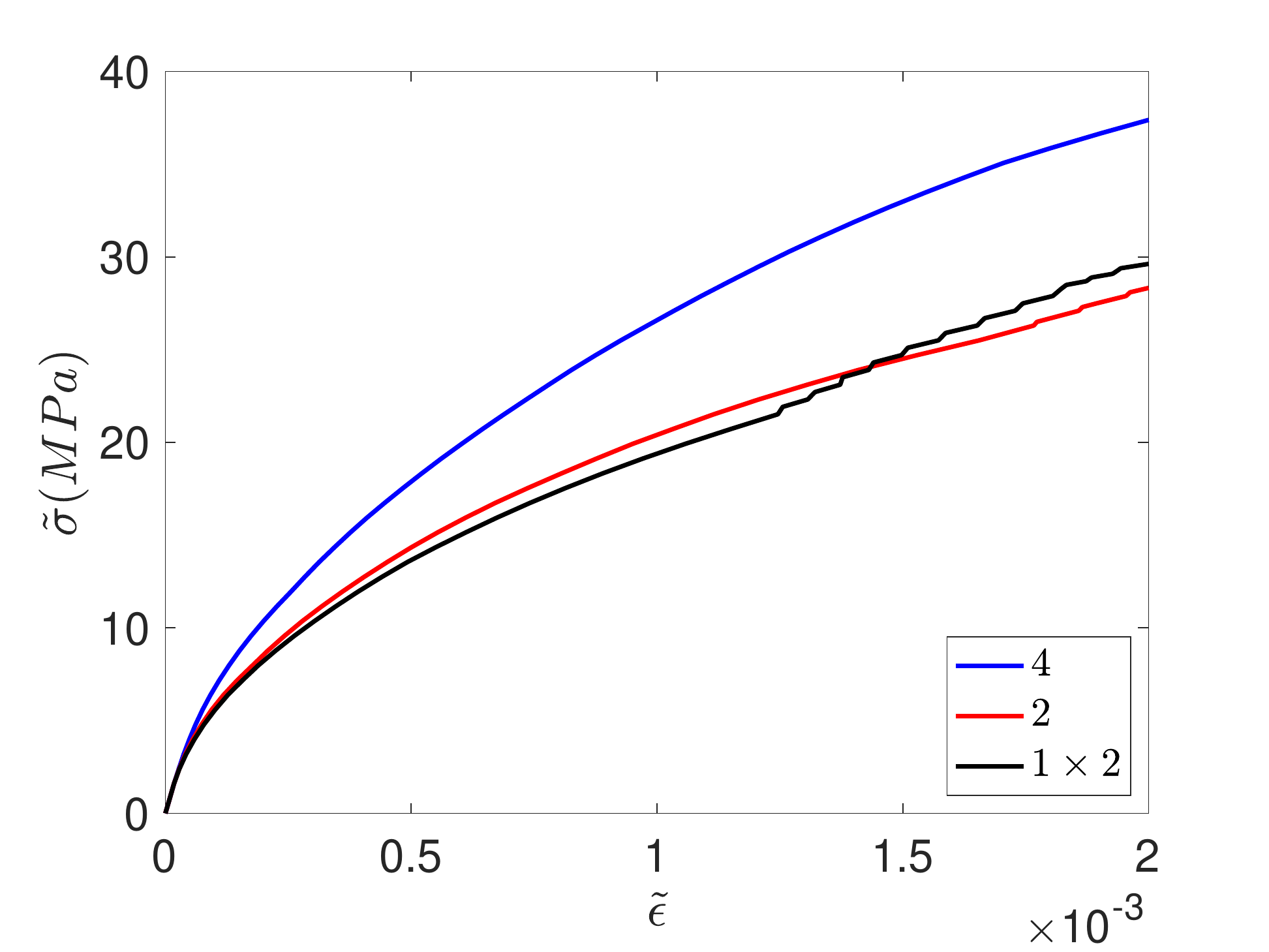}
  \caption{\textit{\small Convergence in effective stress-strain response for $8 \mu m \times 32 \mu m $ sample in pure bending for different stress-averaging sizes \emph{(}the stress-averaging blocks corresponding to the black curve are rectangular with dimensions \emph{(in $\mu m$)}. The others are square with the mentioned edge-lengths \emph{(in $\mu m$)}\emph{)}}.}
  \label{fig:conv_8by32mic_pb}
\end{minipage}%
\hfill
\begin{minipage}{.5\textwidth}
  \centering
  \includegraphics[width=\linewidth]{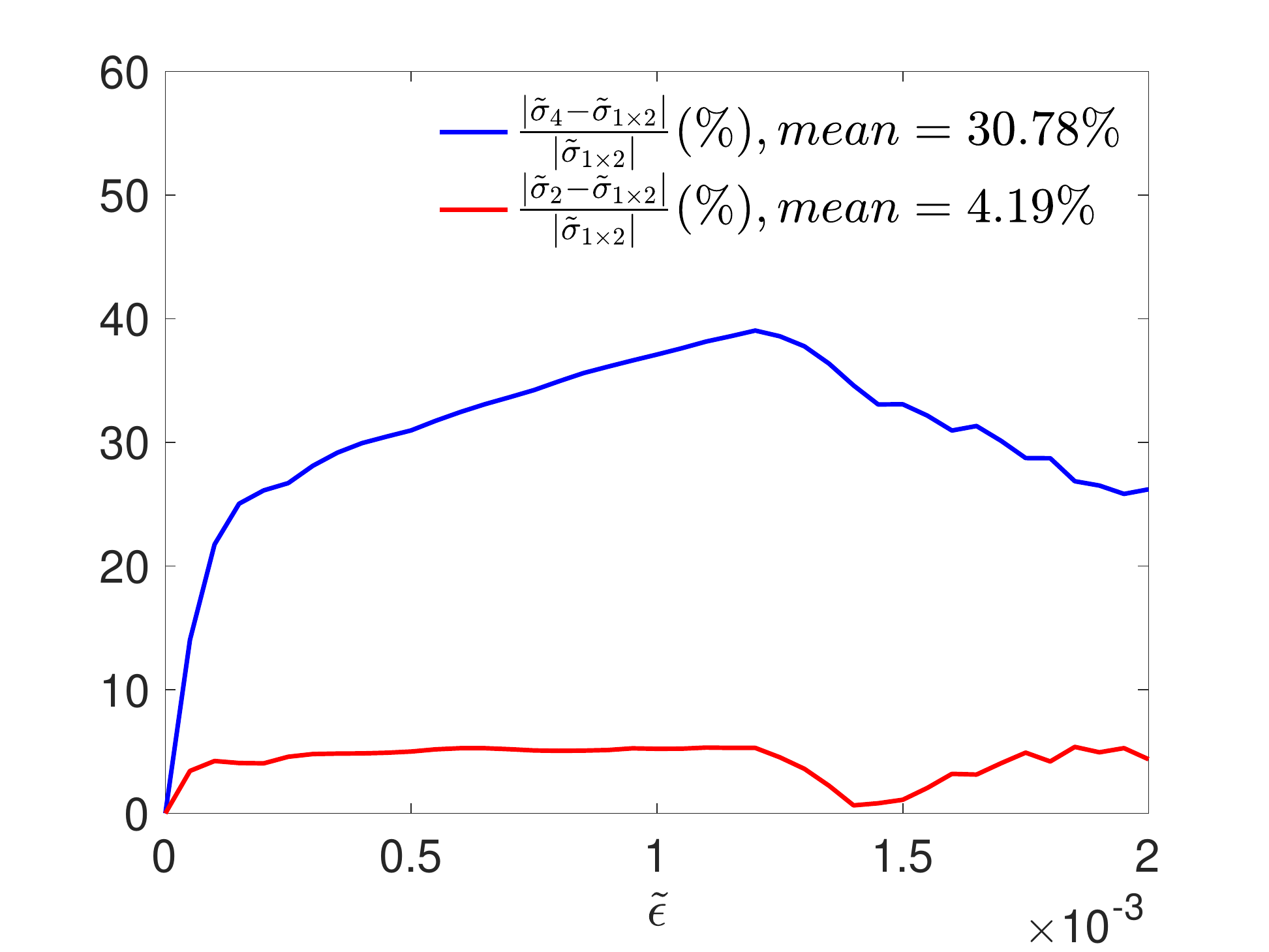}
  \caption{\textit{\small Relative error between the responses in pure bending}.}
  \label{fig:conv_8by32mic_pb_rel_err}
\end{minipage}
\end{figure}

The stress strain curves for the $25 \mu m$ size in tension for different \emph{stress-averaging} sizes (in units of $\mu m$) are shown in Fig. \ref{fig:conv_25mic_t22}. The relative error of the stress strain response is calculated as $\frac{|\sigma_{B_1} (\epsilon)-\sigma_{B_2}(\epsilon)|}{|\sigma_{B_2} (\epsilon)|} \times 100 $, where $\sigma_{B_1} (\epsilon)$ and $\sigma_{B_2} (\epsilon)$ are the stresses corresponding to strain $\epsilon$, for runs with \emph{stress-averaging} sizes of $B_1$ and $B_2$ respectively (where the smaller \emph{stress-averaging} size $B_2$ is taken as the base). The relative error is shown in Fig. \ref{fig:conv_25mic_t22_rel_err} which is very small. We also found that there is no significant size effect as we increase the sample size and the relative error for large samples with large \emph{stress-averaging} size still comes out to be small. For example, the response for a $400 \mu m$ sample with \emph{averaging} sizes of $40 \mu m$ and $80 \mu m$ are very close as shown in Fig. \ref{fig:conv_400mic_t22} and the average relative error is only $1.03 \%$. This shows that for imposed homogeneous boundary conditions, our model works like conventional plasticity (without constitutive assumptions) and the relative error is small for relatively large \emph{stress-averaging} sizes.

We also consider a $8 \mu m \times 16 \mu m \times 1 \mu m$ sample and divide the domain into $2 \times 8$, $4 \times 16$ and $8 \times 16$ blocks with \emph{stress-averaging} size of $4 \mu m \times 4 \mu m$, $2 \mu m \times 2 \mu m$ and $1 \mu m \times 2 \mu m$ respectively and apply the traction $\bft=t_{22} \bfe_2= t_m (1 - 2 \frac{x_1}{H})$ on the top face (where $t_m$ is a constant and $H$ is the size of the sample in the $x_1$ direction), while keeping the bottom face fixed and all other faces free (compare with Fig. \ref{fig:uniaxial_tension_bc}). This corresponds to pure bending of a beam with applied axial force on the top face varying linearly from positive to negative with the bottom face fixed. The effective stress strain response corresponding to the different \emph{stress-averaging} sizes are presented in Fig. \ref{fig:conv_8by32mic_pb} (please note that $\tilde{\sigma}=\frac{M}{b H^2}$ and $\tilde{\epsilon}=\theta \frac{H}{2 L}$ are the effective stress and effective strain respectively, where $M$ is the moment of the applied axial forces about the neutral axis $x_1=H/2$, $\theta$ is the averaged rotation of the top face about the plane $x_2=L$ and $L$ and $b$ are the dimensions of the sample in the $x_2$ and $x_3$ directions respectively). The relative errors between the different responses are shown in Fig. \ref{fig:conv_8by32mic_pb_rel_err}. This shows that in order to see convergence in results for strongly inhomogeneous deformation like in this case, the \emph{stress-averaging} size needs to be approximately limited to the size of the DD box (which is $1 \, \mu m$ in our case). Nevertheless, there are still significant savings due to time averaging, and the `error' between the solution for the $2 \times 2 (\mu m)^2$ and the $1 \times 2 (\mu m)^2$ stress-averaging sizes is quite acceptable.

%%%alfa comparison between uniaxial and pb
%\begin{figure}[!h]
%\centering
%\begin{minipage}{.45\textwidth}
%  \centering
%  \includegraphics[width=\linewidth]{figures_coupling/t22_25mic_25/alfa_0.187.pdf}
%  \caption{\textit{\small The variation in norm of dislocation density for a $25 \mu m$ sample in uniaxial tension with stress averaging size of $5 \mu m$ at $0.277 \%$ strain.}}
%  \label{fig:alfa_25mic_t22_25_0.277}
%\end{minipage}%
%\hfill
%\begin{minipage}{.45\textwidth}
%  \centering
%  \includegraphics[width=\linewidth]{figures_coupling/pureBending/alfa_0.194.pdf}
%  \caption{\textit{\small The variation in norm of dislocation density for a $8 \mu m \times 32 \mu m$ sample in pure bending with stress averaging size of $1 \times 2 (\mu m)^2$ at $0.194 \%$ strain.}}
%  \label{fig:alfa_8by32mic_pb_128_0.194}
%\end{minipage}
%\end{figure}

% alfa comparison between uniaxial and pb
\begin{figure}[!h]
\begin{subfigure}{.45\textwidth}
  \centering
  \includegraphics[width=\linewidth]{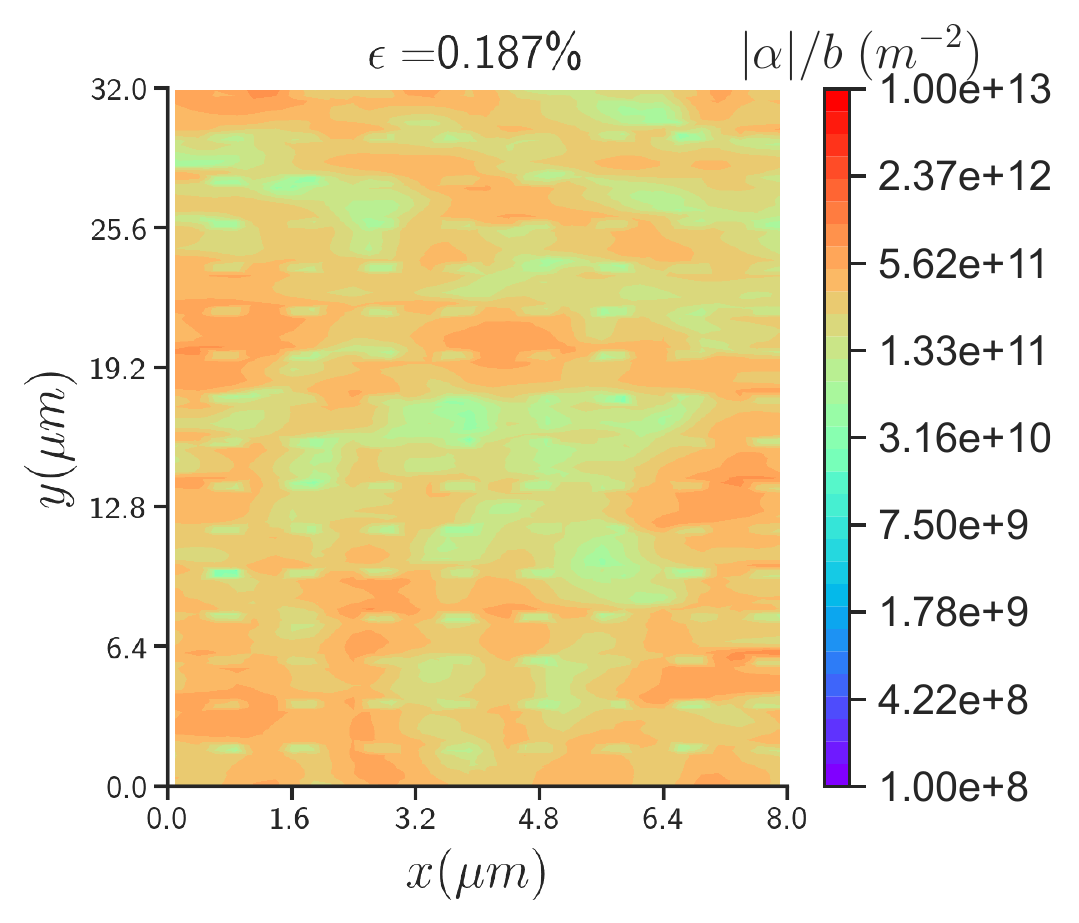}
  \caption{}
  \label{fig:alfa_25mic_t22_25_0.187}
\end{subfigure}%
\hfill
\begin{subfigure}{.45\textwidth}
  \centering
  \includegraphics[width=\linewidth]{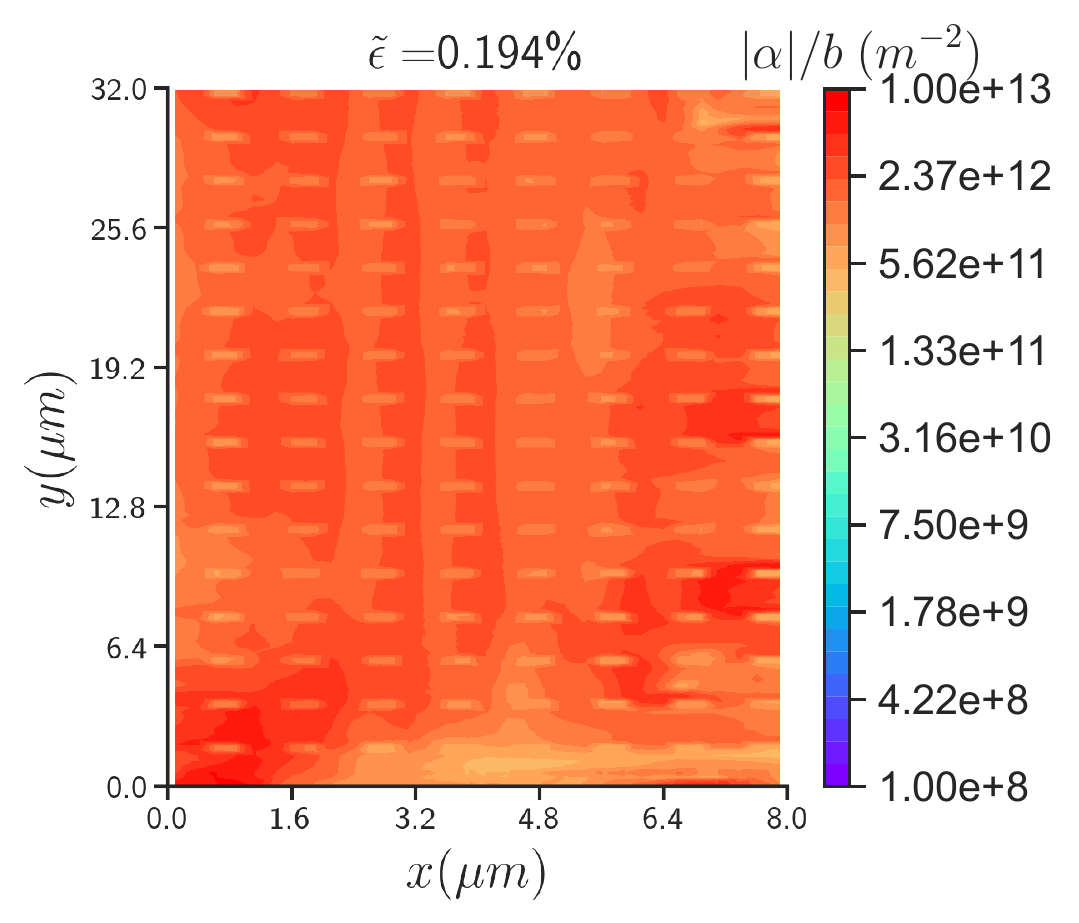}
  \caption{} 
  \label{fig:alfa_8by32mic_pb_128_0.194}
\end{subfigure}
\caption{ \textit{Spatial inhomgeneity measured by the norm of the dislocation density $(|\bfalpha|/b)$ for an $8 \times 32 (\mu m)^2$ sample with stress averaging size of $1 \times 2 (\mu m)^2$ $:$ $($\subref{fig:alfa_25mic_t22_25_0.187}$)$ Uniaxial tension at $0.187 \%$ strain $($\subref{fig:alfa_8by32mic_pb_128_0.194}$)$ Pure bending at $0.194 \%$ strain.}}
\end{figure}

We consider a $8 \times 32 (\mu m)^2$ sample with \emph{stress averaging} size of $1 \times 2 (\mu m)^2$ and subject it to uniaxial tension and pure bending. A field plot of the norm of the dislocation density in uniaxial tension is shown in Fig. \ref{fig:alfa_25mic_t22_25_0.187} while the same for pure bending is shown in Fig. \ref{fig:alfa_8by32mic_pb_128_0.194}, at similar values of strain (approximately $0.19 \%$). The sample in pure bending clearly shows more inhomogeneity (as measured by $|\bfalpha|$) compared to the sample in uniaxial tension. 

%We also choose a domain of $10 \mu m \times 10 \mu m \times 1 \mu m$ and divide it similarly into $2500$ elements each of size $0.5 \mu m \times 0.5 \mu m \times 1 \mu m$. As with the problem of 25 micron sample size, we divide the domain into $5 \times 5$, $7 \times 7$ and $10 \times 10$ blocks and perform DD simulations in each such block (in parallel). The stress strain curves for different number of processors for tension is shown in Fig. \ref{fig:conv_10mic_t22}. The stress strain curves are very close to each other and show convergence. 

\textbf{Microstructure, rate and orientation effects}

The initial state of DD for the simulation is refererred as the initial DD microstructure. The state of the sample obtained from solving the MFDM system is simply called the microstructure. Here, we discuss about the details of the microstructure and various effects that we observe. 

\begin{enumerate}

\item \textbf{Microstructure} We see the variation of the norm of the dislocation density tensor ($|\bfalpha|/b$) and the norm of the deviatoric stress, referred to as $J_2$ here, across the domain for a 25 micron size with \emph{stress averaging} size of $5 \mu m$ in uniaxial tension in Fig. \ref{fig:alfa_25mic_t22_25} and Fig. \ref{fig:j2_25mic_t22_25} respectively. We see that both the dislocation density and stress profiles are heterogeneous at high levels of strain. 

%We also see the corresponding profiles for 100 processors for the tension case in Fig. (25 mic 100 proc tension) and for the shear case in Fig. (25 mic 100 proc shear). 

%We see the evolution of dislocation density and the stress ($J_2$) for the 10 micron size with 25 processors for the tension case in Fig. \ref{fig:al_10mic_t22_25} and Fig. \ref{fig:j2_10mic_t22_25} respectively. For the shear case, we see the evolution of dislocation density and stress in Fig. \ref{fig:dd_10mic_t12_25} and Fig. \ref{fig:j2_10mic_t12_25} respectively. We see similar trends in these results where both the dislocation density and the stress spread out and tend to become homogeneous over the sample with the increase in strain. 

% field plots -25mic tension
\begin{figure}[!h]
\begin{subfigure}{.45\textwidth}
  \centering
  \includegraphics[width=\linewidth]{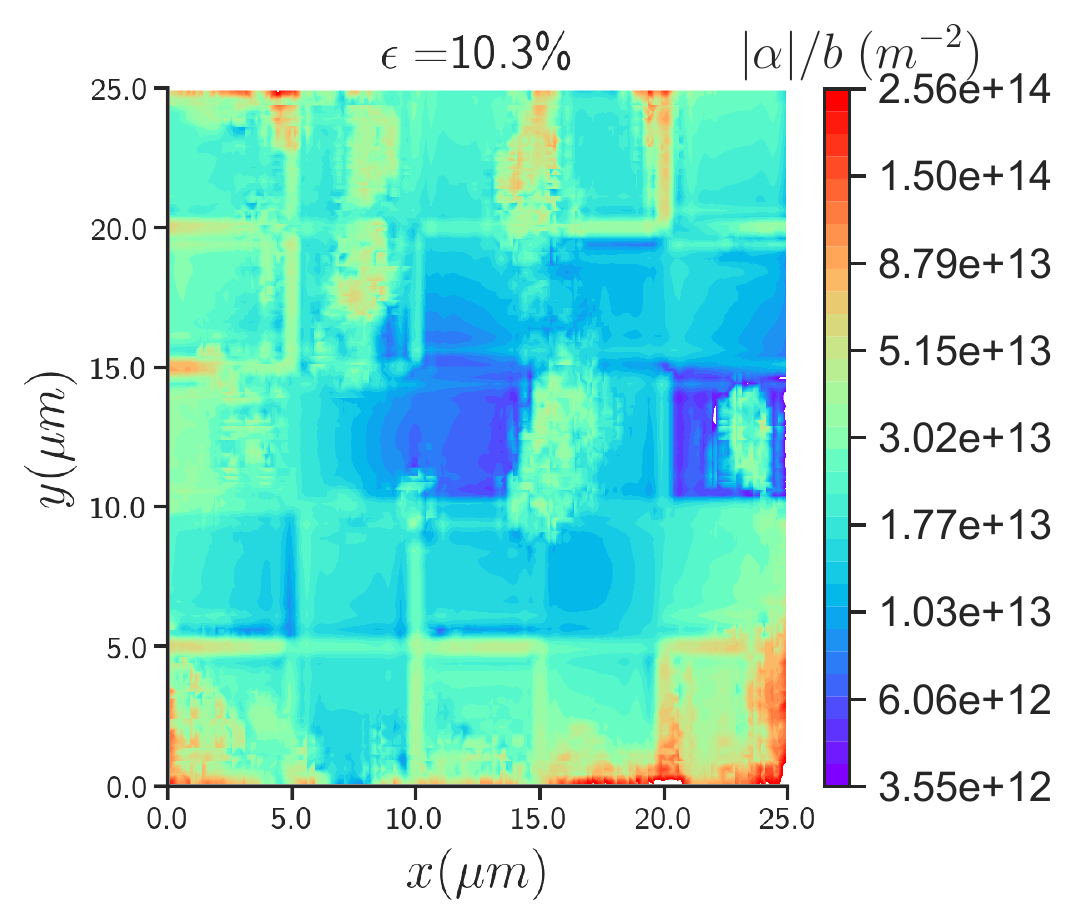}
  \caption{}
  \label{fig:alfa_25mic_t22_25}
\end{subfigure}%
\hfill
\begin{subfigure}{.45\textwidth}
  \centering
  \includegraphics[width=\linewidth]{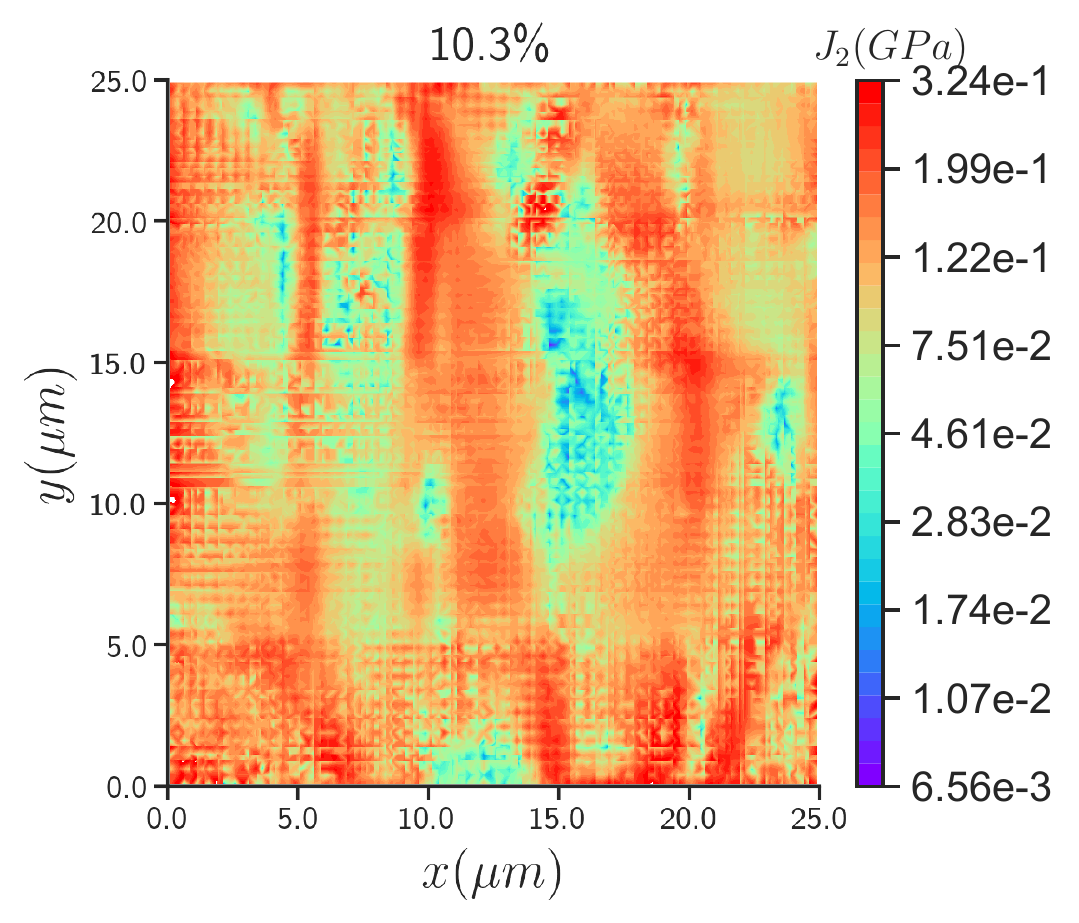}
  \caption{} 
  \label{fig:j2_25mic_t22_25}
\end{subfigure}
\caption{ The microstructure for $25 \mu m$ sample in uniaxial tension with \emph{stress averaging} size of $5 \mu m$ at $10.3 \%$ strain: (\subref{fig:alfa_25mic_t22_25}) The norm of dislocation density ($|\bfalpha|/b$) ; (\subref{fig:j2_25mic_t22_25}) The norm of the deviatoric stress $J_2$. The development of significant heterogeneity can be observed.}
\end{figure}

%% field plots -25mic shear
%\begin{figure}[!h]
%\centering
%\begin{minipage}{.45\textwidth}
%  \centering
%  \includegraphics[width=\linewidth]{figures_coupling/t12_25mic_25/alfa_8.39.pdf}
%  \caption{\textit{$|\bfalpha|$ for $25 ~micron$ sample in simple shear with $5 \times 5$ blocks at $8.39 \%$ strain}}
%  \label{fig:alfa_25mic_t12_25}
%\end{minipage}%
%\hfill
%\begin{minipage}{.45\textwidth}
%  \centering
%  \includegraphics[width=\linewidth]{figures_coupling/t12_25mic_25/j2_8.39.pdf}
%  \caption{\textit{$J_2$ for $25 ~micron$ sample in simple shear with $5 \times 5$ blocks at $8.39 \%$ strain}} 
%  \label{fig:j2_25mic_t12_25}
%\end{minipage}
%\end{figure}

%\subsection{Size effects}
%We see in Fig.\ref{fig:25micVs10mic_25} that for both tension and shear, the stress-strain profile is harder for 10 micron than 25 micron size which demonstrates that smaller sample sizes are harder when compared to larger sizes, %as we observe from experiments. We also see in Fig.\ref{fig:25micVs10mic_25_initYield} that both tension and shear profiles initially yield at at very small values of strain (around 0.001\%) while the profile yields much smoothly for %the 25micron case for both loading cases. 

\item \textbf{Orientation effects} We see in Fig.\ref{fig:orientation_effect} that the stress-strain profile for the 25 $\mu m$ sample is harder for uniaxial tension as compared to simple shear. This is expected, as in the shear case, we have dislocation segments in the primary plane which have a higher Schmid factor, while in the tension case, we have segments in planes which have smaller Schmid factor. The ratio of the sum of the Schmid factors of the active slip systems (denoted as $f_{s,i}$ and defined in \eqref{eq:schmid}) is 1.84. The ratio of the stress response of the uniaxial tension and simple shear as shown in Fig. \ref{fig:orientation_effect} lies between 1.99 and 2.31 with a mean of 2.13. Thus, the difference in the response between the two orientations is in accord with the prediction of the Schmid factor. However, it is to be noted that this is an emergent behavior and there is no ad-hoc \emph{assumption} made here.

\begin{figure}[!h]
\centering
  \includegraphics[width=0.4\linewidth]{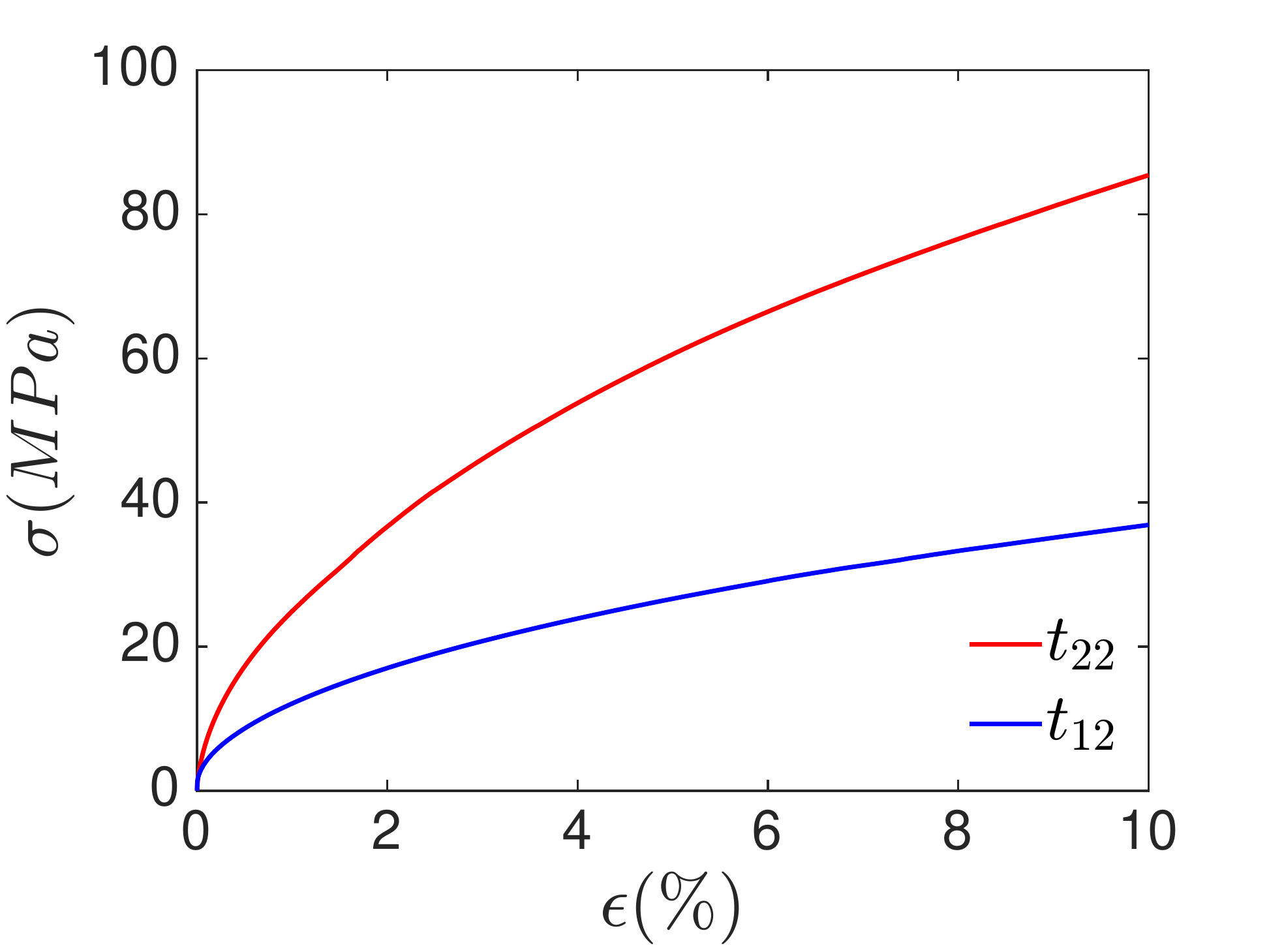}
  \caption{\textit{Orientation effect: stress-strain response for $25~micron$ sample in uniaxial tension ($t_{22}$) and simple shear ($t_{12}$) under load control.} }
  \label{fig:orientation_effect}
\end{figure}
 
\item \textbf{Rate effects}  With the reduction of loading rate, the stress-strain response becomes softer in both tension and shear (as shown in Fig. \ref{fig:rate_effect} for a $25~micron$ sample), as expected, because there is more time for plasticity to happen. The response is appreciably rate dependent for the loading rate of 1 $\textrm{MPa/s}$ and the nominal mobile and sessile dislocation densities (of $1.51\times 10^{12}~ m^{-2}$ and $1.63\times 10^{14}~m^{-2}$ respectively) involved. Rate independence is explored later. 

\begin{figure}[!h]
\centering
  \includegraphics[width=0.4\linewidth]{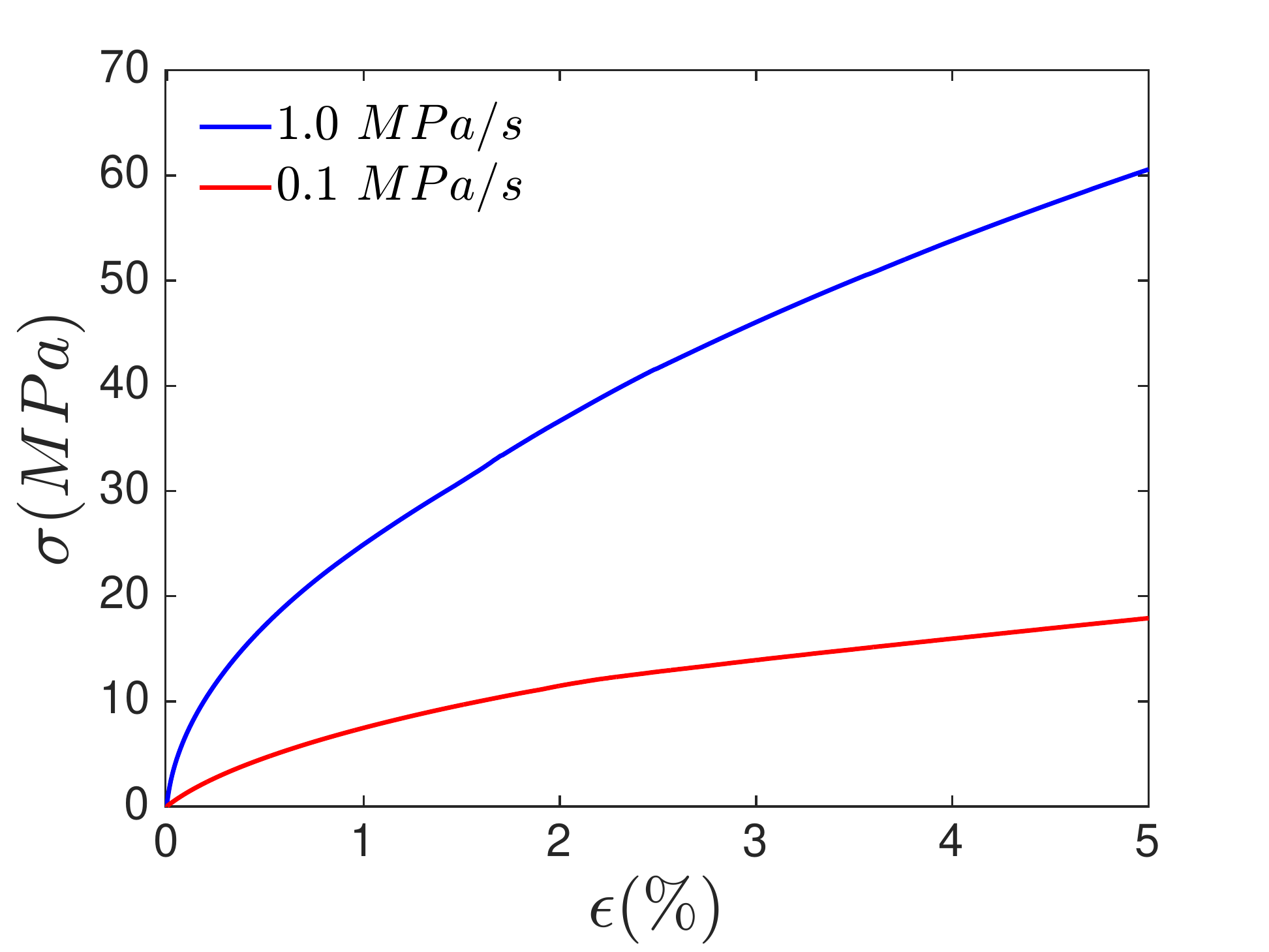}
  \caption{\textit{Rate effect: stress-strain response for $25(\mu m)^2$ sample in uniaxial tension under load control at different rates.} }
  \label{fig:rate_effect}
\end{figure}

\item \textbf{Effect of different initial DD microstructure}
We run a number of simulations with different initial DD microstructures and then take the average of the stress-strain response obtained from these runs. The results are presented in Fig. \ref{fig:diffmc_effect}. 

The response varies with the choice of the initial DD microstructure. In general, for the same $\rho^s$, an increase in $\rho^m$ leads to a softer stress strain response. This is expected as more mobile density means more generation of plastic strain, and hence the curve is supposed to be softer. 

The layout (configuration of the dislocation segments) of the initial DD microstructure also appears to be very important in determining the response of the sample. However, in reality, for macroscopic samples of size greater than 100 $\mu m$, the layout of the initial microstructure does not play such an important role. Thus, this is a limitation of our strategy. One way to address this is to add macroscopic descriptors in MFDM, which will act as sources of feedback, based on which the evolution of the DD microstructure can be controlled.
%\begin{table}
%\centering
%\begin{tabular}{|c| c| c| c| c| c| c| c| c|}
%\hline
%Microstructure &  $s_1(b)$ & $\theta_1$ &    $s_2(b)$ & $\theta_2$ &   $s_3(b)$ & $\theta_3$  &  $s_4(b)$ & $\theta_4$  \\
%\hline
%1     &  327.25     & $116.94^{\circ}$ &  4081.83     & $296.93^{\circ}$ &  3137.63     & $130.27^{\circ}$ &  2483.61    & $310.27^{\circ}$  \\
%2   &  351.27     & $81.72^{\circ}$ &  1584.17     & $261.72^{\circ}$ &  101.08     & $33.91^{\circ}$ &  178.16    & $213.90^{\circ}$  \\
%3   &  65.49     & $196.66^{\circ}$ &  3947.92     & $16.66^{\circ}$ &  825.64     & $312.21^{\circ}$ &  2366.85    & $312.21^{\circ}$  \\
%4   &  77.40     & $59.36^{\circ}$ &  1130.19     & $239.36^{\circ}$ &  3321.34    & $218.14^{\circ}$ &  3064.96   & $38.14^{\circ}$  \\
%5   &  3029.37     & $127.77^{\circ}$ &  370.52     & $307.77^{\circ}$ &  981.59    & $273.06^{\circ}$ &  2746.52    & $93.07^{\circ}$  \\
%\hline
%\end{tabular}
%\caption{Microstructure details.}
%\label{tab:micro_details}
%\end{table}    

%\begin{table}
%\centering
%\begin{tabular}{|c| c| c| c| c| c| c| c| c|}
%\hline
%%Microstructure &  s_1 & \theta_1 &    s_2 & \theta_2 &   s_3 & \theta_3  &  s_4 & \theta_4  \\
%%\hline
%1     &  1     & 1 &  1     & 1 &  1     & 1 &  1     & 1  \\
%2   &  1     & 1 &  1     & 1 &  1     & 1 &  1     & 1  \\
%3   &  1     & 1 &  1     & 1 &  1     & 1 &  1     & 1  \\
%4   &  1     & 1 &  1     & 1 &  1     & 1 &  1     & 1     \\
%5   &  1     & 1 &  1     & 1 &  1     & 1 &  1     & 1  
%\hline
%\end{tabular}
%\caption{Microstructure details.}
%\label{tab:micro_details}
%\end{table}    

\begin{figure}[!h]
\centering
  \includegraphics[width=0.6\linewidth]{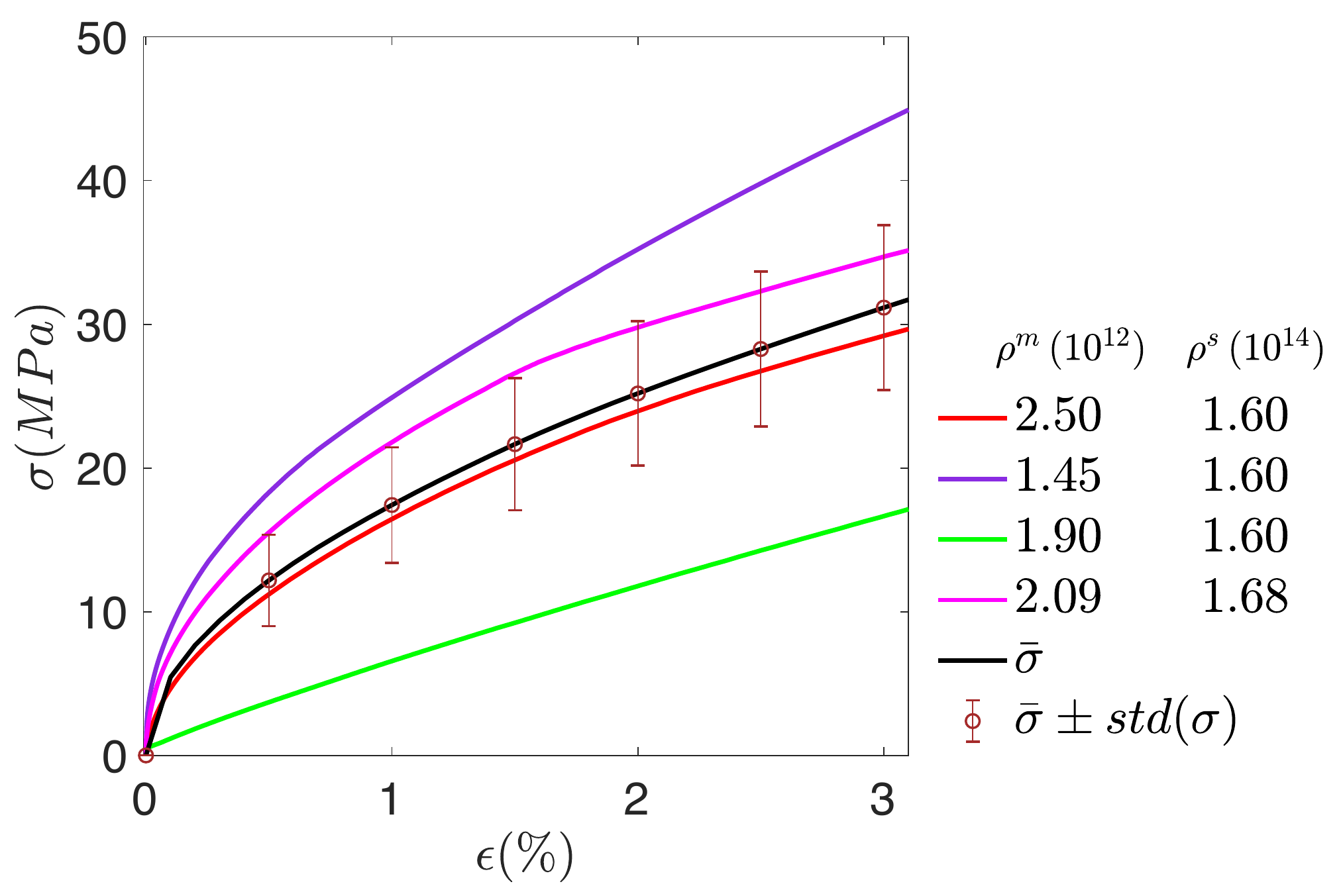}
  \caption{\textit{Stress-strain response for $25 (\mu m)^2$ sample in uniaxial tension with different initial microstructure described by their mobile and sessile densities, their average given by $\bar{\sigma}$ and the upper and lower bounds given by $\bar{\sigma} + std({\sigma})$  and $\bar{\sigma} - std({\sigma})$ respectively, where $std({\sigma})$ is the standard deviation of the stress across all the different initial microstructures. The units of $\rho^m$ and $\rho^s$ is $m^{-2}$.} }
  \label{fig:diffmc_effect}
\end{figure}

\end{enumerate}

\subsubsection{Case 1 with Displacement Control}

We also perform the simulation for the $25~micron$ sample with displacement control. Standard displacement boundary condition to prevent rigid body motion is applied. However, instead of applying the traction $\bft=t_{22} \bfe_2$ on the top face as shown in Fig. \ref{fig:uniaxial_tension_bc}, we apply displacement boundary condition on the top face corresponding to uniaxial tension $\bfx=x_2 \bfe_2$ and the bottom face is kept fixed in the $Y$ direction. The \textit{current} load $x_2$ depends on the strain rate $s$  unless it is set to 0 when the \emph{limit} load is reached. However, one point of difference in the displacement control case from the load control case is that when the load is kept fixed in the load control case, deformation still happens and we are supplying energy to the system, which is not true when we keep the displacement fixed in the displacement control case.  The goal is to be able to run simulations for very slow loading rates upto appreciable values of strain. 

\begin{figure}[!h]
\centering
  \includegraphics[width=0.55\linewidth]{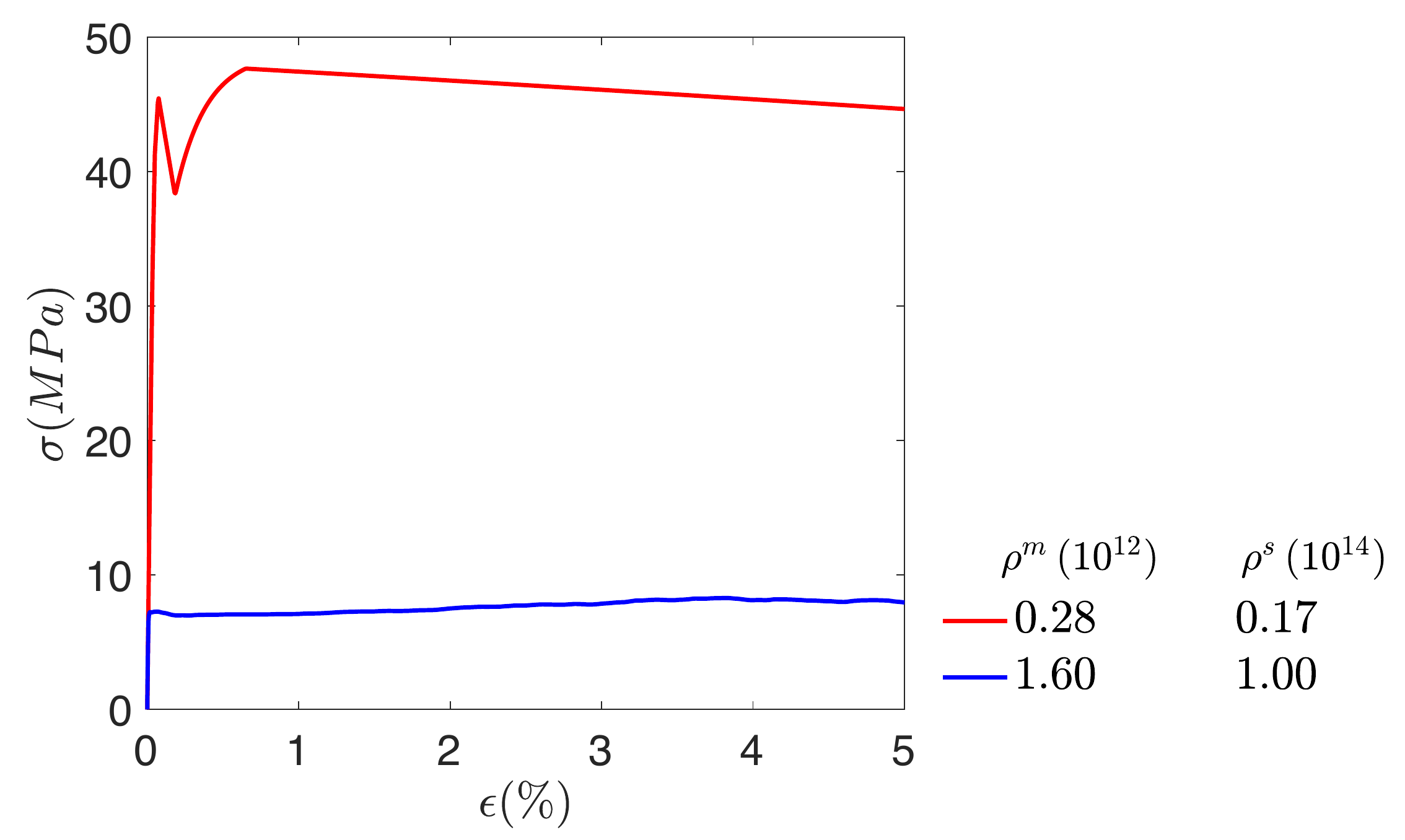}
  \caption{\textit{Stress-strain response for $25~micron$ sample in uniaxial tension under displacement control at applied strain rate of $s=10^{-4}/s$.}}
  \label{fig:t22_disp_ratio}
\end{figure}

The stress-strain response depends on the ratio of mobile segment density ($\rho^m$) to sessile segment density ($\rho^s$). In general, for a particular value of applied strain rate, there appears to exist an approximate ratio $r=\frac{\rho^s}{\rho^m}$, for which the simulations can be performed upto large values of strain, without the occurrence of a collapse (vanishing of the reaction force) in the stress-strain response. For example, we used two microstructures with $\rho^m$ and $\rho^s$ mentioned in Fig. \ref{fig:t22_disp_ratio} ( $\rho^m$ and $\rho^s$ are in units of $m^{-2}$ here and in the results mentioned later). The ratio $r$ comes out to be $590.28$ and $625$ respectively for the two microstructures. Using a ratio of this order for the initial microstructure, the simulations could be performed with an applied strain rate of $s=10^{-4} /s$, without a collapse. The response corresponding to $\rho^m=2.82\times 10^{11}\,m^{-2}$ and $\rho^s=1.7 \times 10^{14}\, m^{-2}$ shows a drop in stress from a strain of $0.07\%$ to a strain of $0.17\%$. The drop in stress at very small strains is a common feature of responses for uniaxial tension using displacement control (see Fig. \ref{fig:orientation_effect_disp} and Fig. \ref{fig:rate_effect_disp}). At small strains and high values of stress, there is increased motion of dislocations, leading to a rise in the plastic strain rate, which causes the drop in stress. This follows with a rise in stress till a strain of $0.63\%$, which is caused by the internal stress fields which affect the Peach-Koehler forces acting on the segments and slows their motion.

% disp control rates

%We see the stress-strain response for different applied loading rates rates in Fig. \ref{fig:t22_disp_rates}. We see that initially there is an elastic response where the stress rises quickly but soon falls and after it reaches a certain value of stress (which is the yield stress), it remains more or less parallel to the $X$ axis. We also see that the value of the yield stress drops as we decrease the applied loading rate, which is also observed in experiments. We also see the variation of dislocation density and stress for this case in Fig. \ref{fig:alfa_25mic_t22_25_disp} and Fig. \ref{fig:j2_25mic_t22_25_disp} respectively, and both of them show heterogeneity like in the load control case. 

Next, we discuss about the orientation and rate effect and the effect of different initial DD microstructures under displacement control. \\
%%{ \color{red} Deleted as it was about Fig 31 and 32 which have been removed. \\ 1. \textbf{Microstructure} We see the variation of the norm of dislocation density tensor ($|\bfalpha|/b$) and the stress ($J_2$) across the domain for a 25 micron size with $5 \times 5$ processors in uniaxial tension for an applied strain rate of $s=10^{-4} /s$ in Fig. \ref{fig:alfa_25mic_t22_25_disp} and Fig. \ref{fig:j2_25mic_t22_25_disp} respectively (the stress-strain curve for this run is shown in Fig. \ref{fig:t22_disp_ratio}). We see that both the dislocation density and stress profiles are very heterogeneous at high levels of strain. }
\begin{enumerate}
%% field plots - displacement control, 25mic tension 
%\begin{figure}[!h]
%\centering
%\begin{minipage}{.45\textwidth}
%  \centering
%  \includegraphics[width=\linewidth]{figures_coupling/t22_25mic_25_disp/alfa_4.98.pdf}
%  \caption{\textit{$|\bfalpha|$ for $25 ~micron$ sample in uniaxial tension with $5 \times 5$ processors at $4.98 \%$ strain}}
%  \label{fig:alfa_25mic_t22_25_disp}
%\end{minipage}%
%\hfill
%\begin{minipage}{.45\textwidth}
%  \centering
%  \includegraphics[width=\linewidth]{figures_coupling/t22_25mic_25_disp/j2_4.98.pdf}
%  \caption{\textit{$J_2$ for $25 ~micron$ sample in uniaxial tension with $5 \times 5$ processors at $4.98 \%$ strain}} 
%  \label{fig:j2_25mic_t22_25_disp}
%\end{minipage}
%\end{figure}

\item \textbf{Orientation effects} We see in Fig. \ref{fig:orientation_effect_disp} that the stress-strain profile for the 25 $\mu m$ sample is harder and has higher yield stress (the value of stress at which the slope reduces significantly from the initial slope of the elastic response) for uniaxial tension as compared to simple shear. This is expected, as in the shear case, we have dislocation segments in the primary plane which have a higher Schmid factor, while in the tension case, we have segments in planes which have smaller Schmid factor. The ratio of the sum of the Schmid factors of the active slip systems (denoted as $f_{s,i}$ and defined by \eqref{eq:schmid}) is 1.84. The ratio of the stress strain response of the uniaxial tension and simple shear, as shown in Fig. \ref{fig:orientation_effect_disp}, lies between 2.87 and 3.49, for strain higher than 1 $\%$ (which is maximum value of strain at which the response for both the loading cases show yielding).

%orientation effect
\begin{figure}[!h]
\centering
  \includegraphics[width=0.4\linewidth]{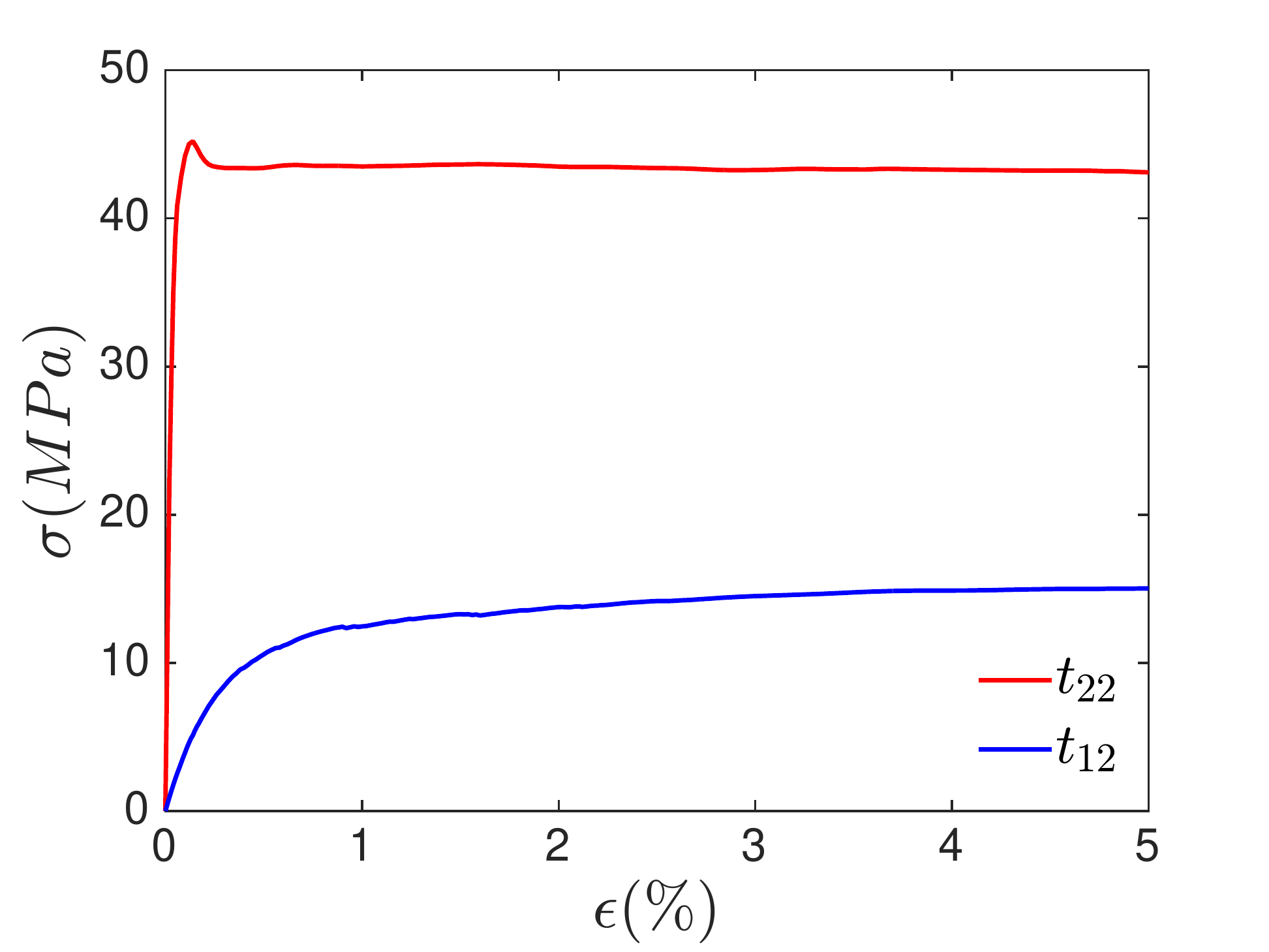}
  \caption{\textit{Orientation effect: stress-strain response for $25 (\mu m)^2$ sample in uniaxial tension ($t_{22}$) and simple shear ($t_{12}$) under displacement control.} }
  \label{fig:orientation_effect_disp}
\end{figure}

\item \textbf{Rate effects} We see in Fig. \ref{fig:rate_effect_disp} that with the reduction of loading rate, the stress-strain response becomes softer and has a lower yield stress in uniaxial tension. This is expected, as for lower strain rate, there is more time for plastic deformation to happen. The response is rate dependent for a loading rate of $10^{-3} /s$ and mobile and sessile dislocation densities of $3.73 \times 10^{11}~ m^{-2}$ and $1.67\times 10^{14}~m^{-2}$ respectively.

The response is harder for $s=10^{-4} /s$ (Fig.\! \ref{fig:disp_rate_2}) compared to $s=2 \times 10^{-5} /s$ for $\rho^s=1.7 \times 10^{14} ~m^{-2}$. However, for $\rho^s=10^{15} ~ m^{-2}$, the response is rate independent for $s=10^{-4} /s$ compared to $s=2 \times 10^{-5} /s$, till a strain of $0.2 \%$. For higher strains, the response for $s=2 \times 10^{-5} /s$ is harder compared to that for $s=10^{-4} /s$. The response for $s=2 \times 10^{-5} /s$ shows Stage I hardening till a strain of 0.2 $\%$. Then it rises steeply till a strain of 0.35 $\%$, which is characteristic of Stage II hardening. The average slope of the stress-strain curve in this part is $17.71 \, \textrm{GPa}$, which is much higher than $\frac{\mu}{200}=0.24 \, \textrm{GPa}$ (where $\mu$ is the shear modulus, whose value has been provided in Table \ref{tab:simulation_details}), which is the slope observed in Stage II hardening in macroscopic samples. 

This follows with a decrease in the slope (Stage III hardening). It is observed that $\rho^m$ does not appreciably increase (while $\rho^s$ is fixed), so the hardening is not caused by an increase in the density of dislocation segments. This strongly implies that the internal stress field affects the Peach-Koehler force  acting on the segments and causes the hardening. 

The response with different initial microstructures having approximately same $\rho^m$ ($\approx 1.5 \times 10^{12} m^{-2}$), $\rho^s$ ($\approx 10^{15} m^{-2}$) and with loading rate $s=2 \times 10^{-5} s^{-1}$ are shown in Fig. \ref{fig:disp_rate_2e-5}. We see variation in Stage I and Stage II hardening in these responses.
%rate effect
\begin{figure}[!h]
\centering
  \includegraphics[width=0.4\linewidth]{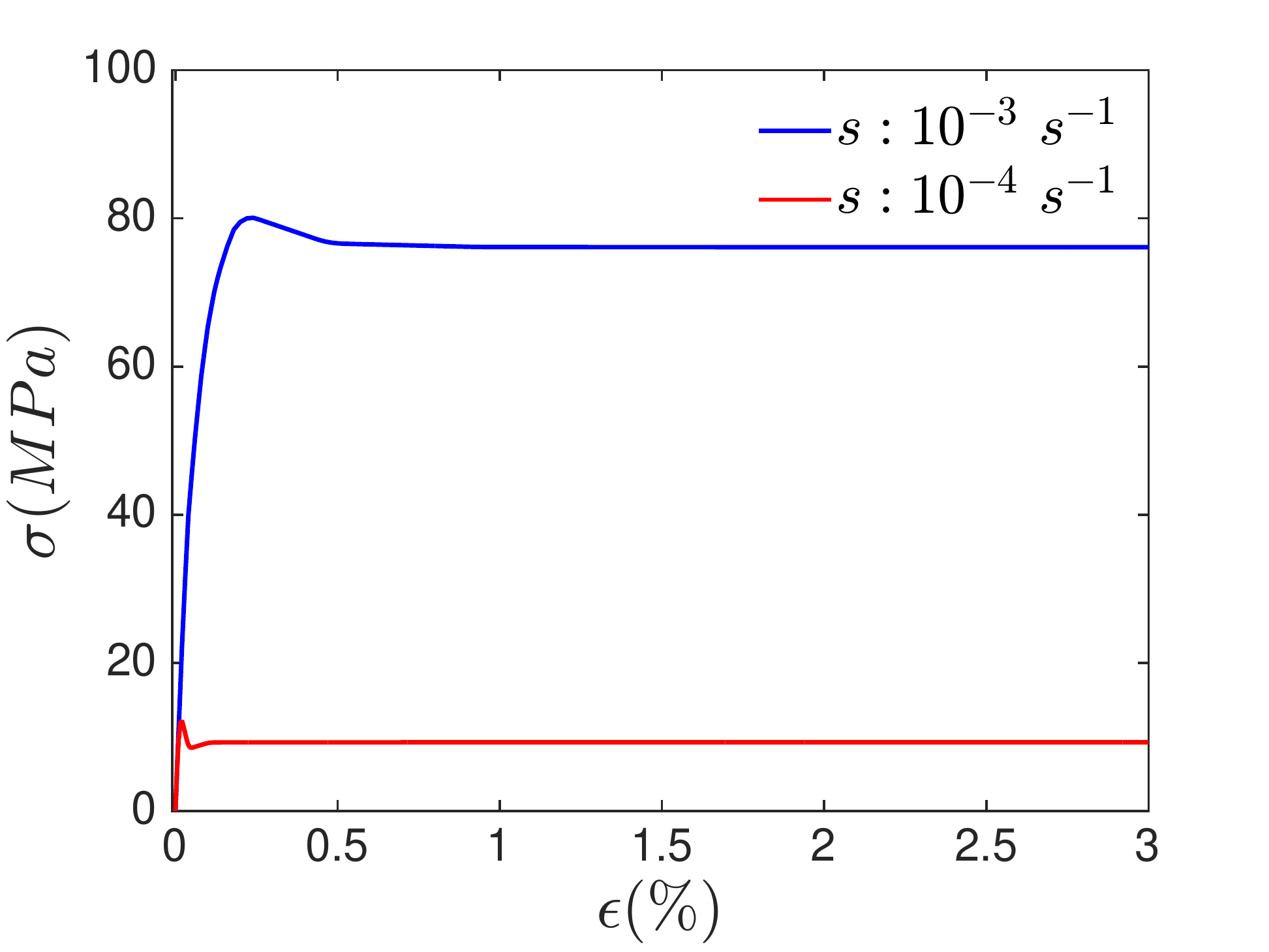}
  \caption{\textit{Rate effect: stress-strain response for $25 (\mu m)^2$ sample in uniaxial tension under displacement control at different rates.} }
  \label{fig:rate_effect_disp}
\end{figure}

\begin{figure}[!h]
\centering
  \includegraphics[width=0.65\linewidth]{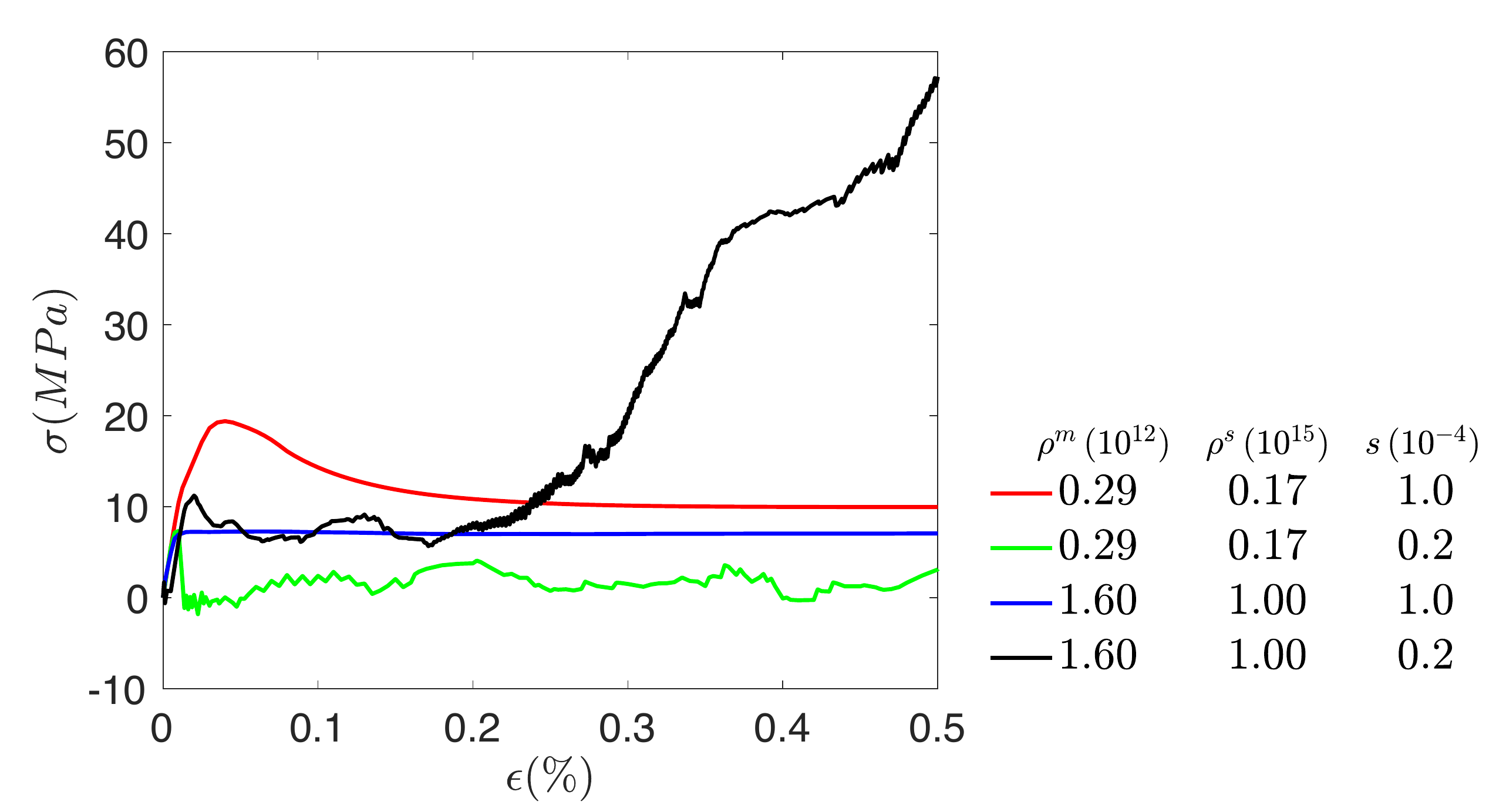}
  \caption{\textit{Rate effect: stress-strain response for $25 (\mu m)^2$ sample in uniaxial tension under displacement control at different rates. The strain rate $s$ is in units of $sec.^{-1}$.}}
  \label{fig:disp_rate_2}
\end{figure}

\begin{figure}[!h]
\centering
  \includegraphics[width=0.65\linewidth]{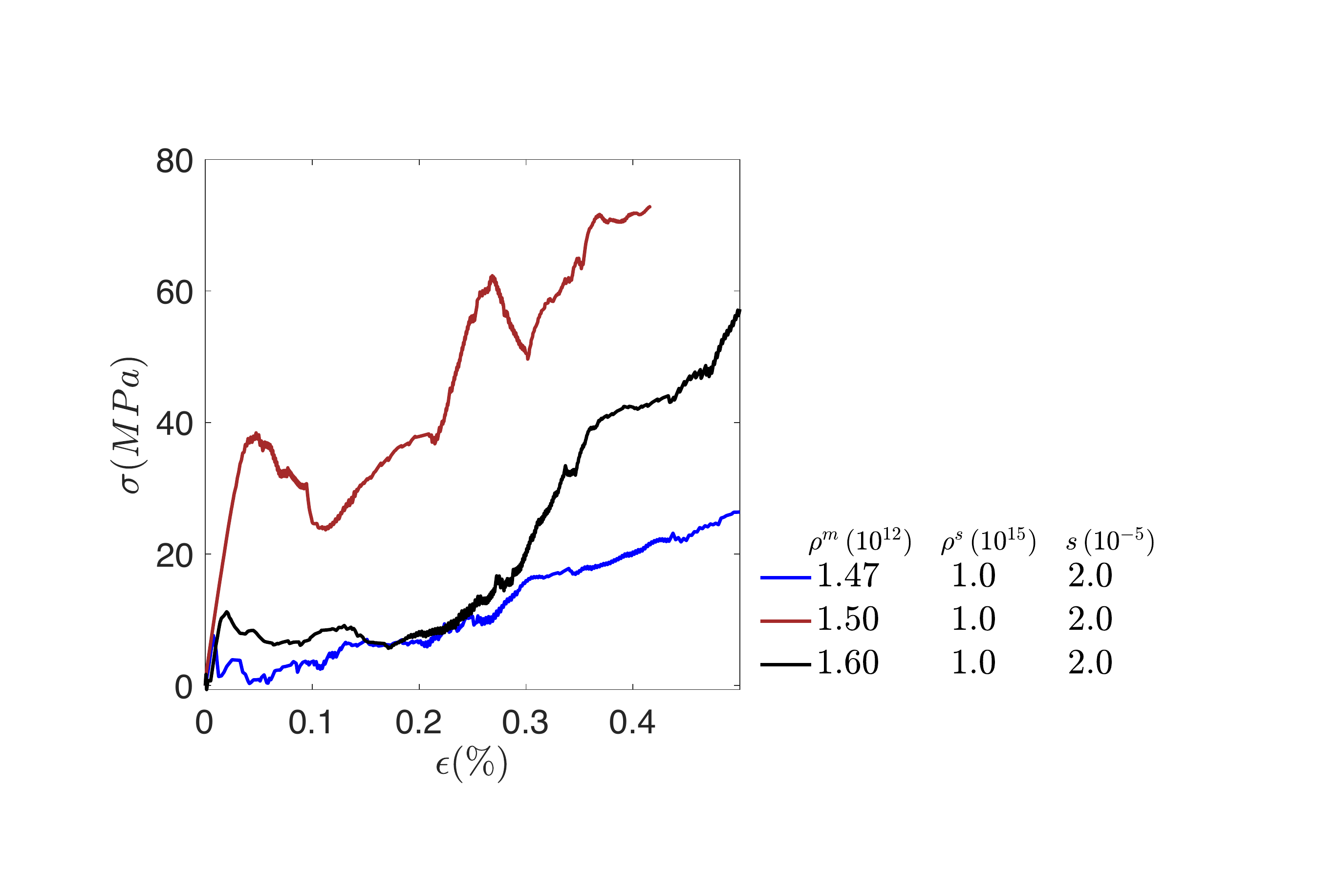}
  \caption{\textit{Rate effect: The strain rate $s$ is in units of $sec.^{-1}$.}}
  \label{fig:disp_rate_2e-5}
\end{figure}

\item \textbf{Effect of different initial microstructure}
We run a number of simulations with different initial microstructures and then take the average of the stress-strain response obtained from these runs. %The mobile ($\rho^m$) and sessile  ($\rho^s$) segment density for the microstructures are provided in the legend. %The microstructures are described by the set of pairs $\{s_n, \theta_n\}$ where the definition of $s_n$ and $\theta_n$ are provided in \emph{Construction of initial microstructure} in subsection \ref{subsec:dd_setup}. 

The results are presented in Fig. \ref{fig:diffmc_effect_disp}. We see that there the response varies with the choice of the initial DD microstructure. In general, for the same $\rho^s$, higher the $\rho^m$, the softer the stress strain response is. This is expected as more mobile density means more generation of plastic strain, and hence the curve is supposed to be softer. 

\begin{figure}[!h]
\centering
  \includegraphics[width=0.6\linewidth]{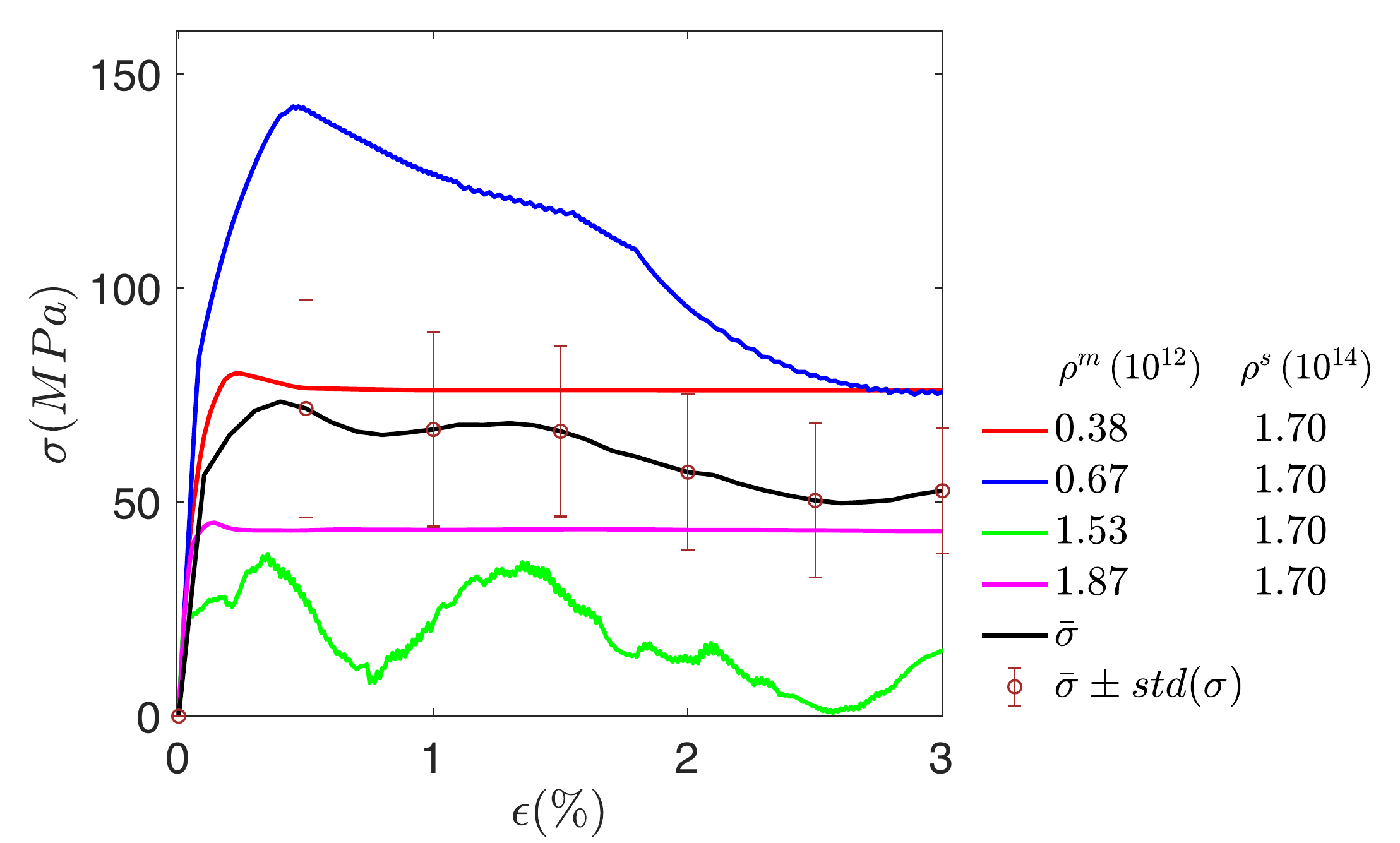}
  \caption{\textit{Stress-strain response for $25~micron$ sample in uniaxial tension with different initial microstructure described by their mobile and sessile densities, their average given by $\bar{\sigma}$ and the upper and lower bounds given by $\bar{\sigma} + std({\sigma})$  and $\bar{\sigma} - std({\sigma})$, where $std({\sigma})$ is the standard deviation of the stress across all the different initial microstructures.} }
  \label{fig:diffmc_effect_disp}
\end{figure}

\item \textbf{Initial yield} In Fig. \ref{fig:orientation_effect_disp}, the intial yield stress (the value of stress at which the response deviates from being elastic) of the response corresponding to uniaxial tension is approximately 35 $\textrm{MPa}$ while that for simple shear is approximately 10 $\textrm{MPa}$. Thus, the ratio between the yield stresses for the two cases is around 3.5. The ratio of the Schmid factors corresponding to the primary planes of the orientations for these two loading cases (as described in Section \ref{sec:uniaxial_tension} and \ref{sec:simple_shear} respectively) is 2.45. This is a prediction of the coupled DD-MFDM strategy, without any ad-hoc \emph{assumption} put in by hand. 
\end{enumerate}

\subsubsection{Case 2} 

In this case, the Burgers vector of the sessile segments lie in the slip plane. Thus, this is a more physically appropriate case. We present some results for this case to show how it compares with Case 1. 

\begin{figure}[h!]
\centering
\begin{minipage}{.55\textwidth}
  \centering
  \includegraphics[width=\linewidth,height=5cm]{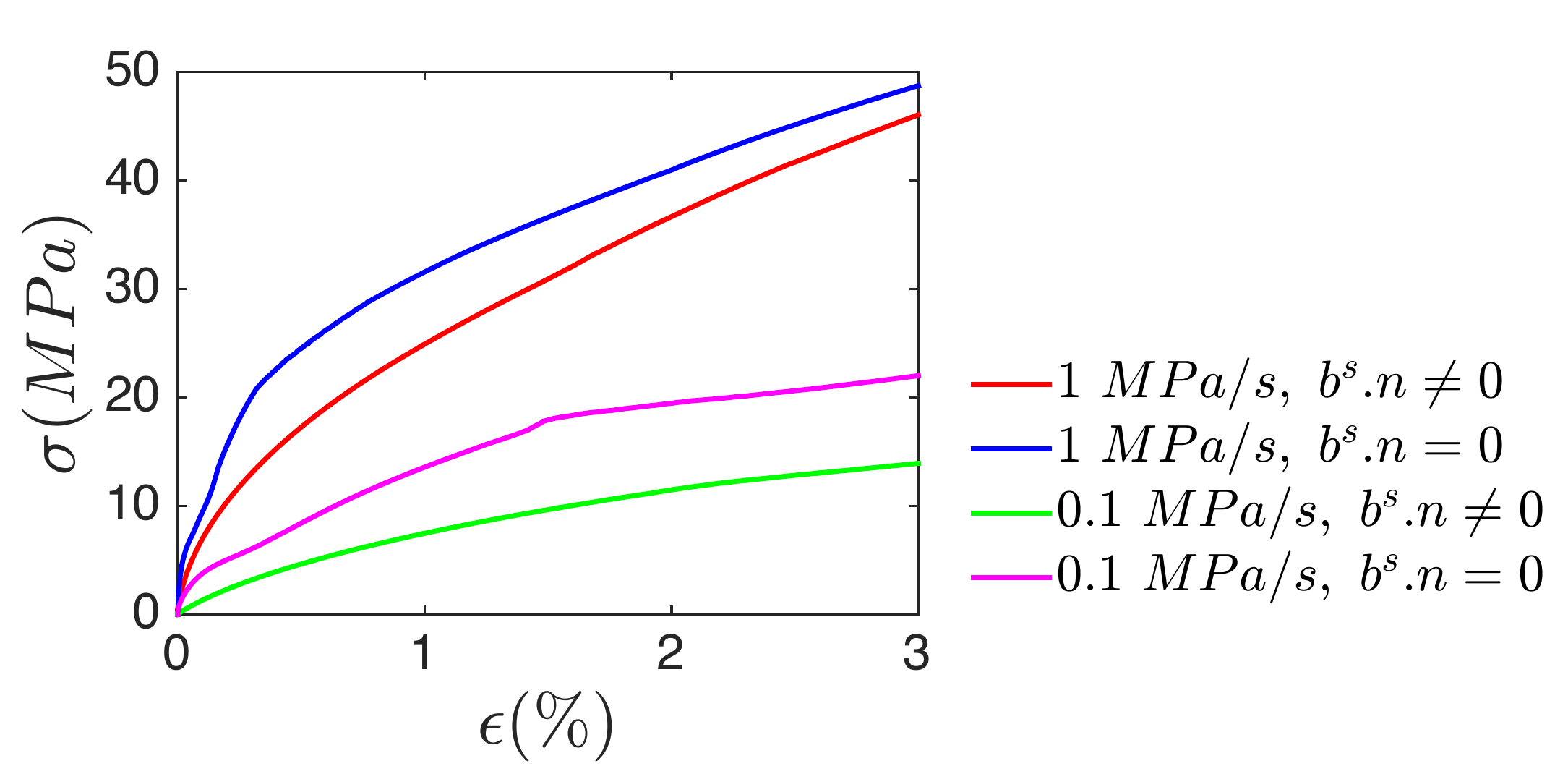}
  \caption{\textit{Stress strain response of a 25 micron sample in uniaxial tension at different loading rates under load control for Case 1 {\normalfont (}$\bfb^s \cdot \bfn \neq 0${\normalfont)} and Case 2 {\normalfont(}$\bfb^s \cdot \bfn=0${\normalfont)}.}}
  \label{fig:sessinplane_load}
\end{minipage}%
\hfill
\begin{minipage}{.4\textwidth}
  \centering
  \includegraphics[width=\linewidth,height=5cm]{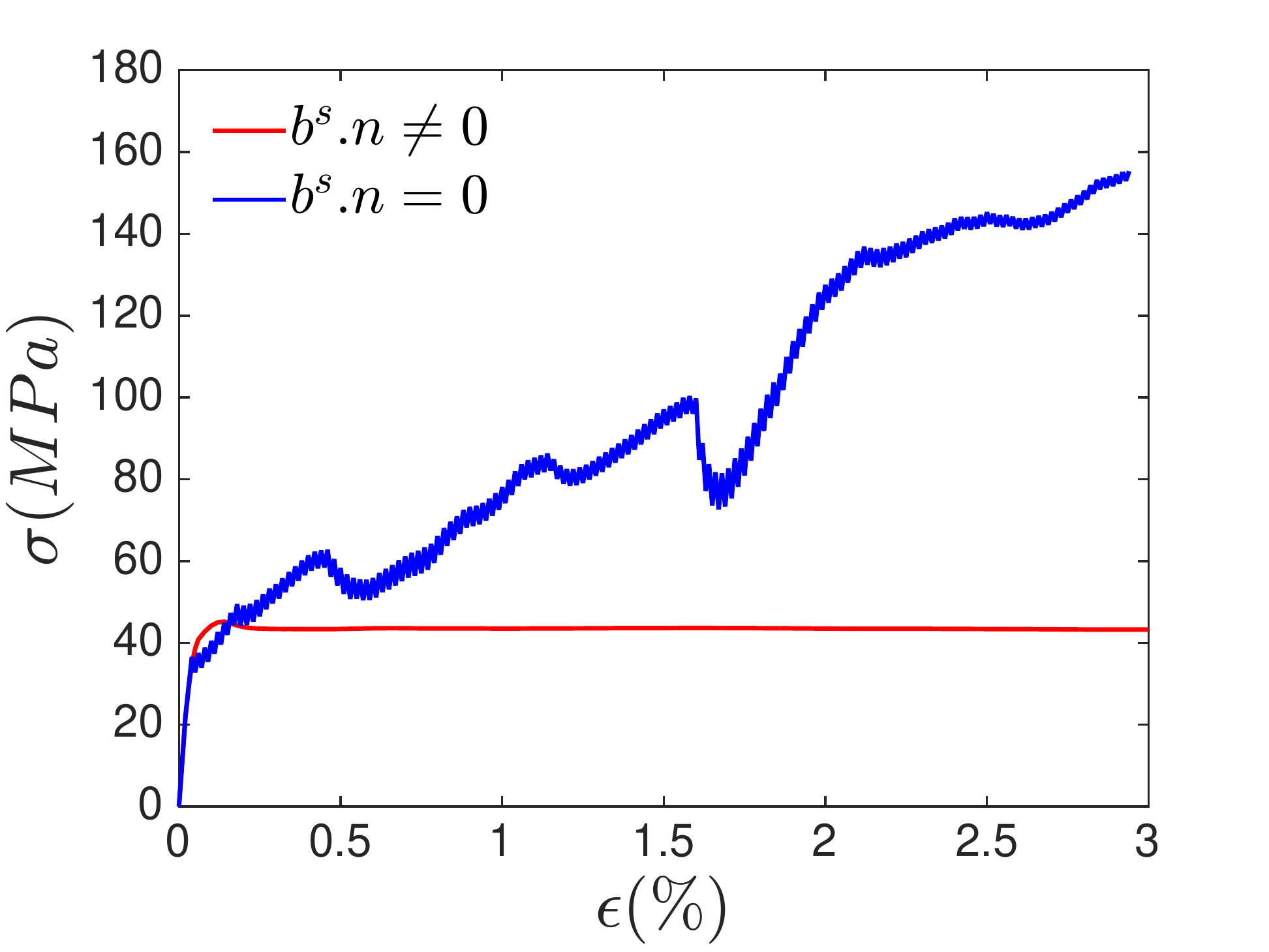}
  \caption{\textit{Stress strain response of a 25 micron sample in uniaxial tension at strain rate of $10^{-3}~s^{-1}$ under displacement control for Case 1 {\normalfont(}$\bfb^s \cdot \bfn \neq 0${\normalfont)} and Case 2 {\normalfont(}$\bfb^s \cdot \bfn=0${\normalfont)}.}} 
  \label{fig:sessinplane_disp}
\end{minipage}
\end{figure}

The stress strain response of a 25 $\mu m$ sample in uniaxial tension, under load control, at loading rates of 1 MPa/s and 0.1 MPa/s is shown in Fig. \ref{fig:sessinplane_load}. Case 1 is represented as $\bfb^s \cdot \bfn \neq 0$.  Case 2 is represented as $\bfb^s \cdot \bfn = 0$. 

The stress strain response of a 25 $\mu m$ sample in uniaxial tension, under displacement control, at a strain rate of $10^{-3}~s^{-1}$ is shown in Fig. \ref{fig:sessinplane_disp}. 

This important physical idealization appears to suggest (as evident in Fig. \ref{fig:sessinplane_load} and Fig. \ref{fig:sessinplane_disp}) that the response is harder when the Burgers vector of the sessile segments lie in the slip plane, when compared to the case where they lie outside the slip plane. The Burgers vector distribution of the sessile segments affect the Peach-Koehler force driving the motion of each segment, thus affecting the overall plasticity in the block. These preliminary results suggest that, even after averaging, this is a significant effect.

%diff mc effect
%\begin{figure}[!h]
%\centering
%  \includegraphics[width=0.4\linewidth]{figures_coupling/diffmc_effect_tension_ddbox2.pdf}
%  \caption{\textit{Stress-strain response for $25~micron$ sample in uniaxial tension with different initial microstructure denoted by the numbers, their average given by $\mu(\sigma)$ and the upper and lower bounds given by $\mu(\sigma) + std({\sigma})$  and $\mu(\sigma) - std({\sigma})$, where $std({\sigma})$ is the standard deviation of the stress across all the different initial microstructures.} }
%  \label{fig:diffmc_effect}
%\end{figure}

\subsubsection{Speedup}
The speedup in compute time for a single Gauss point case, for a $1 \mu m$ DD box, as mentioned in Section \ref{sec:sgp_results}, is around $1000$. So, for a $25~ \mu m$ sample, the speedup is around 
\[
\frac{25}{1} \times \frac{25}{1} \times 1000 = 6.25 \times 10^5.
\]
This is a very conservative estimate since we are not considering the interactions that would exist between these $1 ~\mu m$ boxes throughout the sample of $25 ~\mu m$. But even for such a conservative estimate, the speedup is very high when compared to conventional DD, which shows the advantage of our DD-MFDM coupling strategy. 

\section{Summary and Conclusions}\label{sec:coupling_disc}

A novel concurrent, multiscale approach to meso/macroscale plasticity has been implemented using a carefully designed coupling of MFDM with space-time averaged inputs from DD simulations. Stress-strain response at \emph{realistic} slow loading rates for large sample sizes and with significant speedup in compute time (around $10^5$ using a conservative estimate) have been obtained, showing the advantage of our coupled approach compared to conventional DD. 

We demonstrate a strong dependence of the results on
\begin{itemize}
\item the orientation of the microstructure (for the two loading cases of simple shear and uniaxial tension)
\item the loading rate, and 
 \item the ratio of mobile to sessile segment density,
\end{itemize}
in both load and displacement controlled simulations. There appears to be a limiting \emph{stress-averaging} size for imposed inhomogeneous deformation 
%% which is approximately equal to the size of the DD box, 
for which converged stress-strain response may be obtained. The collective behavior of dislocations accounting for their stress interactions in detail is demonstrated. The effect of internal stresses, which control the Peach-Koehler forces acting on the segments and affect their motion, is visible in the computed stress-strain response. 

The only constitutive assumption used in this coupled strategy is a simplified adaptation of the thermal activation of dislocation motion past obstacles \cite{kocks_mecking_2003}, which is described in Section \ref{sec:thermal_activation}. However, the order of the timescale set by the plastic strain rate obtained in our simulations is very different from the timescale set by the junction breaking time.

We point out the (current) limitations of our approach. These are 
\begin{itemize}
\item The dislocation content that is allowed to be mobile does not grow in density to the extent that is observed in reality. In a well annealed crystal, the total dislocation density grows by around 8 orders of magnitude and a large fraction of the mobile segments becomes sessile. Our simulations are currently incapable of representing such growth of the dislocation density. To account for this deficiency, we adopt the physical picture of Kocks-Mecking \cite{kocks_mecking_2003} and work with an a-priori assumption of a sessile distribution of dislocations in each DD box and a separate mobile population, the latter being allowed to evolve and grow (or diminish), with full interaction within itself as well as with the sessile population. 

\item The polar dislocation velocity is negligible. 
\item The response is highly dependent on the configuration of the segments in the initial DD microstructure. In reality, for macroscopic samples, it is generally observed that the response does not vary so much based on the state of the initial microstructure. Whether our simulations bear out this expectation for larger sample sizes needs to be explored. A difficulty associated with performing our simulations for large sample sizes is the computational expense. However, this is not a fundamental difficulty but a practical one, which can be addressed with more sophisticated parallelization algorithms and implementation than this first effort.

\end{itemize}
Immediate partial remedial measures for these limitations are expected to be the accounting of the mobile density in DD boxes in accord with the averaged dislocation density $\overline{\rho}$ (as discussed in Section \ref{sec:key_changes}) and the GND density $\overline{\bfalpha}$ suggested by MFDM, in the coupled DD-MFDM strategy. These descriptors will act as feedback for the initialization of the DD microstructure at discrete time steps. 

\begin{appendices}
\addcontentsline{toc}{section}{Appendices}
%%COUPLING
\section*{Appendix: Some details of crystallographic setup}\label{app:c2g}
Let an orthonormal basis (in which crystallographic vectors of a lattice are most easily represented in components) be called a  crystal basis $\{ \hat{\bfe}_i\}$. Let a  global orthonormal basis representing a laboratory frame be $\{ \bfe_i\}$.  The crystal-to-global (C2G) transformation matrix $A$ is defined as the transformation rule expressing components of any vector on the
global basis in terms of its components in the crystal basis. Thus, writing any vector $\bfv$ as $\bfv=c_i \hat{\bfe}_i= g_i \bfe_i$, we have  $g_k=(\bfe_k \cdot \hat{\bfe}_i)c_i $, and therefore, $A_{ki} = \bfe_k \cdot \hat{\bfe}_i$.  

Now suppose we do not have information on the crystal basis vectors but instead know a set of orthonormal crystallographic directions $\{ \bfc_j \}$ that coincide with the global basis vectors, i.e. $\bfc_j=\bfe_j$, where $\bfc_j=C_{ij} \hat{\bfe}_i$ \emph{and the $C_{ij}$ are known by hypothesis}. Then,
\begin{align*}
\bfe_j=\bfc_j \implies \delta_{mj}=(\bfe_m \cdot \hat{\bfe}_i) C_{ij} =A_{mi} C_{ij}  \implies C^{-1}_{mp}= A_{mp}.
\end{align*}
But the matrix $C$ has for  columns the components of an orthonormal basis expressed on the basis $\{  \hat{\bfe}_i  \}$. Thus, $C$ is an orthogonal matrix and its transpose is its inverse. Then, $A$ has as rows the components of the basis $\{ \bfc_j \}$ expressed in the crystal basis $\{ \hat{\bfe}_i \}$. 

In the symmetric double slip orientation used for the uniaxial tension in Section \ref{sec:uniaxial_tension}, the crystal is rotated such that the crystallographic direction $\frac{1}{\sqrt{2}} [0 \bar{1} 1]$ is along the global $X$ axis and the crystallographic direction $\frac{1}{\sqrt{6}} [2 1 1]$ is along the global $Y$ axis. The C2G transformation matrix corresponding to this orientation is 
\[
A=
\begin{bmatrix}
 0 & -\frac{1}{\sqrt 2} & \frac{1}{\sqrt 2} \\
 \frac{2}{\sqrt 6} & \frac{1}{\sqrt 6} & \frac{1}{\sqrt 6} \\
 -\frac{1}{\sqrt 3} & \frac{1}{\sqrt 3} & \frac{1}{\sqrt 3} 
\end{bmatrix}.
\]

The C2G transformation matrix corresponding to the simple shear orientation (as described in Section \ref{sec:simple_shear}) is 
\[
A=
\begin{bmatrix}
 0 & \frac{1}{\sqrt 2}  & \frac{1}{\sqrt 2} \\
 -\frac{1}{\sqrt 3} & \frac{1}{\sqrt 3} & -\frac{1}{\sqrt 3} \\
 -\frac{2}{\sqrt 6} & -\frac{1}{\sqrt 6} & \frac{1}{\sqrt 6} 
\end{bmatrix}.
\]

\end{appendices}

\section*{Acknowledgment}
Support from NSF grant NSF-CMMI-1435624 is gratefully acknowledged.

\bibliography{testbibliog}
\bibliographystyle{alpha}

\end{document}